\documentstyle[12pt,epsf,eqsection,subeqnarray,epic,eepicemu]{article}

\begin{document}

\bibliographystyle{plain}

\title{\vspace*{-2.6cm}Wolff-Type Embedding Algorithms for
  General Nonlinear $\sigma$-Models{}{}\footnote{Submitted to Nuclear
  Physics B.}}
\author{
  {\small Sergio Caracciolo}              \\[-0.2cm]
  {\small\it Scuola Normale Superiore and INFN -- Sezione di Pisa}  \\[-0.2cm]
  {\small\it Piazza dei Cavalieri}        \\[-0.2cm]
  {\small\it Pisa 56100, ITALIA}          \\[-0.2cm]
  {\small Internet: {\tt CARACCIO@UX1SNS.SNS.IT}}     \\[-0.2cm]
  {\small Bitnet:   {\tt CARACCIO@IPISNSVA.BITNET}}   \\[-0.2cm]
  {\small Hepnet/Decnet:   {\tt 39198::CARACCIOLO}}   \\[-0.2cm]
  \\[-0.1cm]  \and
  {\small Robert G. Edwards}              \\[-0.2cm]
  {\small\it Supercomputer Computations Research Institute}         \\[-0.2cm]
  {\small\it  Florida State University}   \\[-0.2cm]
  {\small\it Tallahassee, FL 32306 USA}   \\[-0.2cm]
  {\small Internet: {\tt EDWARDS@MAILER.SCRI.FSU.EDU}} \\[-0.2cm]
  \\[-0.1cm]  \and
  {\small Andrea Pelissetto}              \\[-0.2cm]
  {\small\it Dipartimento di Fisica}      \\[-0.2cm]
  {\small\it Universit\`a degli Studi di Pisa}        \\[-0.2cm]
  {\small\it Pisa 56100, ITALIA}          \\[-0.2cm]
  {\small Internet: {\tt PELISSET@SUNTHPI1.DIFI.UNIPI.IT}}   \\[-0.2cm]
  {\small Bitnet:   {\tt PELISSET@IPISNSVA.BITNET}}   \\[-0.2cm]
  {\small Hepnet/Decnet:   {\tt 39198::PELISSETTO}}   \\[-0.2cm]
  {\protect\makebox[5in]{\quad}}  
  \\[-0.1cm]  \and
  {\small Alan D. Sokal}                  \\[-0.2cm]
  {\small\it Department of Physics}       \\[-0.2cm]
  {\small\it New York University}         \\[-0.2cm]
  {\small\it 4 Washington Place}          \\[-0.2cm]
  {\small\it New York, NY 10003 USA}      \\[-0.2cm]
  {\small Internet: {\tt SOKAL@ACF3.NYU.EDU}}        \\[-0.2cm]
  {\protect\makebox[5in]{\quad}}  
  \\
}

\maketitle
\thispagestyle{empty}   

\begin{abstract}
We study a class of Monte Carlo algorithms for the nonlinear $\sigma$-model,
based on a Wolff-type embedding of Ising spins into the target manifold $M$.
We argue heuristically that,
at least for an asymptotically free model,
such an algorithm can have dynamic critical exponent $z \ll 2$
only if the embedding is based on an (involutive) isometry
of $M$ whose fixed-point manifold has codimension 1.  Such an isometry
exists only if the manifold is a discrete quotient of a product of spheres.
Numerical simulations of the idealized codimension-2 algorithm for the
two-dimensional $O(4)$-symmetric $\sigma$-model yield
$z_{int,{\cal M}^2} = 1.5 \pm 0.5$ (subjective 68\% confidence interval),
in agreement with our heuristic argument.
\end{abstract}

\clearpage

\newcommand{\be}{\begin{equation}}
\newcommand{\ee}{\end{equation}}
\newcommand{\<}{\langle}
\renewcommand{\>}{\rangle}
\newcommand{\para}{\|}
\renewcommand{\perp}{\bot}

\def\spose#1{\hbox to 0pt{#1\hss}}
\def\ltapprox{\mathrel{\spose{\lower 3pt\hbox{$\mathchar"218$}}
 \raise 2.0pt\hbox{$\mathchar"13C$}}}
\def\gtapprox{\mathrel{\spose{\lower 3pt\hbox{$\mathchar"218$}}
 \raise 2.0pt\hbox{$\mathchar"13E$}}}
\def\inapprox{\mathrel{\spose{\lower 3pt\hbox{$\mathchar"218$}}
 \raise 2.0pt\hbox{$\mathchar"232$}}}

\def\half{ {{1 \over 2 }}}
\def\scra{{\cal A}}
\def\scrc{{\cal C}}
\def\scre{{\cal E}}
\def\scrf{{\cal F}}
\def\scrm{{\cal M}}
\def\scrs{{\cal S}}
\def\scrt{{\cal T}}
\def\scrw{{\cal W}}
\def\tauss{\tau_{int,\,\scrm^2}}
\def\taux{\tau_{int,\,{\cal M}^2}}
\def\taue{\tau_{int,\,{\cal E}}}
\newcommand{\imag}{\mathop{\rm Im}\nolimits}
\newcommand{\real}{\mathop{\rm Re}\nolimits}
\newcommand{\tr}{\mathop{\rm tr}\nolimits}
\newcommand{\codim}{\mathop{\rm codim}\nolimits}
\def\textprime{{${}^\prime$}}
\newcommand{\longto}{\longrightarrow}
\def\var{ \hbox{var} }
\newcommand{\gtilde}{ {\widetilde{G}} }
\newcommand{\USp}{ \hbox{\it USp} }
\newcommand{\CP}{ \hbox{\it CP\/} }
\newcommand{\QP}{ \hbox{\it QP\/} }
\def\hboxscript#1{ {\hbox{\scriptsize\em #1}} }

\newcommand{\plotdot}{\makebox(0,0){$\bullet$}}
\newcommand{\plotsmalldot}{\makebox(0,0){{\footnotesize $\bullet$}}}

\def\bsigma{\mbox{\protect\boldmath $\sigma$}}
\def\btau{\mbox{\protect\boldmath $\tau$}}
\def\br{{\bf r}}

\newcommand{\reff}[1]{(\ref{#1})}

\font\specialroman=msym10 scaled\magstep1  
\font\sevenspecialroman=msym7              
\def\zed{\hbox{\specialroman Z}}
\def\szed{\hbox{\sevenspecialroman Z}}
\def\R{\hbox{\specialroman R}}
\def\sR{\hbox{\sevenspecialroman R}}
\def\N{\hbox{\specialroman N}}
\def\C{\hbox{\specialroman C}}
\def\Q{\hbox{\specialroman Q}}
\renewcommand{\emptyset}{\hbox{\specialroman ?}}

\font\german=eufm10 scaled\magstep1	
\def\germang{\hbox{\german g}}
\def\germansu{\hbox{\german su}}

\font\amssymbol=msxm10 scaled \magstep1  
\def\transversal{\hbox{\amssymbol t}}  

\newtheorem{theorem}{Theorem}[section]
\newtheorem{corollary}[theorem]{Corollary}
\newtheorem{lemma}[theorem]{Lemma}
\def\proof{\bigskip\par\noindent{\sc Proof.\ }}
\def\qed{\hbox{\hskip 6pt\vrule width6pt height7pt depth1pt \hskip1pt}\bigskip}

%
%
\newenvironment{sarray}{
          \textfont0=\scriptfont0
          \scriptfont0=\scriptscriptfont0
          \textfont1=\scriptfont1
          \scriptfont1=\scriptscriptfont1
          \textfont2=\scriptfont2
          \scriptfont2=\scriptscriptfont2
          \textfont3=\scriptfont3
          \scriptfont3=\scriptscriptfont3
        \renewcommand{\arraystretch}{0.7}
        \begin{array}{l}}{\end{array}}

\newenvironment{scarray}{
          \textfont0=\scriptfont0
          \scriptfont0=\scriptscriptfont0
          \textfont1=\scriptfont1
          \scriptfont1=\scriptscriptfont1
          \textfont2=\scriptfont2
          \scriptfont2=\scriptscriptfont2
          \textfont3=\scriptfont3
          \scriptfont3=\scriptscriptfont3
        \renewcommand{\arraystretch}{0.7}
        \begin{array}{c}}{\end{array}}

%
%

\section{Introduction}  \label{s1}

Wolff \cite{Wolff_89a,Wolff_89b,Wolff_89c} has recently
proposed an extraordinarily efficient collective-mode Monte Carlo algorithm
for simulating the nonlinear $\sigma$-model
taking values in the sphere $S^{N-1}$ and having symmetry group $O(N)$
[also called the $N$-vector model].
Numerical tests of the dynamic critical behavior of this algorithm
show the complete or almost complete absence of critical slowing-down
($z \ltapprox 0.1$)
for two-dimensional models with $N=2,3,4$
\cite{Wolff_89a,Wolff_89b,Wolff_89c,Edwards_89,Hasenbusch_89a},
and a small but apparently nonzero critical slowing-down
($z \approx 0.25-0.5$)
for the three-dimensional $XY$ model \cite{Hasenbusch_89b,Janke_90}.\footnote{%
But, regarding \cite{Hasenbusch_89b,Janke_90}, see our discussion in
Section \ref{s5} below.}

In this paper\footnote{A preliminary version of this work was reported at the
Lattice '90    conference \cite{CEPS_LAT90}.}

we consider generalizations of Wolff's algorithm
to $\sigma$-models taking values in manifolds other than spheres.
(We see this as a first step toward generalizing the Wolff algorithm
to lattice gauge theories.)
Our conclusion is somewhat surprising:
we argue that the generalized Wolff algorithm can work well
(i.e.\ have $z \ll 2$) {\em only}\/ if the manifold is a
discrete quotient of a product of spheres (e.g.\ real projective space).
This conclusion is based on a combination of a heuristic argument
and a rigorous mathematical theorem;
it is supported by a numerical test in one prototypical case.

Wolff's algorithm for the $N$-vector model
is based on an embedding of Ising spins $\{\varepsilon\}$
into $N$-component continuous spins $\{\bsigma\}$ according to
\be
  \bsigma_x   \;=\;   \bsigma_x^\perp  \,+\,
    \varepsilon_x |\bsigma_x^\para| \, \br     \;,
\label{embedding}
\ee
where $\br$ is a unit vector chosen randomly on $S^{N-1}$,
$\bsigma_x^\perp \,\equiv\, \bsigma_x - (\bsigma_x\cdot\br)\br$
and $\bsigma_x^\para \,\equiv\, (\bsigma_x \cdot \br)\br$
are the components of $\bsigma_x$ perpendicular and parallel to $\br$, and
$\varepsilon_x \equiv {\rm sgn}(\bsigma_x\cdot\br) = \pm 1$.
Flipping the Ising spin $\varepsilon_x$
corresponds to a reflection of $\bsigma_x$ in
the hyperplane perpendicular to $\br$.
With $\{\bsigma^\perp\}$ and $\{|\bsigma^\para|\}$ held fixed,
the $\sigma$-model Hamiltonian
\be
   H(\{\bsigma\})   \;=\;  - \beta \sum_{\< xy \>}  \bsigma_x \cdot \bsigma_y
\ee
reduces to the ferromagnetic random-bond Ising model
(in zero magnetic field) defined by
\be
  H(\{\varepsilon\})  \;=\;  - \sum_{\< xy \>}
    J_{xy}\varepsilon_x\varepsilon_y   \,+\,   {\rm const}     \;,
\label{Ising_ham}
\ee
where $J_{xy} = \beta |\bsigma_x\cdot\br| |\bsigma_y\cdot\br|$.
The Ising model \reff{Ising_ham} can then
be simulated by any legitimate Monte Carlo algorithm, such as the
the Swendsen-Wang (SW) algorithm \cite{Swendsen_87,Edwards_88} or
its single-cluster variant \cite{Wolff_89a,Wolff_89d,Tamayo_90,Baillie_91}.

The dynamic critical behavior of Wolff-type algorithms
is determined by the combined effect of two {\em completely distinct}\/ issues:
\begin{itemize}
 \item[i)]  How well the embedding \reff{embedding}
     succeeds in ``encoding'' the important large-scale collective modes
     of the $\sigma$-model into the Ising variables $\{\varepsilon\}$.
 \item[ii)]  How well any given algorithm for Ising models
     (e.g.\ standard SW or single-cluster SW)
     succeeds in updating the spins $\{\varepsilon\}$
     governed by the (random) Hamiltonian \reff{Ising_ham}.
\end{itemize}
We wish to emphasize the importance of studying these questions
{\em separately}\/.
If the physically relevant large-scale collective motions of the $\sigma$-model
cannot be obtained by varying the $\{\varepsilon\}$ at fixed
$\{\bsigma^\perp, \, |\bsigma^\para|\}$,
then the Wolff algorithm will have severe critical slowing-down
{\em no matter what}\/ method is used to update the Ising variables
$\{\varepsilon\}$.
On the other hand, if the Wolff algorithm with a {\em particular}\/ choice
of Ising-updating method shows severe critical slowing-down,
this does {\em not}\/ necessarily mean that the embedding \reff{embedding}
works badly:  the poor performance might be due to slow decorrelation in the
Ising-updating subroutine, and could possibly be remedied by switching to
a better Ising algorithm.

In this paper our goal is to study the {\em embedding}\/
defined by \reff{embedding} or its generalizations,
independently of the question of how the induced Ising model
is to be updated.
To address this question, it is conceptually useful to consider the
{\em idealized Wolff algorithm\/},
which for the $N$-vector model goes as follows:
a vector $\br$ is chosen randomly from the unit sphere $S^{N-1}$;
and a new configuration of Ising spins $\{\varepsilon\}$,
{\em independent of the old configuration\/},
is generated with probabilities given by
the Hamiltonian \reff{Ising_ham}.
Of course, such an algorithm is not practical,
but that is not its role.
Rather, it serves as a standard of comparison
(and presumed lower bound on the autocorrelation time)
for {\em all}\/ algorithms based on the embedding \reff{embedding}.
If the idealized Wolff algorithm performs badly, then so must any
algorithm based on the given embedding.  On the other hand, if the idealized
Wolff algorithm performs well, then it is clearly worthwhile to seek
(if necessary) new Ising-model algorithms capable of simulating efficiently
the induced Ising Hamiltonian.

To approximate in practice the idealized Wolff algorithm,
we update the $\{\varepsilon\}$ configuration
by $N_{hit}$ hits of some chosen Ising-model algorithm
(e.g.\ standard SW) and extrapolate to $N_{hit}=\infty$.\footnote{Preferably
we would perform simulations for successively increasing
values of $N_{hit}$ until the autocorrelation time is
constant within error bars.  However, this may not always be feasible.}
To be sure, this test procedure can be very time-consuming.
But it is essential if we wish to obtain {\em physical insight}\/
into the embedding.

The extraordinary performance of the Wolff algorithm for $N$-vector models
has spurred attempts (so far unsuccessful) to generalize it to lattice gauge
theories.  Gauge theories differ from $N$-vector models in two ways:
\begin{itemize}
   \item[a)]  The field takes values in a {\em group}\/ rather than a sphere.
       [$U(1)$ and $SU(2)$ are spheres, but higher Lie groups are not.]
   \item[b)]  The field is a 1-form rather than a 0-form, i.e.\ it lives
       on {\em links}\/ rather than sites.  Correspondingly, the energy
       is the {\em curl}\/ of the field rather than its gradient,
       and it lives on {\em plaquettes}\/ rather than links.
       As a result, the theory has a {\em local gauge invariance}\/
       rather than just a global symmetry.
\end{itemize}
The deep physical difference between gauge and spin models is, of course,
item (b).  The fact of gauge invariance, and the transverseness of
physical excitations in a gauge theory, will impose severe constraints,
we believe, on the as-yet-unknown analogue of the embedding \reff{embedding}
[if indeed such an analogue exists].\footnote{The same issue arises in devising
multi-grid algorithms for gauge theories \cite[Section V]{MGMC_1}.}
At present we have little to say in this direction
(though some insight might possibly be gleaned from the
Swendsen-Wang algorithm for Potts lattice gauge theories
\cite{Ben-Av_90,Brower_90}).
In this paper we address the less profound, but still highly nontrivial,
problem (a).  To do this, we ask whether the embedding \reff{embedding}
can be generalized to nonlinear $\sigma$-models with values in
{\em manifolds other than spheres}\/ --- such as $SU(N)$ for $N \ge 3$ ---
and, if so, what is the dynamic critical behavior of the corresponding
idealized Wolff algorithm.

Our approach is as follows:  First we ask what are the fundamental
properties of the embedding \reff{embedding} that cause the Wolff algorithm
to work so well.  Then we ask whether embeddings having these properties
exist also in other Riemannian manifolds $M$;  this is a question in
differential geometry to which we are able to give a fairly complete answer.
Finally, we perform a numerical study to test (in one case)
whether our theoretical reasoning is correct.
The conclusion of this analysis is quite surprising:
roughly speaking, we find that a generalized Wolff algorithm can work well
(i.e.\ have $z \ll 2$) {\em only}\/ if the manifold $M$
is either a sphere, a product of spheres,
or the quotient of such a space by a discrete group
(for example, real projective space $RP^{N-1}$).
If correct, this conclusion is quite disappointing,
and lends renewed impetus to other classes of collective-mode algorithms
such as multi-grid Monte Carlo \cite{MGMC_PRL,MGMC_1,MGMC_XY,MGMC_O4}
and Fourier acceleration \cite{Parisi_84,Batrouni_85,Dagotto_87a,Dagotto_87b}.

The plan of this paper is as follows:
In Section \ref{s2} we define generalized Wolff-type embedding algorithms
for $\sigma$-models taking values in a Riemannian manifold $M$.
We argue heuristically that such an algorithm will work well
--- at least for asymptotically free $\sigma$-models ---
only if the embedding is based on an (involutive) isometry of $M$
whose fixed-point manifold has codimension 1.
This argument is the key physical idea of this paper.
In Section \ref{s3} we study the conditions under which
such an isometry can exist.
In Section \ref{s4} we test our heuristic argument with a large-scale
numerical study of the codimension-2 idealized embedding algorithm
for the two-dimensional $N$-vector model with $N=4$.
In Section \ref{s5} we discuss our results.
Appendix \ref{appendix_A} collects some results
from topology and differential geometry that are essential to our argument
in Sections \ref{s2} and \ref{s3}.
Some of these results are difficult to find in the mathematical literature,
and at least a few appear to be new.
Appendix \ref{appendix_B} contains a complete classification of the
involutive isometries for the physically important case of $SU(N)$.
Much to our surprise, we have been unable to find this classification
anywhere in the mathematical or physical literature.

\section{Generalized Embedding Algorithms}   \label{s2}
\subsection{Wolff algorithm for the $N$-vector model}   \label{s2.1}

The single-spin configuration space of the $N$-vector model is the sphere
\be
   S^{N-1}   \;=\;   \{ \bsigma \in \R^N \colon\;  |\bsigma| = 1 \}  \;,
\ee
and the key object of the Wolff algorithm is the map
of {\em reflection in the equator}\/:
\be
   T(\sigma^{1},\sigma^{2},\ldots,\sigma^{N})  \;=\;
   (-\sigma^{1},\sigma^{2},\ldots,\sigma^{N})  \;.
 \label{eqwolff2.2}
\ee
(This corresponds to $\br = (1,0,\ldots,0)$; obviously the case
of arbitrary $\br$ can be obtained by conjugating with a rotation.)
The map $T$ has two key properties:
\begin{itemize}
   \item[a)]  $T$ is an isometry (i.e.\ it preserves distances).
   \item[b)]  The fixed-point set
     ${\rm Fix}(T) \equiv \{ \bsigma\in S^{N-1} \colon\; T\bsigma=\bsigma \}$
     --- which is of course the equator --- has codimension 1.
\end{itemize}
We argue that these are the crucial properties that make the Wolff algorithm
work so well.

Because $T$ is an {\em isometry}\/,
a global application of $T$ is an exact symmetry (zero change in energy),
and the application of $T$ in a large but bounded region
costs only a {\em surface}\/ energy.
(If $T$ were not an isometry, then the application of $T$ in a region
would cost a {\em bulk}\/ energy, and the idealized Wolff update
would probably be unable to make any significant large-scale change in the
$\{\bsigma\}$ configuration.)

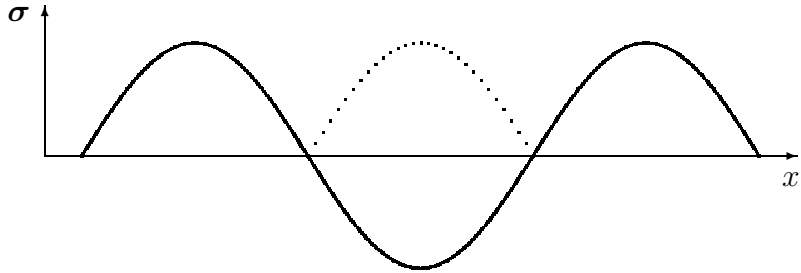
\begin{figure}
\unitlength = 1.0mm
\begin{center}
\begin{normalsize}
\begin{picture}(90,30)(0,0)

\thinlines
\put(-5,15){\vector(1,0){100}}
\put(93,11){$x$}
\put(-5,15){\vector(0,1){20}}
\put(-10,33){$\bsigma$}

\thicklines
\begin{drawjoin}
\jput(0.000000,15.000000){\ }
\jput(1.000000,16.567926){\ }
\jput(2.000000,18.118675){\ }
\jput(3.000000,19.635254){\ }
\jput(4.000000,21.101049){\ }
\jput(5.000000,22.500000){\ }
\jput(6.000000,23.816778){\ }
\jput(7.000000,25.036959){\ }
\jput(8.000000,26.147173){\ }
\jput(9.000000,27.135254){\ }
\jput(10.000000,27.990381){\ }
\jput(11.000000,28.703182){\ }
\jput(12.000000,29.265848){\ }
\jput(13.000000,29.672215){\ }
\jput(14.000000,29.917828){\ }
\jput(15.000000,30.000000){\ }
\jput(16.000000,29.917828){\ }
\jput(17.000000,29.672215){\ }
\jput(18.000000,29.265848){\ }
\jput(19.000000,28.703182){\ }
\jput(20.000000,27.990381){\ }
\jput(21.000000,27.135256){\ }
\jput(22.000000,26.147173){\ }
\jput(23.000000,25.036961){\ }
\jput(24.000000,23.816780){\ }
\jput(25.000000,22.500002){\ }
\jput(26.000000,21.101051){\ }
\jput(27.000000,19.635256){\ }
\jput(28.000000,18.118677){\ }
\jput(29.000000,16.567928){\ }
\jput(30.000000,15.000002){\ }
\jput(31.000000,13.432076){\ }
\jput(32.000000,11.881327){\ }
\jput(33.000000,10.364747){\ }
\jput(34.000000,8.898952){\ }
\jput(35.000000,7.500002){\ }
\jput(36.000000,6.183223){\ }
\jput(37.000000,4.963043){\ }
\jput(38.000000,3.852829){\ }
\jput(39.000000,2.864747){\ }
\jput(40.000000,2.009620){\ }
\jput(41.000000,1.296819){\ }
\jput(42.000000,0.734153){\ }
\jput(43.000000,0.327787){\ }
\jput(44.000000,0.082172){\ }
\jput(45.000000,0.000000){\ }
\jput(46.000000,0.082171){\ }
\jput(47.000000,0.327785){\ }
\jput(48.000000,0.734151){\ }
\jput(49.000000,1.296817){\ }
\jput(50.000000,2.009617){\ }
\jput(51.000000,2.864743){\ }
\jput(52.000000,3.852825){\ }
\jput(53.000000,4.963038){\ }
\jput(54.000000,6.183218){\ }
\jput(55.000000,7.499997){\ }
\jput(56.000000,8.898947){\ }
\jput(57.000000,10.364741){\ }
\jput(58.000000,11.881321){\ }
\jput(59.000000,13.432069){\ }
\jput(60.000000,14.999996){\ }
\jput(61.000000,16.567923){\ }
\jput(62.000000,18.118671){\ }
\jput(63.000000,19.635250){\ }
\jput(64.000000,21.101046){\ }
\jput(65.000000,22.499996){\ }
\jput(66.000000,23.816774){\ }
\jput(67.000000,25.036955){\ }
\jput(68.000000,26.147169){\ }
\jput(69.000000,27.135252){\ }
\jput(70.000000,27.990379){\ }
\jput(71.000000,28.703180){\ }
\jput(72.000000,29.265846){\ }
\jput(73.000000,29.672213){\ }
\jput(74.000000,29.917828){\ }
\jput(75.000000,30.000000){\ }
\jput(76.000000,29.917830){\ }
\jput(77.000000,29.672215){\ }
\jput(78.000000,29.265850){\ }
\jput(79.000000,28.703184){\ }
\jput(80.000000,27.990383){\ }
\jput(81.000000,27.135258){\ }
\jput(82.000000,26.147177){\ }
\jput(83.000000,25.036963){\ }
\jput(84.000000,23.816784){\ }
\jput(85.000000,22.500006){\ }
\jput(86.000000,21.101055){\ }
\jput(87.000000,19.635260){\ }
\jput(88.000000,18.118681){\ }
\jput(89.000000,16.567932){\ }
\jput(90.000000,15.0){\ }
\end{drawjoin}
%
\jput(30.000000,15.000000){\picsquare}
\jput(31.000000,16.567924){\picsquare}
\jput(32.000000,18.118673){\picsquare}
\jput(33.000000,19.635252){\picsquare}
\jput(34.000000,21.101048){\picsquare}
\jput(35.000000,22.499998){\picsquare}
\jput(36.000000,23.816776){\picsquare}
\jput(37.000000,25.036957){\picsquare}
\jput(38.000000,26.147171){\picsquare}
\jput(39.000000,27.135254){\picsquare}
\jput(40.000000,27.990379){\picsquare}
\jput(41.000000,28.703180){\picsquare}
\jput(42.000000,29.265846){\picsquare}
\jput(43.000000,29.672213){\picsquare}
\jput(44.000000,29.917828){\picsquare}
\jput(45.000000,30.000000){\picsquare}
\jput(46.000000,29.917830){\picsquare}
\jput(47.000000,29.672215){\picsquare}
\jput(48.000000,29.265848){\picsquare}
\jput(49.000000,28.703184){\picsquare}
\jput(50.000000,27.990383){\picsquare}
\jput(51.000000,27.135258){\picsquare}
\jput(52.000000,26.147175){\picsquare}
\jput(53.000000,25.036963){\picsquare}
\jput(54.000000,23.816782){\picsquare}
\jput(55.000000,22.500004){\picsquare}
\jput(56.000000,21.101053){\picsquare}
\jput(57.000000,19.635258){\picsquare}
\jput(58.000000,18.118679){\picsquare}
\jput(59.000000,16.567930){\picsquare}
\jput(60.000000,15.0){\picsquare}
\end{picture}
\end{normalsize}
\end{center}
\caption{
   Action of the Wolff algorithm on a long-wavelength spin wave.
   For simplicity, both spin space ($\bsigma$) and physical space ($x$)
   are depicted as one-dimensional.
}
\label{swwon_figure_a}
\end{figure}

To see the importance of the {\em codimension-1 property}\/,
let us review our heuristic understanding \cite{Edwards_89}
of why the Wolff algorithm works so well.
Consider a slowly varying spin configuration $\{\bsigma\}$.
Since $J_{xy} = \beta |\bsigma_x\cdot\br| |\bsigma_y\cdot\br|$,
the Hamiltonian \reff{Ising_ham} {\em almost decouples}\/ along the surfaces
where $\bsigma\cdot\br \approx 0$,
i.e.\ where $\bsigma$ is on or near the equator.
These surfaces (= codimension-1 submanifolds)
divide $x$-space into disconnected regions $R_i$ in which
$\bsigma\cdot\br > 0$ or $\bsigma\cdot\br < 0$.
Provided that we are working in an {\em non-magnetized}\/ phase
(i.e.\ there is no preferred orientation for the spins),
the largest of the regions $R_i$ will presumably have linear size on
the order of the correlation length $\xi$.
In particular, if $1 \ll \xi \ltapprox L/2$
(here $L$ is the lattice linear size),
then there will be, with high probability,
at least two disjoint large regions $R_i$.
Now let us apply an an idealized Wolff update:
the spins $\{\varepsilon\}$ will be given new values, and these values will be
{\em almost independent}\/ in distinct regions $R_i$.
That is, the choice to reflect or not reflect a spin $\bsigma$
will have been made almost independently in distinct regions $R_i$
(but very coherently within each $R_i$).
If the original configuration $\{\bsigma\}$ is a long-wavelength spin wave,
then these almost-independent reflections correspond roughly to a
long-wavelength collective-mode change in $\{ \bsigma \}$
(Figure \ref{swwon_figure_a}).
We predict, therefore, that the {\em idealized}\/ Wolff algorithm will
have dynamic critical exponent $z$ much less than 2,
and quite possibly $z \approx 0$.
This prediction has been confirmed numerically
in the two-dimensional $N$-vector models with $N=2,3,4$
\cite{Wolff_89a,Wolff_89b,Wolff_89c,Edwards_89,Hasenbusch_89a},
and in the three-dimensional model with $N=2$
\cite{Hasenbusch_89b,Janke_90}.\footnote{Regarding
\cite{Hasenbusch_89b,Janke_90}, see our discussion in
   Section \ref{s5} below.}

%
%

The assumption in this argument that $\{\bsigma\}$ is slowly varying
requires some clarifi\-ca\-tion.\footnote{ We did not fully understand
these subtleties when we wrote   \cite{Edwards_89,CEPS_LAT90}.
(Quite possibly we still don't.)}
Consider first the case in which the critical point is at zero temperature
($\beta_c = \infty$);  this case corresponds to asymptotically free theories
such as the two-dimensional $N$-vector models for $N \ge 3$.
For such theories, the typical spin configurations become
smooth as $\beta \to \beta_c = \infty$:
more precisely, for a nearest-neighbor pair of sites $x,y$,
we have $\< \bsigma_x \cdot \bsigma_y \> \approx 1 - c/\beta$
for a (calculable) constant $c$.
That is, the typical angle between nearest-neighbor spins is of order
$1/\sqrt{\beta}$;  so on bonds $\< xy \>$ where $\bsigma_x$ and $\bsigma_y$
are near the equator, we typically have
$J_{xy} \equiv \beta |\bsigma_x\cdot\br| |\bsigma_y\cdot\br|
 \sim \beta (1/\sqrt{\beta})^2 \sim 1$.
Note that this behavior occurs not only on the bonds $\< xy \>$ where $\bsigma$
{\em crosses}\/ the equator
[i.e.\ for which $(\bsigma_x\cdot\br) (\bsigma_y\cdot\br) < 0$],
but also on the nearby bonds where $\bsigma$ remains {\em near}\/
the equator.
On the other hand, on bonds $\< xy \>$ where $\bsigma_x$ and $\bsigma_y$
are far from the equator (i.e.\ bonds deep inside one of the regions $R_i$),
we typically have $J_{xy} \sim \beta$.
Thus, the couplings $\{ J_{xy} \}$ of the induced Ising model have typically
the following structure:  there are regions $R_i$ of linear size $\sim \xi$
within which the couplings are very strong (of order $\beta$),
surrounded by transition layers (of width probably $\sim 1$)
where the couplings are of moderate strength (of order 1).

Now, from this information alone we cannot determine the phase structure
of the induced Ising model.  For example, if the couplings $\{ J_{xy} \}$
in the transition layers --- which are ``of order 1'' ---
were typically greater than the critical coupling $J_\hboxscript{c,Ising}$
for a $d$-dimensional translation-invariant Ising ferromagnet,
then the induced Ising model would surely exhibit long-range order.
(The strongly ferromagnetic couplings $J_{xy} \sim \beta$ {\em within}\/
the regions $R_i$ would only strengthen this long-range order.)
On the other hand, if the transition-layer couplings were sufficiently
smaller than $J_\hboxscript{c,Ising}$,
then the induced Ising model could exhibit
exponentially-decaying correlations in spite of the large couplings
$J_{xy} \sim \beta$ within the regions $R_i$.  The distinction between
these two scenarios seems to depend on the precise values of the
transition-layer couplings $\{ J_{xy} \}$, as well as on the widths
of the transition layers --- quantities which our order-of-magnitude
argument is too crude to predict.

On the other hand, we know that the original $N$-vector model is in
a phase with exponentially decaying correlations;  in particular,
we expect that
\be
 \label{eq2.1.star1}
   \< {\rm sgn}(\bsigma_x \cdot \br) \, {\rm sgn}(\bsigma_y \cdot \br) \>
   \;\sim\;   e^{-|x-y|/\xi}
\ee
as $|x-y| \to\infty$.  [This is not the usual correlation function
$\< \bsigma_x \cdot \bsigma_y \>$, but it should exhibit the same
correlation length.]
On the other hand, the unconditional expectation \reff{eq2.1.star1}
is equal to the average over $\{\bsigma^\perp, \, |\bsigma^\para|\}$
(with Boltzmann-Gibbs weight) of the {\em conditional}\/ expectations
\be
 \label{eq2.1.star2}
   \< {\rm sgn}(\bsigma_x \cdot \br) \, {\rm sgn}(\bsigma_y \cdot \br)
   \,|\, \{\bsigma^\perp, |\bsigma^\para|\}   \>
   \;\equiv\;   \< \varepsilon_x \varepsilon_y \> _{ \{ J_{xy} \} }
   \;.
\ee
Now, the conditional expectations \reff{eq2.1.star2} are always
{\em nonnegative}\/, since the induced Ising model is purely ferromagnetic.
So there is no possibility of ``cancellations'' in forming
\reff{eq2.1.star1} from \reff{eq2.1.star2}.
We conclude that the induced-Ising-model correlations \reff{eq2.1.star2}
should also exhibit exponential (or faster-than-exponential) decay,
on a characteristic scale $\xi$ (or smaller),
for ``nearly all'' configurations $\{\bsigma^\perp, \, |\bsigma^\para|\}$.

This exponential decay is clearly not due to the large couplings
$J_{xy} \sim \beta$ within the regions $R_i$;
it must be due to the moderate couplings $J_{xy} \sim 1$
in the transition layers.  In other words, the transition layers {\em do}\/
succeed in (almost-)decoupling neighboring (or at least almost-neighboring)
regions $R_i$.  This reasoning justifies, in the case of an asymptotically
free model, our proposed mechanism for the good dynamic critical behavior
($z \approx 0$) of the {\em idealized}\/ Wolff algorithm.

Next consider the case of a critical point at finite temperature
($\beta_c < \infty$), as in the two-dimensional $XY$ model or
the $N$-vector models (all $N$) in dimension $d \ge 3$.
Here the spin configurations for $\beta \approx \beta_c$
are {\em rough}\/ at short distance scales:
for a nearest-neighbor pair of sites $x,y$, we have
$\< \bsigma_x \cdot \bsigma_y \> \le C < 1$
uniformly in $\beta$ ($\approx \beta_c$) and $L$.
Thus, the distinction between ``regions $R_i$'' and ``transition layers''
is no longer so clear.  The couplings $\{ J_{xy} \}$ will be highly irregular,
but we may expect $J_{xy} \sim \beta_c \sim 1$ more or less everywhere.
On the other hand, for $\beta < \beta_c$ it remains true that the
correlations \reff{eq2.1.star1} decay exponentially;
so we may again conclude that the induced-Ising-model correlations
\reff{eq2.1.star2} decay exponentially on scale $\xi$ for
``nearly all'' $\{\bsigma^\perp, \, |\bsigma^\para|\}$.
In this case, the {\em mechanism}\/ of the decay is not the decoupling
of regions $R_i$, but simply the fact that the ``average'' $J_{xy}$
is less than $J_\hboxscript{c,Ising}$.  Irrespective of the mechanism, however,
it remains true that an {\em independent}\/ resampling of the
induced Ising model will produce a significant collective change on
scale $\xi$ (and all smaller scales).  We may thus continue to expect
that $z \approx 0$ for the {\em idealized}\/ Wolff algorithm.

\medskip

{\em Remarks.}\/
1.  For the two-dimensional $XY$ model near the Kosterlitz-Thouless
transition, spin waves are not the only important large-scale
collective modes:  there are also large-scale collective modes associated
with vortices.  Therefore, to explain heuristically why $z \ll 2$,
it is necessary to show also that the (idealized) Wolff updates are capable of
destroying a widely-separated vortex-antivortex pair.
This has been done in \cite{Edwards_89}.

2.  The foregoing reasoning does not apply to a {\em magnetized}\/ phase
(except in the scaling region near the critical point, where the
magnetization is very small):
the trouble is that the spins will be aligned in some particular
direction\footnote{In finite volume with free or periodic boundary conditions,
   this direction of alignment is {\em random}\/, and will vary
   from configuration to configuration.  Nevertheless, in each
   individual configuration, the spins will be rather strongly aligned
   in {\em some}\/ direction.},
and nearly all the spins will lie in a single cluster $R_i$
unless the direction of alignment happens to be almost perpendicular to $\br$.
Therefore, in order for the Wolff algorithm to work well in a magnetized phase,
it may be necessary to choose the direction vector $\br$
with a probability distribution that
{\em depends on the current spin configuration $\{ \bsigma \}$}\/.
For example, one possible algorithm would be to choose at random a site $x$,
and then choose $\br$ to be a random vector orthogonal to $\bsigma_x$.
Since $\bsigma_x$ will be unchanged under the subsequent move,
this algorithm is easily seen to satisfy detailed balance.\footnote{The
   same logic also applies if $\br$ is chosen {\em parallel}\/ to
   $\bsigma_x$, as $\bsigma_x$ will be unchanged up to a sign
   (and a change of sign of $\br$ is irrelevant).
   However, we do not know of any useful application for this algorithm.}
(If $N=2$, this choice of $\br$ given $\bsigma_x$ is deterministic,
and the algorithm is nonergodic:  for example, if the spins $\bsigma_x$ are
all aligned along one axis, they will forever remain so.  The ergodicity
can be restored by admixing, with a nonzero probability, moves of the
usual type with a randomly chosen $\br$.)
The hope here is that $\br$ would most of the time be chosen approximately
perpendicular to the current total magnetization
${\cal M} = \sum_x \bsigma_x$,
so that the Goldstone spin waves would be sliced as in
Figure \ref{swwon_figure_a}.
(One might inquire about the algorithm in which $\br$ is chosen to be
perpendicular to ${\cal M}$ itself.  Unfortunately,
this algorithm fails to satisfy detailed balance, and we strongly suspect
that it fails to leave invariant the correct Gibbs measure.)

3.  In a preliminary version of this work \cite{CEPS_LAT90},
we emphasized an additional property of $T$, namely that it is
{\em involutive}\/ (i.e.\ its square is the identity map).
This property ensures that the induced Ising model has zero magnetic field
(which is convenient although perhaps not essential).
In fact, it turns out that an isometry possessing the codimension-1 property
is {\em automatically}\/ involutive:
see Theorem \ref{thmA.9} in Appendix \ref{appendix_A}.

\bigskip

Next we would like to argue that the {\em practical}\/ Wolff algorithm
based on standard SW (or single-cluster SW) updates of the induced
Ising model (with $N_{hit} = 1$) should also have $z \ll 2$.
We can give two alternative lines of reasoning:

1)  We have already argued that the idealized Wolff algorithm has
$z \approx 0$.  On the other hand, the induced Ising model is ferromagnetic,
and the SW (or 1CSW) algorithm is known to perform well for Ising
ferromagnets:  the exponential autocorrelation time behaves as
$\tau_{exp,SW} \sim \min(\xi,L)^{z_{exp,SW}}$,
where the best currently available estimates of $z_{exp,SW}$
(standard SW algorithm) for translation-invariant nearest-neighbor
ferromagnets are
\be
   z_{exp,SW}   \;\approx\;
   \cases{
      0 - 0.3     & for $d=2$
 \protect\cite{Swendsen_87,Wolff_89d,Heermann_90,Baillie_91,Coddington_private}
                                                                            \cr
      0.35 - 0.75 & for $d=3$
 \protect\cite{Swendsen_87,Wolff_89d,Wang_90,Coddington_private}            \cr
      1           & for $d=4$  \protect\cite{Klein_89,Coddington_private}   \cr
         }
\ee
Of course, our induced Ising model is a {\em non-translation-invariant}\/
ferromagnet, but the performance is likely to be qualitatively similar.
Now, $N_{hit}$ hits of the SW algorithm are at least
$1 - e^{-N_{hit}/\tau_{exp,SW}}$ times as effective as independent sampling
in generating a ``new'' Ising configuration.
Thus, we expect
\begin{subeqnarray}
   \tau_\hboxscript{Wolff,practical}   & \ltapprox &
   \tau_\hboxscript{Wolff,idealized} / (1 - e^{-N_{hit}/\tau_{exp,SW}})   \\
       & \sim &   \tau_\hboxscript{Wolff,idealized} \tau_{exp,SW} / N_{hit}
                  \qquad \hbox{if }   N_{hit} \ltapprox \tau_{exp,SW}
\end{subeqnarray}
--- hence $z_\hboxscript{Wolff,practical} \le
           z_\hboxscript{Wolff,idealized} + z_{exp,SW} \ll 2$.

2)  Let us consider the bond configuration that will be generated
in the {\em first}\/ SW (or 1CSW) update.
If the nearest-neighbor spins $\bsigma_x$ and $\bsigma_y$
belong to opposite hemispheres ---
that is, if $(\bsigma_x \cdot \br) (\bsigma_y \cdot \br)  \le 0$ ---
then the bond $\< xy \>$ is guaranteed to be unoccupied.
It follows that if $\bsigma_z$ and $\bsigma_{z'}$ are {\em any}\/
pair of spins belonging to opposite hemispheres,
then the sites $z$ and $z'$ are guaranteed to belong to {\em different}\/
SW clusters.
On the other hand, {\em within}\/ each connected region $R_i$
consisting of spins belonging to the same hemisphere,
we expect that the bond occupation probabilities
$p_{xy} \equiv 1 - e^{-2 J_{xy}}$
will mostly lie beyond the percolation threshold, so that most of the
region $R_i$ will belong to the {\em same}\/ SW cluster.
[This is clear in the asymptotically free case, where
$J_{xy} \sim \beta$ and $p_{xy} \approx 1$.
In the finite-$\beta_c$ case, however, the $J_{xy}$ are of order 1,
and there is a nonzero probability (uniformly in $\beta \approx \beta_c$
and $L$) for a small cluster of spins to become detached from its
surroundings.  Thus, each region $R_i$ would have the structure
of Swiss cheese:  there would be one or a few large SW clusters
in which there live a nonzero density of small SW clusters.
Nevertheless, it is reasonable to expect that the large clusters
will still be of linear size $\sim \xi$, i.e.\ that the small clusters
will not be so numerous as to disconnect the large clusters into pieces
of size $\ll \xi$.]
Thus, each region $R_i$ will be reflected more or less coherently
--- but distinct regions $R_i$ will be reflected {\em independently}\/ ---
in the first SW update.
This corresponds to a long-wavelength collective mode
(Figure \ref{swwon_figure_a}).

\subsection{Wolff algorithm for the $RP^{N-1}$ model}  \label{s2.2}

Let us now consider a second example of a Wolff-type embedding algorithm,
namely the case of the nonlinear $\sigma$-model taking values in the
real projective space $RP^{N-1}$, where $N \ge 3$.
This example differs from the $N$-vector model in two principal ways:

1)  The equator in $S^{N-1}$ not only has codimension 1, but has the
stronger property of dividing the sphere into two {\em disconnected}\/
regions (the northern and southern hemispheres).  That is,
\begin{itemize}
  \item[b\textprime)]    The complement of ${\rm Fix}(T)$ is disconnected.
\end{itemize}
This property implies the codimension-1 property, but is strictly stronger
(see Theorems \ref{thmA.1} and \ref{thmA.8} in Appendix \ref{appendix_A}).
It is not clear, at first glance, whether our heuristic argument
in the preceding section was based on (b) or on (b\textprime).
Analysis of the $RP^{N-1}$ model, which satisfies (b) but not (b\textprime),
will clarify this issue.

2)  In the Wolff embedding algorithm for the $N$-vector model, the induced
Ising model is {\em unfrustrated}\/.
[When written in the form \reff{embedding}/\reff{Ising_ham},
it is manifestly ferromagnetic;  when written in the alternate form
\reff{embedding_gen}--\reff{Ising_ham_gen2}, it is equivalent
via a $Z_2$ gauge transformation to a ferromagnetic model.]
In the embedding algorithm for the $RP^{N-1}$ model, by contrast,
the induced Ising model is generically {\em frustrated}\/.
It is important to know what difference, if any, this makes.

The real projective space $RP^{N-1}$
is, by definition, the sphere $S^{N-1}$ with antipodal
points identified, i.e.\ $RP^{N-1} = S^{N-1} / Z_2$.
One could now proceed to pick a parametrization of $RP^{N-1}$
(e.g.\ by embedding it diffeomorphically
 as a submanifold of some $\R^m$ \cite{James_71});
but it is simpler and more convenient to consider instead spins taking values
in the sphere $S^{N-1}$, subject to the condition that the Hamiltonian
and all physical observables must be invariant under the $Z_2$ local
gauge transformation $\bsigma_x \to \eta_x \bsigma_x$ ($\eta_x = \pm 1$).
The simplest lattice Hamiltonian for this model is therefore
\be
  H(\{\bsigma\})   \;=\;
    -\, {\beta \over 2} \sum_{\< xy \>} (\bsigma_x \cdot \bsigma_y)^2
  \;.
\ee
We use the same embedding as before, namely \reff{embedding},
but the coefficients $\{ J_{xy} \}$ in the
induced Ising Hamiltonian \reff{Ising_ham} are now
\be
  J_{xy}   \;=\;  \beta (\bsigma_x^\perp \cdot \bsigma_y^\perp)
                      |\bsigma_x\cdot\br| |\bsigma_y\cdot\br|
  \;.
 \label{Ising_ham_RP}
\ee
(For $N \ge 3$, this is generically a {\em frustrated}\/ Ising model.)
Since the reflection in the equator on $S^{N-1}$ commutes with the map
$\bsigma \to -\bsigma$, this reflection also induces a well-defined map $T$
on the quotient space $RP^{N-1}$.
Its fixed-point set ${\rm Fix}(T)$
is the union of two disjoint components:
one component consists of a single point, namely the pole
($\equiv$ the image of the north or south pole in $S^{N-1}$
under the canonical projection $\pi\colon\; S^{N-1} \to RP^{N-1}$);
the other component consists of the ``equator'' in $RP^{N-1}$
($\equiv$ the image of the equator in $S^{N-1}$
under the canonical projection $\pi$),
which is isometric to $RP^{N-2}$.
Note that the equator in $RP^{N-1}$ has codimension 1,
but (unlike what happens in $S^{N-1}$)
it does {\em not}\/ disconnect the space $RP^{N-1}$:
this is because in $RP^{N-1}$ there is no distinction between
the northern and southern hemispheres.
Therefore, the map $T$ on $RP^{N-1}$ satisfies
property (b) but {\em not}\/ (b\textprime).

We now wish to analyze the behavior of the idealized Wolff algorithm
for the $RP^{N-1}$ model,
by heuristic arguments analogous to those used for the $N$-vector model.
Consider, therefore, a slowly varying spin configuration $\{ \bsigma \}$.
For such a configuration, we can imagine the lattice $\zed^d$
to be replaced by a continuous space $\R^d$,
and we can imagine the configuration $\{ \bsigma \}$ to be a smooth
map from $\R^d$ into $RP^{N-1}$.
Then the set of points in $x$-space where $\bsigma\cdot\br = 0$
is again (generically) a codimension-1 submanifold (= hypersurface) of $\R^d$
(see Theorems \ref{thmA.4} and \ref{thmA.5}).
Since $\R^d$ is simply connected, this hypersurface divides $\R^d$ into
at least two disconnected regions $R_i$ (Theorem \ref{thmA.2}).
[{\em Alternative argument:}\/  Since $\R^d$ is simply connected,
the map $\bsigma\colon\; \R^d \to RP^{N-1}$
lifts to a continuous map $\widetilde{\bsigma} \colon\; \R^d \to S^{N-1}$
satisfying $\bsigma = \pi \circ \widetilde{\bsigma}$.
Now make the same argument as for the $N$-vector model,
but using $\widetilde{\bsigma}$ in place of $\bsigma$.]
Therefore, by the same arguments as in the $N$-vector case,
the induced Ising Hamiltonian \reff{Ising_ham_RP}
almost decouples along the surfaces where $\bsigma$
(or equivalently $\widetilde{\bsigma}$) lies on or near the equator,
and under an idealized Wolff update the regions $R_i$ will be updated
almost independently.
We thus predict that the {\em idealized}\/ Wolff algorithm for the
$RP^{N-1}$ model (in a non-magnetized phase)
will have dynamic critical exponent $z \ll 2$,
and quite possibly $z \approx 0$.

{\em Remark.}\/
If $x$-space is not simply connected (e.g.\ $X$ is the torus $T^d$ if
periodic boundary conditions are used), then the foregoing argument
fails:  a single hypersurface need not disconnect $X$.  However, {\em a
few}\/ hypersurfaces suffice to disconnect $X$ (see Theorem
\ref{thmA.3}); and if $1 \ll \xi \ltapprox L/4$, then it is natural to
expect that, with reasonable probability, the set where
$\bsigma\cdot\br = 0$ will have {\em several}\/ connected components.
So we expect the algorithm to have the same dynamic critical exponent
irrespective of the boundary conditions used.  At worst, the
autocorrelation time might be larger by a {\em constant factor}\/ for
periodic as compared with free boundary conditions, for the same
lattice size.

As in the $N$-vector case, it is necessary to inquire more closely
into the assumption that $\{\bsigma\}$ is slowly varying.
Consider, as before, the two cases:

\medskip

(a) {\em The critical point is at zero temperature.}\/
This case corresponds to asymptotically free theories;
low-temperature perturbation theory suggests that
the two-dimensional $RP^{N-1}$ models for $N \ge 3$
are asymptotically free, but the Monte Carlo evidence is ambiguous
(see \cite{CEPS_LAT91} for references).
As $\beta \to \beta_c = \infty$,
the typical spin configurations become
smooth modulo a sign:
for a nearest-neighbor pair of sites $x,y$,
we have $\< (\bsigma_x \cdot \bsigma_y)^2 \> \approx 1 - c/\beta$
for a (calculable) constant $c$.
We now claim that the couplings $\{ J_{xy} \}$ have the following properties:
\begin{itemize}
  \item[(i)]  In regions where $\bsigma$ is not near the equator or the pole
     (i.e.\ deep inside one of the regions $R_i$),
     the couplings $\{ J_{xy} \}$ are of order $\beta$ and
     {\em globally unfrustrated}\/.
  \item[(ii)]  In regions where $\bsigma$ is near the equator
     (i.e.\ the transition layers between the regions $R_i$),
     the couplings $\{ J_{xy} \}$ are of order 1 and
     {\em locally unfrustrated}\/, though they
     may be {\em globally frustrated}\/.
  \item[(iii)]  In regions where $\bsigma$ is near the pole
     (these are isolated patches),
     the couplings $\{ J_{xy} \}$ are of order 1 and
     may be {\em frustrated}\/.
\end{itemize}
The statements concerning the typical magnitudes of $J_{xy}$ are
demonstrated exactly as in the $N$-vector case, using \reff{Ising_ham_RP}.
The statements concerning local frustration are proven as follows:
Let $x_1, \ldots, x_n, x_{n+1} \equiv x_1$ be a (small) closed path
of nearest-neighbor sites such that the spins
$\bsigma_{x_1}, \ldots, \bsigma_{x_n}$ are {\em all}\/ close to each other
and not near the pole, i.e.\ there exists $C > {1 \over 2}$ such that
$| \bsigma_{x_i} \cdot \bsigma_{x_j} | \ge C$ and
$| \bsigma_{x_i} \cdot \br | \le \sqrt{C}$
for all $i,j = 1,\ldots,n$.
Then
$J_{cycle} \equiv J_{x_1 x_2} J_{x_2 x_3} \cdots J_{x_{n-1} x_n} J_{x_n x_1}
 \ge 0$.\footnote{{\em Proof:}\/  Note first that changing $\bsigma_{x_i}$
   to $-\bsigma_{x_i}$ leaves $J_{cycle}$ unchanged;  therefore we may choose
  the signs of $\bsigma_{x_2},\ldots,\bsigma_{x_n}$ so that
   $\bsigma_{x_1} \cdot \bsigma_{x_i} \ge C > 0$ for $i=2,\ldots,n$.
   Since $C > {1 \over 2}$, this means that the angle between
   $\bsigma_{x_1}$ and $\bsigma_{x_i}$ is less than $60^\circ$;
   so the angle between $\bsigma_{x_i}$ and $\bsigma_{x_j}$ is less than
   $120^\circ$, i.e.\ $\bsigma_{x_i} \cdot \bsigma_{x_j} > - {1 \over 2}$.
   But $| \bsigma_{x_i} \cdot \bsigma_{x_j} | \ge C > {1 \over 2}$
   by hypothesis, so $\bsigma_{x_i} \cdot \bsigma_{x_j} \ge C$.
   Then $\bsigma_{x_i}^\perp \cdot \bsigma_{x_j}^\perp  =
         \bsigma_{x_i} \cdot \bsigma_{x_j}  -
         (\bsigma_{x_i} \cdot \br) (\bsigma_{x_j} \cdot \br)  \ge
         C - (\sqrt{C})^2 = 0$.
   So each coupling $J_{x_i x_{i+1}}$ is $\ge 0$.}

To study the global frustration, define
$\widetilde{J}_{xy} = \beta (\bsigma_x^\perp \cdot \bsigma_y^\perp)
                      (\bsigma_x\cdot\br) (\bsigma_y\cdot\br)$
[these couplings are equivalent to $\{ J_{xy} \}$ via a $Z_2$ gauge
transformation].
Now draw a bond between any nearest-neighbor pair of sites $\< xy \>$
for which $\widetilde{J}_{xy} > 0$.  This will occur, in particular,
if $\bsigma_x$ and $\bsigma_y$ are close to each other and not near
the pole or the equator.\footnote{More precisely, suppose that there
   exists $\epsilon \ge 0$ such that
   $| \bsigma_x \cdot \bsigma_y | \ge 1-\epsilon$ and
   $\sqrt{\epsilon} < |\bsigma_x\cdot\br|, |\bsigma_y\cdot\br| <
    \sqrt{1-\epsilon}$.
   Then we claim that $\widetilde{J}_{xy} > 0$.
   {\em Proof:}\/  Since the substitutions $\bsigma_x \to -\bsigma_x$
   and $\bsigma_y \to -\bsigma_y$ leave $\widetilde{J}_{xy}$ unchanged,
   we can assume without loss of generality that
   $\bsigma_x\cdot\br, \bsigma_y\cdot\br > 0$.
   Then $\bsigma_x \cdot \bsigma_y = \bsigma_x^\perp \cdot \bsigma_y^\perp
           + (\bsigma_x\cdot\br) (\bsigma_y\cdot\br)
           > -1 + (\sqrt{\epsilon})^2 = -(1-\epsilon)$,
   so we must have $\bsigma_x \cdot \bsigma_y \ge 1-\epsilon$.
   Then $\bsigma_x^\perp \cdot \bsigma_y^\perp = \bsigma_x \cdot \bsigma_y
           - (\bsigma_x\cdot\br) (\bsigma_y\cdot\br)
           > (1-\epsilon) - (\sqrt{1-\epsilon})^2 = 0$.
   So $\widetilde{J}_{xy} > 0$.}
The connected components formed by the bonds thus drawn correspond roughly
to the regions $R_i$;  it is reasonable to expect that they are of linear
size $\sim \xi$, although they could in principle be larger.\footnote{%
   There is no {\em topological}\/ obstacle to these connected components
   being of size much larger than $\xi$ (e.g.\ spanning the whole lattice).
   Indeed, {\em any}\/ two spin values
   $\bsigma_x, \bsigma_{x'}$ not lying on the equator or the pole
   can potentially be connected by a chain of spins
   $\bsigma_x \equiv \bsigma_{x_0}, \bsigma_{x_1}, \ldots,
    \bsigma_{x_k} \equiv \bsigma_{x'}$
   such that
   $\widetilde{J}_{x_i x_{i+1}}  \equiv
    (\bsigma_{x_i}^\perp \cdot \bsigma_{x_{i+1}}^\perp)
    (\bsigma_{x_i}\cdot\br) (\bsigma_{x_{i+1}}\cdot\br)  > 0$
   for all $i$.
   {\em Proof:}\/
   Assume without loss of generality that $\bsigma_x$ and $\bsigma_{x'}$
   both lie in the northern hemisphere, with
   ${\rm longitude}(\bsigma_x) = 0^\circ$.
   Then, if $|{\rm longitude}(\bsigma_{x'})| < 90^\circ$,
   we can take $k=1$;
   if $90^\circ \le |{\rm longitude}(\bsigma_{x'})| < 180^\circ$,
   we can take $k=2$;
   and if $|{\rm longitude}(\bsigma_{x'})| = 180^\circ$,
   we can take $k=3$.}
Within each such connected component, the induced Ising model is
{\em globally unfrustrated}\/:  indeed, for any closed path of
nearest-neighbor bonds
$(x_1,x_2),\, (x_2,x_3),\, \ldots,\, (x_{n-1},x_n),\, (x_n,x_{n+1} \equiv x_1)$
such that $\widetilde{J}_{x_i x_{i+1}} > 0$ for all $i$,
we trivially have
$J_{cycle} \equiv \prod_{i=1}^n J_{x_i x_{i+1}} =
                  \prod_{i=1}^n \widetilde{J}_{x_i x_{i+1}} > 0$.

What is the phase structure of this induced Ising model?
Here we cannot argue as we did in the $N$-vector case
[cf.\ \reff{eq2.1.star1}/\reff{eq2.1.star2}],
because the global frustration of the $\{ J_{xy} \}$ in the transition layers
could cause ``cancellations''.  Instead, we can distinguish at least
three {\em a priori}\/ possibilities:
\begin{itemize}
   \item[(i)]  The induced Ising model exhibits exponentially decaying
      correlations on a characteristic scale $\xi$ (or smaller).
   \item[(ii)]  The induced Ising model lies in a spin-glass phase of
      Parisi-M\'ezard-Sourlas-Toulouse-Virasoro type
      \cite{Parisi,MPSTV,Mezard-Parisi-Virasoro,noi},
      i.e.\ there are many ``dominant'' regions of configuration space
      which differ on long as well as short length scales
      (``rough free-energy landscape'').
   \item[(iii)]  The induced Ising model lies in a spin-glass phase of
      Fisher-Huse type \cite{Fisher-Huse}, i.e.\ there are only two
      ``dominant'' regions of configuration space, related to each other
      by a global spin flip.
\end{itemize}
In cases (i) and (ii), an {\em independent}\/ resampling of the
induced Ising model will produce a significant collective change on
scale $\xi$ (and all smaller scales),
so we expect $z \approx 0$ for the {\em idealized}\/ Wolff algorithm.
In case (iii), by contrast, the idealized Wolff updates
do essentially nothing (merely a global spin flip),
and the performance should be poor.

In dimension $d=2$, it is believed that there is no spin-glass phase
for the ordinary random-bond Ising model
\cite{no_spin_glass_phase_in_d=2}.
That does not imply anything, of course, for our induced Ising model,
in which the structure of the couplings is quite different,
but it does {\em suggest}\/ that scenarios (ii) and (iii) are unlikely,
and that scenario (i) is probably the correct one.
We therefore expect that $z \approx 0$ for the {\em idealized}\/
Wolff algorithm.
This prediction has been confirmed in a preliminary study \cite{CEPS_LAT91}
of the two-dimensional $RP^2$ model.

\medskip

(b) {\em The critical point is at finite temperature.}\/
This case is expected to hold in the $RP^{N-1}$ models (all $N$)
in dimension $d \ge 3$, whenever the phase transition is continuous
(as opposed to first-order).  Unfortunately, in this case we have
little to say:  the typical spin configurations for $\beta \approx \beta_c$
are {\em rough}\/ at short distance scales, and the induced Ising model
will be highly frustrated.  Any one of the scenarios (i)--(iii) just discussed
--- and possibly others as well --- could occur.
So we are unable to predict with confidence whether the idealized Wolff
algorithm will exhibit $z \approx 0$ or not.

\bigskip

The behavior of the {\em practical}\/ Wolff algorithm based on
standard SW (or single-cluster SW) updates is even less clear.
Although the induced Ising Hamiltonian \reff{Ising_ham_RP} is frustrated,
this does not necessarily mean that the SW algorithm
will perform poorly for it.
To see this, note first that if
$\widetilde{J}_{xy} \equiv \beta (\bsigma_x^\perp \cdot \bsigma_y^\perp)
 (\bsigma_x\cdot\br) (\bsigma_y\cdot\br)  \le 0$,
then the bond $\< xy \>$ is guaranteed to be unoccupied.
If such bonds are numerous enough to divide the lattice into at least two
large clusters $R_i$ of size $\sim \xi$,
then the SW algorithm will flip these clusters independently
--- and thus make a long-wavelength collective-mode change in $\{\bsigma\}$ ---
{\em irrespective of the frustration within each $R_i$}\/.
On the other hand, there is no topological guarantee that the lattice will
be divided in this way
(unlike what happens in the $N$-vector case);
the clusters $R_i$ might have size much larger than $\xi$,
in which case the flips would not be of much use.\footnote{In the
   $N$-vector model, spins $\bsigma_x$ and $\bsigma_{x'}$
   belonging to different hemispheres
   [i.e.\ for which $(\bsigma\cdot\br) (\bsigma_{x'}\cdot\br)  \le 0$]
   are guaranteed to belong to distinct SW clusters.
   In the $RP^{N-1}$ model, by contrast, {\em any}\/ two spins
   $\bsigma_x, \bsigma_{x'}$ not lying on the equator or the pole
   can potentially be connected by a chain of spins
   $\bsigma_x \equiv \bsigma_{x_0}, \bsigma_{x_1}, \ldots,
    \bsigma_{x_k} \equiv \bsigma_{x'}$
   such that
   $\widetilde{J}_{x_i x_{i+1}}  \equiv
    (\bsigma_{x_i}^\perp \cdot \bsigma_{x_{i+1}}^\perp)
    (\bsigma_{x_i}\cdot\br) (\bsigma_{x_{i+1}}\cdot\br)  > 0$
   for all $i$:
   see the preceding footnote.
   Thus, $\bsigma_x$ and $\bsigma_{x'}$ could belong to the same SW cluster.}
This whole question deserves further study \cite{CEPS_LAT91}.

\subsection{Generalized Wolff-type embedding algorithms}  \label{s2.3}

It is now clear how to generalize this reasoning to a nonlinear $\sigma$-model
taking values in an arbitrary compact Riemannian manifold $M$.
Let us assume that the lattice Hamiltonian is of the form
\be
   H(\{\bsigma\})   \;=\;  \beta \sum_{\< xy \>}  E(\bsigma_x,\bsigma_y)
\ee
where $E(\bsigma,\bsigma') = E(\bsigma',\bsigma)
       \sim {\rm const} + d(\bsigma,\bsigma')^2$
as $\bsigma' \to \bsigma$ (here $d$ is geodesic distance in $M$).\footnote{%
   More generally, we could sum over {\em oriented}\/ bonds $(x,y)$
   and allow $E$ to be non-symmetric
   [i.e.\ $E(\bsigma,\bsigma') \neq E(\bsigma',\bsigma)$]:
   this situation arises, for example, in an $N$-vector model in a fixed
   external gauge field (in particular, in a frustrated $N$-vector model).
   However, only the symmetric part of $E$ contributes to the induced
   Ising Hamiltonian [see \reff{Ising_ham_gen2} below].}
Now let $T\colon\; M \to M$ be any map satisfying:
\begin{itemize}
   \item[a)]  $T$ is {\em energy-preserving}\/, i.e.\
     $E(T\bsigma,T\bsigma') = E(\bsigma,\bsigma')$
     for all $\bsigma,\bsigma' \in M$.
     [In particular, by taking $\bsigma' \to \bsigma$ it follows that
     $T$ preserves the metric tensor, and is therefore an isometry.]
\end{itemize}
We can then define an embedding of Ising spins $\{ \varepsilon \}$
by the rule
\be
 \label{embedding_gen}
 \begin{array}{lcl}
   \varepsilon_x = +1 & \Longrightarrow & \bsigma_x^{new} = \bsigma_x^{old}  \\
   \varepsilon_x = -1 & \Longrightarrow & \bsigma_x^{new} = T\bsigma_x^{old}
 \end{array}
\ee
so that the induced Ising Hamiltonian is
\be
  H(\{\varepsilon\})  \;=\;
    - \sum_{\< xy \>} J_{xy}\varepsilon_x\varepsilon_y
    \,-\, \sum_{\< xy \>} h_{xy} (\varepsilon_x - \varepsilon_y)
    \,+\, {\rm const}
 \label{Ising_ham_gen}
\ee
where
\begin{subeqnarray}
 \label{Ising_ham_gen2}
   J_{xy}   & = &   {\beta \over 4}
      \left[  E(T\bsigma_x,\bsigma_y) + E(\bsigma_x,T\bsigma_y)
              - 2 E(\bsigma_x,\bsigma_y) \right]                  \\
   h_{xy}   & = &   {\beta \over 4}
      \left[  E(T\bsigma_x,\bsigma_y) - E(\bsigma_x,T\bsigma_y) \right]
\end{subeqnarray}
(For the original Wolff algorithm, this definition of $\{\varepsilon\}$
differs from \reff{embedding} by a $Z_2$ gauge transformation.)
In particular, if $T$ is involutive,
then $E(T\bsigma_x,\bsigma_y) = E(\bsigma_x,T\bsigma_y)$,
so that the induced Ising Hamiltonian has zero magnetic field:
\begin{subeqnarray}
 \label{Ising_ham_gen3}
   J_{xy}   & = &  {\beta \over 2}
      \left[  E(\bsigma_x,T\bsigma_y) - E(\bsigma_x,\bsigma_y) \right]  \\
   h_{xy}   & = &  0
\end{subeqnarray}
We remark that in general the couplings (\ref{Ising_ham_gen3}a)
are frustrated;
indeed, in Appendix \ref{appendix_A} we prove that non-frustration occurs
essentially {\em only}\/ in the codimension-1 algorithm for the
$N$-vector model, i.e.\ the original Wolff algorithm.
(See Theorem \ref{thmA.10} and Corollaries \ref{corA.11} and \ref{corA.12}
for details.)

A {\em Wolff-type embedding algorithm}\/ is now specified by:
\begin{itemize}
  \item[(i)]  a collection $\scrt$ of energy-preserving maps of $M$;
     and
  \item[(ii)]  a probability distribution $\rho$ on $\scrt$.
\end{itemize}
One step of the algorithm consists of the following operations:
\begin{itemize}
  \item[(i)]  Choose randomly (with probability distribution $\rho$)
     a map $T \in \scrt$.
  \item[(ii)] Initialize $\{\varepsilon\} \equiv +1$.
  \item[(iii)]  Update the induced Ising model
     \reff{Ising_ham_gen}/\reff{Ising_ham_gen2},
     using any chosen Monte Carlo algorithm (e.g.\ standard SW).
  \item[(iv)]  Update the $\{ \bsigma \}$ according to \reff{embedding_gen}.
\end{itemize}
If the group generated by $\scrt$ acts transitively on $M$,
then (under mild conditions on $\rho$) this algorithm will be ergodic.
If not, then the foregoing moves must be supplemented by other types
of updates (e.g.\ a local heat-bath or Metropolis update of the
$\{ \bsigma \}$) so as to make the combined algorithm ergodic.

The {\em idealized}\/ embedding algorithm is, by definition,
the one in which step (iii) consists of obtaining an {\em independent}\/
sample from the induced Ising model.

Typically $\scrt$ is obtained from a single map $T$ by conjugating it
with isometries of $M$, and $\rho$ is taken to be an invariant measure;
this is the case, for example, in the original Wolff algorithm, with its
random choice of the unit vector $\br$.
We shall henceforth assume $\scrt$ to be of this form,
and shall refer to ``the'' map $T$ as a shorthand for
``any one of the maps $T \in \scrt$''.

In summary, the embedding algorithm can be {\em defined}\/
for any collection $\scrt$ of maps which are energy-preserving.
But we claim that the algorithm will be {\em successful}\/ in eliminating
or at least radically reducing the critical slowing-down ---
i.e.\ have a dynamic critical exponent $z$ significantly smaller than the
$z \approx 2$ typical of local algorithms ---
only if a second property is satisfied:
\begin{itemize}
   \item[b)]  The fixed-point set
     ${\rm Fix}(T) \equiv \{ \bsigma\in M \colon\; T\bsigma=\bsigma \}$
     has codimension 1.
\end{itemize}
Let us now set out, in detail, our reasoning supporting this claim:
\begin{itemize}
   \item[1)]  In order for an algorithm to do a good job of beating
critical slowing-down (i.e.\ have dynamic critical exponent $z \ll 2$),
it is necessary that the algorithm be capable of making quickly
(i.e.\ in one or a very few time steps)
a significant change in all of the large-scale collective modes
that are relevant for the (static) critical behavior of the given model.
   \item[2)]  Among the important large-scale collective modes in the
nonlinear $\sigma$-models are the long-wavelength spin waves.
(There may also be other important modes, such as vortices in the
two-dimensional $XY$ model; but there are in any case {\em at least}\/
spin waves.)
   \item[3)]  It follows that for the algorithm to have $z \ll 2$,
it is {\em necessary}\/ that it do a good job of handling
long-wavelength spin waves.  (This may not be {\em sufficient}\/,
if other types of collective modes are also important.  For example,
the multi-grid Monte Carlo algorithm \cite{MGMC_PRL,MGMC_1}
does an excellent job of handling long-wavelength spin waves,
but does less well in handling vortices;  as a result, it eliminates
critical slowing-down in the low-temperature (spin-wave) phase of the
two-dimensional $XY$ model, but not on the high-temperature side of
criticality \cite{MGMC_XY}.  The same appears to be true of Fourier
acceleration \cite{Dagotto_87a}.)
   \item[4)]  An idealized Wolff-type embedding algorithm does a good job
of handling spin waves {\em if and only if}\/ it causes $x$-space to
be divided into two or more large disconnected regions $R_i$
which are almost decoupled from each other in the induced Ising Hamiltonian.
The same holds true for a practical Wolff-type embedding algorithm,
with ``if and only if'' replaced by ``only if''.
   \item[5)]  The induced Ising Hamiltonian decouples along the manifold
$\{x\colon\; \bsigma(x) \inapprox {\rm Fix}(T) \} =
 \{x\colon\; \bsigma(x) \approx T\bsigma(x) \}$,
{\em and only there}\/.\footnote{{\em Proof:}\/
   Assume that $E(\bsigma,\bsigma')$ has a {\em unique}\/
   minimum for fixed $\bsigma$ when $\bsigma' = \bsigma$
   (and hence also for fixed $\bsigma'$ when $\bsigma = \bsigma'$).
   Then consider a bond $\< xy \>$:  we have $\bsigma_x \approx \bsigma_y$,
   since the original configuration is assumed smooth.
   Then $J_{xy} \approx 0$ if and only if
   $E(T\bsigma_x,\bsigma_y)$ and $E(\bsigma_x,T\bsigma_y)$
   are both $\approx E(\bsigma_x,\bsigma_y) \approx \min E$.
   It follows that $T\bsigma_x \approx \bsigma_x$
   (and $T\bsigma_y \approx \bsigma_y$)
   by the uniqueness of the minimum of $E$.}

   \item[6)]  For a manifold to divide $x$-space into disconnected regions,
it is {\em necessary}\/ that the manifold be
of codimension 0 or 1 (see Theorem \ref{thmA.1} in Appendix \ref{appendix_A}).
   \item[7)]  For a ``generic'' smooth spin configuration
$\bsigma\colon\; \R^d \to M$,
the codimension of the set $\{x\colon\; \bsigma(x) \in {\rm Fix}(T) \}$
in $x$-space equals the codimension of ${\rm Fix}(T)$ in the
target manifold $M$ (see Theorems \ref{thmA.4} and \ref{thmA.5}).
   \item[8)]  ${\rm Fix}(T)$ cannot have codimension 0 except
in the trivial case ${\rm Fix}(T) = M$ (see Theorem \ref{thmA.8}).
   \item[9)]  {\bf Conclusion:}  For a Wolff-type embedding algorithm
to have $z \ll 2$, it is {\em necessary}\/ (though perhaps not sufficient)
that ${\rm Fix}(T)$ have codimension 1 in the target manifold $M$.
\end{itemize}

\smallskip

In our opinion, the only truly questionable step in this reasoning
--- at least as regards asymptotically free models ---
is Step 4.  We have explained previously why
an idealized Wolff-type embedding algorithm does a good job
of handling spin waves {\em if}\/ it causes $x$-space to be divided
into large disconnected regions which are almost decoupled from each other.
It is a substantial --- and quite possibly unjustified --- leap
to assert that this is the {\em only}\/ mechanism by which
an idealized Wolff-type embedding algorithm can do a good job
of handling spin waves.
For example, suppose that the induced Ising Hamiltonian were to lie
in a highly frustrated phase characterized by several ``dominant''
regions of configuration space which differ on long as well as short
length scales.
Then the corresponding {\em idealized}\/ Wolff-type algorithm
would, by definition, make frequent transitions between these regions;
in particular, it would make frequent {\em global}\/ changes
in the $\{ \varepsilon \}$ configuration, and hence presumably also
in the $\{ \bsigma \}$ configuration.
These global changes might be sufficient for the algorithm to do a good job
of handling spin waves.\footnote{We thank Ferenc Niedermayer for
   pointing out this possibility to us.}
(Of course, a {\em practical}\/ Wolff-type algorithm might
have great difficulty surmounting the barriers between these dominant
regions of configuration space.  But that is not our concern at present.)

Since our reasoning is based on considering ``smooth'' spin configurations,
it is also questionable when applied to non-asymptotically-free models,
as already discussed in Sections \ref{s2.1} and \ref{s2.2}.
Possibly our reasoning {\em does}\/ become valid when reinterpreted
as referring to a suitably coarse-grained spin field;  but possibly not.
Some insight into this problem could perhaps be obtained by comparing
the (idealized) codimension-1 and codimension-2 algorithms
[cf.\ \reff{eq2.3} for $r=1,2$] for the {\em three}\/-dimensional
$N$-vector model.  However, it will be difficult to disentangle the
effects of codimension from those of frustration, as the two phenomena
go together (codimension 1 is unfrustrated, while codimension $> 1$
is frustrated).  In this regard the three-dimensional $RP^{N-1}$ model
(if its phase transition is indeed continuous)
could serve as a valuable comparison, since in this case {\em both}\/
codimension 1 and codimension $> 1$ lead to frustrated induced Ising models.

All in all, we consider our Conclusion to be a {\em plausible conjecture}\/,
which needs to be tested carefully.  In Section \ref{s4} we test it
in one asymptotically free model.

\section{Classification of (Involutive) Isometries}  \label{s3}

The reasoning of the preceding section leads us to the following problem
in differential geometry:  Given a Riemannian manifold $M$,
classify all the isometries of $M$ according to the codimension(s)
of their fixed-point manifolds.  In particular, if we are interested
in fixed-point manifolds of codimension 1, we can restrict attention to
{\em involutive}\/ isometries (see Theorem \ref{thmA.9}).

If $T$ is an isometry (resp.\ involutive isometry) of $M$,
and $f$ is any isometry of $M$,
then $f \circ T \circ f^{-1}$ is also an isometry (resp.\ involutive isometry)
of $M$; we say that it is {\em conjugate to $T$}\/ via the isometry $f$.
Obviously, conjugacy is an equivalence relation;  for any given manifold $M$,
our goal will be to classify the (involutive) isometries of $M$,
modulo conjugacy.
Note that if $p$ is a fixed point of $T$, then $f(p)$ is a fixed point of
$f \circ T \circ f^{-1}$.  In particular, if ${\rm Fix}(T)$ is nonempty and
the isometry group ${\rm Isom}(M)$ acts transitively on $M$,
then the conjugating map $f$
can be chosen so that ${\rm Fix}(f \circ T \circ f^{-1})$
passes through any desired point $q \in M$
(though not necessarily in any desired {\em direction}\/).
So we can restrict attention, if desired, to (involutive) isometries $T$
that leave fixed some chosen point $p$.

\subsection{Some Examples}   \label{s3.1}

\quad\par\indent
{\bf Example 1.}  $M=S^{N-1}$.  All involutive isometries of $S^{N-1}$
are conjugate (via an orthogonal transformation) to
\be
 \label{eq2.3}
   T(\sigma^{1},\ldots,\sigma^{N})  \;=\;
   (-\sigma^{1},\ldots,-\sigma^{r},\sigma^{r+1},\ldots,\sigma^{N})
\ee
for some $r$ ($0 \le r \le N$).\footnote{{\em Proof:}\/
   Every isometry of $S^{N-1}$ is induced by an orthogonal
   matrix $T \in O(N)$.  An involutive isometry has $T^2 = I$.
   It follows that $T$ is symmetric as well as orthogonal, and hence
   has eigenvalues $\pm 1$.  Any such matrix can be diagonalized by a
   rotation; and by a further permutation of coordinates (which is also
   an orthogonal transformation) we can make all the $-1$ eigenvalues
   come first.  Thus, there exists a matrix $R \in O(N)$ such that
   $RTR^{-1} = I_r \equiv
   {\rm diag}( \underbrace{-1,\ldots,-1}_{\displaystyle{r} \hbox{ times}},
               \underbrace{+1,\ldots,+1}_{\displaystyle{N-r} \hbox{ times}} )$
   for some $r$ ($0 \le r \le N$).}
Henceforth we shall write this transformation more compactly as
$\bsigma \to I_r \bsigma$, where
\be
   I_r   \;=\;
   {\rm diag}( \underbrace{-1,\ldots,-1}_{\displaystyle{r} \hbox{ times}},
               \underbrace{+1,\ldots,+1}_{\displaystyle{N-r} \hbox{ times}}  )
   \;.
\ee
For $0 \le r \le N-1$ the fixed-point manifold has codimension $r$;
for $r=N$ the fixed-point manifold is empty.
(More precisely, for $0 \le r \le N-2$ the fixed-point manifold is
isometric to $S^{N-r-1}$; for $r=N-1$ it consists of two points.)
The standard Wolff reflection \reff{eqwolff2.2} corresponds to the case $r=1$.

\bigskip
{\bf Example 2.}  $M=RP^{N-1}$.
Every isometry $T$ of
$RP^{N-1} \equiv S^{N-1}/ \{ \pm I \}$
``lifts'' (nonuniquely) to an isometry $\widetilde{T}$ of $S^{N-1}$
[hence a matrix $\widetilde{T} \in O(N)$];
and conversely, every isometry of $S^{N-1}$ commutes with $\pm I$
and hence induces a (unique) isometry on $RP^{N-1}$.
Moreover, $T$ is involutive if and only if $\widetilde{T}^2 = \pm I$.
We must therefore analyze two cases:

(a) If $\widetilde{T}^2 = +I$, then $\widetilde{T}$ is conjugate
[via an orthogonal transformation $R \in O(N)$]
to a transformation of the form \reff{eq2.3} for some $r$.
Moreover, since $R$ commutes with $\pm I$, this conjugacy acts
also on $RP^{N-1}$, i.e.\ $T$ is conjugate to the map induced on $RP^{N-1}$
by \reff{eq2.3}.

(b) If $\widetilde{T}^2 = -I$, then $N$ must be even
[since $\det(\widetilde{T}^2) = (\det \widetilde{T})^2 = +1$];
and $\widetilde{T}$ is easily seen to be conjugate
[via some $R \in O(N)$]
to the map $\bsigma \to J \bsigma$, where
\be
   J   \;=\;   \left(  \begin{array}{cc}  0 & I \\ -I & 0 \end{array} \right)
\ee
and $I$ is an $(N/2) \times (N/2)$ identity matrix.\footnote{{\em Proof:}\/
   Since $\widetilde{T}^T \widetilde{T} = I$ and
   $\widetilde{T}^2 = -I$, we conclude that $\widetilde{T}$ is
   antisymmetric as well as orthogonal, and hence has eigenvalues $\pm i$.
   Any such matrix can be brought into $2 \times 2$ block-diagonal form
   by a rotation;  and this is equivalent, by a permutation of coordinates,
   to $J$.  Thus, there exists $R \in O(N)$ such that
   $R \widetilde{T} R^{-1} = J$.}

Thus, $T$ is conjugate to the map induced on $RP^{N-1}$ by one of the
following:
\begin{itemize}
   \item[(a${}_r$)]  $\bsigma \to I_r \bsigma$.
      [Note that (a${}_r$) and (a${}_{N-r}$) are the same transformation.]
      The fixed-point set is the union
      of two disjoint connected manifolds, one isometric to $RP^{N-r-1}$
      and the other isometric to $RP^{r-1}$;
      they have codimensions $r$ and $N-r$, respectively.
   \item[(b)]  [Only for $N$ even]  $\bsigma \to J \bsigma$.
      The fixed-point manifold is empty.
\end{itemize}

\bigskip
{\bf Example 3.}  $M= \CP^{N-1}$.
The complex projective space $\CP^{N-1}$ is, by definition,
the set of unit vectors in $\C^N$ modulo the
equivalence relation
\be
 z \simeq z'   \;\Longleftrightarrow\;
     z = e^{i\theta} z' \hbox{ for some real number } \theta   \;.
\ee
There is a unique (up to multiples) Riemannian metric on $\CP^{N-1}$
that is invariant under the natural action of $U(N)$, namely the
Fubini-Study metric \cite[vol.\ II, pp.\ 159--160, 169]{Kobayashi_69}.
The geodesic distance associated with this Riemannian metric is\footnote{This
   is easily demonstrated by integrating the Fubini-Study metric
   along the geodesic curves in $\CP^{N-1}$
   \cite[vol.\ II, p.\ 277]{Kobayashi_69},
   and appealing to unitary invariance.}
\be
  d(z,w)  \;=\;    2 \cos^{-1} \left| \sum_{i=1}^N \overline z_i w_i \right|
\ee
(where, by abuse of language, we do not distinguish between an equivalence
class $z \in \CP^{N-1}$ and a representative unit vector $z \in \C^N$).
The isometries of $\CP^{N-1}$ ($N \ge 2$) are fully classified by
{\em Wigner's theorem}\/:  every such isometry arises from a map
$\C^N \to \C^N$ which is either unitary or antiunitary.\footnote{See
   e.g.\ \cite{Bargmann_64} or \cite[pp.\ 305--306]{Wolf_77}.}
Thus, any isometry of $\CP^{N-1}$ is of one of the forms
\begin{itemize}
  \item[(a)] $z \to Uz$ with $U \in U(N)$
  \item[(b)] $z \to U \bar{z}$ with $U \in U(N)$
\end{itemize}
A little analysis
then shows that every {\em involutive}\/ isometry of $\CP^{N-1}$
is conjugate to one of the following\footnote{{\em Proof:}\/
   Let $T$ be an involutive isometry of $\CP^{N-1}$.
   In case (a), we must have $U^2 = e^{i\alpha} I$,
   and by redefining the phases we can require $U^2 = I$.
   It follows that the eigenvalues of $U$ are $\pm 1$,
   so there exists $V\in SU(N)$ such that $U = V I_r V^\dagger$
   for some $r$ ($0 \le r \le N$).
   Thus $T$ is conjugate to the map $z \to I_r z$.
   In case (b), we must have $U \bar{U} = e^{i\alpha} I$,
   hence $U = e^{i\alpha} U^T$.  Since $U \neq 0$, the only possibilities are
   $e^{i\alpha} = \pm 1$;  and by taking determinants we see that
   $e^{i\alpha} = -1$ can occur only for $N$ even.
   If $U \bar{U} = I$, then $U = XX^T$ for some $X \in U(N)$
   [see the proof of case (c1) in Theorem \ref{thmB.4}];
   but then, taking $f(z) = X^\dagger z$, we have
   $(f \circ T \circ f^{-1})(z) = \bar{z}$.
   Similarly, if $U \bar{U} = -I$, then $U = XJX^T$ for some $X \in U(N)$
   [see the proof of case (c2) in Theorem \ref{thmB.4}];
   but then, taking $f(z) = X^\dagger z$, we have
   $(f \circ T \circ f^{-1})(z) = J \bar{z}$.

   Let us now compute the fixed-point manifolds $F$:
   {\em Case (a):}\/   $F$ consists of points $z$ satisfying
   $z = e^{i\alpha} I_r z$ for some $\alpha$.
   For $\alpha \neq 0,\pi$ there are no $z \neq 0$ which satisfy this equation.
   Thus $F$ consists of $z$ which satisfy either $z=I_r z$ or $z= -I_r z$.
   These equations define disjoint connected manifolds isometric to
   $CP^{N-r-1}$ and $CP^{r-1}$, respectively.
   {\em Case (b1):}\/  $F$ consists of points $z$ satisfying
   $z = e^{i\alpha} \bar{z}$ for some $\alpha$, or equivalently
   $z e^{-i\alpha/2} = \overline{ze^{-i\alpha/2}}$.
   Redefining phases, we can assume that $\alpha=0$.
   So $F$ is a connected manifold isometric to $RP^{N-1}$.
   {\em Case (b2):}\/  $F$ consists of points $z$ satisfying
   $z = e^{i\alpha} J \bar{z}$ for some $\alpha$.
   It follows that $\bar{z} = e^{-i\alpha} J z = J^2 \bar{z} = - \bar{z}$,
   which is clearly impossible for $z \neq 0$.
   So the fixed-point manifold is empty.

   See also \cite[Theorem 7]{Leung_73} and
   \cite[Theorem 4.2]{Leung_74} (case AIII with $p=N-1$, $q=1$)
   for an alternate derivation of this list, based on a Lie-algebraic analysis.
}:
\begin{itemize}
   \item[(a${}_r$)]  $z \to I_r z$.  [Note that (a${}_r$) and (a${}_{N-r}$)
      are the same transformation.]  The fixed-point set is the union
      of two disjoint connected manifolds, one isometric to $\CP^{N-r-1}$
      and the other isometric to $\CP^{r-1}$;
      they have codimensions $2r$ and $2(N-r)$, respectively.
   \item[(b1)]  $z \to \bar{z}$.
      The fixed-point manifold is connected and isometric to $RP^{N-1}$;
      it has codimension $N-1$.
   \item[(b2)]  [Only for $N$ even]  $z \to J \bar{z}$.
      The fixed-point manifold is empty.
\end{itemize}
We conclude that {\em in $\CP^{N-1}$ ($N \ge 3$), there do not exist
involutive isometries of codimension 1.}
(For $N=2$, $\CP^1$ is isometric to the sphere $S^2$,
so an involutive isometry of codimension 1 of course does exist,
namely $z \to \bar{z}$.)
An even stronger nonexistence result is mentioned in Further Remarks 3
at the end of Appendix \ref{secA.3}.

\bigskip
{\bf Example 4.}  $M=SU(2)$.  Since $SU(2)$ is isometric to $S^3$ via the map
\begin{eqnarray}
   (\sigma^0,\sigma^1,\sigma^2,\sigma^3)   & \mapsto &
   \sigma^0 I + i(\sigma^1 \btau^1 + \sigma^2 \btau^2 + \sigma^3 \btau^3)
                                                         \nonumber \\
   & = &   \left(  \begin{array}{cc}
                      \sigma^0 + i\sigma^3   &   \sigma^2 + i\sigma^1   \\
                     -\sigma^2 + i\sigma^1   &   \sigma^0 - i\sigma^3
                   \end{array}
           \right)
\end{eqnarray}
where $\btau^1,\btau^2,\btau^3$ are the usual Pauli matrices,
this is a special case of Example 1.
But it is useful to translate those results into the $SU(2)$ matrix language,
as a hint toward generalizations to $SU(N)$.  Recalling the definitions
$I_1 = \left( \begin{array}{cc}  -1 & 0 \\ 0 & 1 \end{array} \right)$ and
$J = \left( \begin{array}{cc}  0 & 1 \\ -1 & 0 \end{array} \right)$,
here is a (redundant) list of some involutive isometries of $SU(2)$:
\begin{center}
\begin{tabular}{llll}
   Codimension 0:  & $A \to A \; (= -J \bar{A} J)$   &
     Fixed points: & All $SU(2)$   \\[0.5cm]
   Codimension 1:  & $A \to -A^\dagger$   &
     Fixed points: & $\sigma^0 = 0$  \\
                   & $A \to I_1 A^T I_1$   &
     Fixed points: & $\sigma^1 = 0$  \\
                   & $A \to A^T$   &
     Fixed points: & $\sigma^2 = 0$  \\
                   & $A \to I_1 A^\dagger I_1$   &
     Fixed points: & $\sigma^3 = 0$  \\[0.5cm]
   Codimension 2:  & $A \to \bar{A}$   &
     Fixed points: & $\sigma^1 = \sigma^3 = 0$  \\
                   & $A \to I_1 A I_1$   &
     Fixed points: & $\sigma^1 = \sigma^2 = 0$  \\
                   & \multicolumn{1}{c}{etc.} &   &   \\[0.5cm]
   Codimension 3:  & $A \to A^\dagger \; (= -J A^T J)$   &
     Fixed points: & $\sigma^1=\sigma^2=\sigma^3=0$   \\
                   & \multicolumn{1}{c}{etc.} &   &
\end{tabular}
\end{center}

\bigskip
{\bf Example 5.}  $M=SU(N)$.
In Appendix \ref{appendix_B} we carry out a complete classification
(modulo conjugation by an isometry) of the involutive isometries of $SU(N)$.
[Surprisingly, we have been unable to find this classification
anywhere in the mathematical or physical literature.]
In particular, the involutive isometries of $SU(N)$ having
nonempty fixed-point manifold are the following\footnote{See also
   \cite[Theorem 3.3]{Leung_74} for an alternate derivation
   of this list, based on a Lie-algebraic analysis. }:
\bigskip

(a${}_{r,r}$)   $A \to I_r A I_r$.
[Note that (a${}_{r,r}$) and (a${}_{N-r,N-r}$) are the same transformation.]
The fixed points are matrices of the form
$\left(\!\! \begin{array}{cc} B & 0 \\ 0 & C \end{array} \!\!\right)$
with $B \in U(r)$, $C \in U(N-r)$ and $(\det B)(\det C) = 1$.
We call this space $S(U(r)\times U(N-r))$; it is a subgroup of $SU(N)$,
and has codimension $2r(N-r)$.
\medskip

(b1)   $A \to A^\dagger$.  The fixed-point manifold is
$$\bigcup_{\begin{scarray}  0 \le r \le N  \\ r \hbox{\scriptsize\ even}
           \end{scarray}
          }
 F_r  \;,$$
where $F_r \equiv \{ U I_r U^\dagger \colon\;  U \in SU(N) \}$
is the coset space $SU(N)/S(U(r)\times U(N-r))$;
it has codimension $r^2 + (N-r)^2 -1$.
\medskip

(b2) [Only for $N$ even]   $A \to e^{2\pi i/N} A^\dagger$.
The fixed-point manifold is
$$\bigcup_{\begin{scarray}  0 \le r \le N  \\ r \hbox{\scriptsize\ odd}
           \end{scarray}
          }
 F_r  \;,$$
where $F_r \equiv \{ e^{\pi i/N} U I_r U^\dagger \colon\;  U \in SU(N) \}$
is isometric to the coset space $SU(N)/S(U(r)\times U(N-r))$;
it has codimension $r^2 + (N-r)^2 -1$.
\medskip

(c1)  $A \to \bar{A}$.  The fixed points are real matrices $A \in SO(N)$.
This space is a subgroup of $SU(N)$, and has codimension $\half (N^2+N-2)$.
\medskip

(c2a)  [Only for $N$ even]  $A \to -J \bar{A} J$.
The fixed points are matrices $A$ that are symplectic ($A^T J A = J$)
as well as unitary.  We call this space\footnote{See e.g.\ \cite[pp.\
   347--349]{Miller_72}.
   In the differential-geometry literature
   (e.g.\ \cite[p.\ 340]{Helgason_78}) this group is often called $Sp(N/2)$.}
$\USp(N/2) \equiv Sp(N/2,\C) \cap U(N)$;
it is a subgroup of $SU(N)$, and has codimension $\half (N^2-N-2)$.
\medskip

(d1)  $A \to A^T$.  The fixed points are matrices $A = U U^T$
with $U \in SU(N)$.
This is the coset space $SU(N)/SO(N)$;  it has codimension $\half(N^2-N)$.
\medskip

(d2)  [Only for $N$ even]  $A \to - A^T$.
The fixed points are matrices $A = U J U^T$ or $A = e^{2\pi i/N} U J U^T$
with $U \in SU(N)$.  These two disjoint manifolds are each isometric to the
coset space $SU(N)/\USp(N/2)$;  they have codimension $\half (N^2+N)$.
\bigskip

\noindent
Therefore, in $SU(3)$ the involutive isometries
(other than the identity map) having nonempty fixed-point manifold are:
\begin{center}
\begin{tabular}{llll}
   Codimension 3:  & $A \to A^T$   &
     Fixed points: & $SU(3)/SO(3)$   \\[0.5cm]
   Codimension 4:  & $A \to I_1 A I_1$   &
     Fixed points: & $S(U(1)\times U(2))$   \\
                   & $A \to A^\dagger$   &
     Fixed points: & $SU(3)/S(U(1)\times U(2))$   \\[0.5cm]
   Codimension 5:  & $A \to \bar{A}$   &
     Fixed points: & $SO(3)$
\end{tabular}
\end{center}
In $SU(4)$ the non-identity involutive isometries of smallest codimension are
\begin{center}
\begin{tabular}{llll}
   Codimension 5:  & $A \to -J \bar{A} J$   &
     Fixed points: & $\USp(2)$               \\[0.5cm]
   Codimension 6:  & $A \to I_1 A I_1$   &
     Fixed points: & $S(U(1)\times U(3))$   \\
                   & $A \to A^T$   &
     Fixed points: & $SU(4)/SO(4)$
\end{tabular}
\end{center}
In $SU(N)$, $N \ge 5$, the non-identity involutive isometry of smallest
codimension is $A \to I_1 A I_1$, with codimension $2N-2$.

We conclude that {\em in $SU(N)$, $N \ge 3$, there do not exist
involutive isometries of codimension 1}\/
(or even codimension 2).

\subsection{General Theory}   \label{s3.2}

Now let us try to find all Riemannian manifolds
(within a certain class to be defined below)
admitting an involutive isometry with fixed-point manifold of codimension 1.
(Such an involutive isometry is sometimes called a {\em mirror}\/.)
We need a few definitions and facts
from differential geometry \cite{Kobayashi_69,Helgason_78,Wolf_77}:

A Riemannian manifold is called {\em homogeneous}\/
if it possesses a transitive group of isometries
[i.e.\ if for all $p,q \in M$
there exists an isometry $f$ such that $f(p) = q$].
Every homogeneous Riemannian manifold is complete.

A Riemannian manifold is called {\em symmetric}\/
(or {\em globally symmetric}\/, or a {\em symmetric space}\/)
if for each $p \in M$ there exists an involutive isometry of $M$
having $p$ as an isolated fixed point.
Every Riemannian symmetric space is homogeneous,
and the Riemannian symmetric spaces have been completely classified
\cite[Theorems 8.11.4 and 8.3.12]{Wolf_77}.
Virtually all the target manifolds arising in physical applications
are symmetric spaces \cite{Pisarski_79,Brezin_80}.
(The only exception we know of arises in the
$\sigma$-model approach to string theory \cite{Friedan_85,Lott_86},
in which general target spaces are needed.)

Let $M$ be a Riemannian manifold with metric tensor $g$ and
curvature tensor $R$.  Then, for any point $p \in M$ and any
two-dimensional vector subspace $S$ of the tangent space $T_p M$,
the {\em sectional curvature at $p$ along $S$}\/ is defined to be
\be
   K(p,S)   \;=\;   { g_p(R_p(v,w)w,v)
                      \over
                      g_p(v,v) g_p(w,w) - g_p(v,w)^2
                    }
   \;,
\ee
where $v$ and $w$ are any two linearly independent vectors in $S$.
If $K(p,S) = k$ for all $p \in M$ and all planes $S \subset T_p M$,
then $M$ is called a
{\em space of constant curvature $k$}\/.\footnote{In fact,
   if $\dim M \ge 3$ it suffices that $K(p,S)$ be independent of $S$
   for each $p \in M$;  then a theorem of Schur \cite[Theorem 2.2.7]{Wolf_77}
   ensures that $K(p,S)$ is constant also as a function of $p$.
}%
${}^{,}$\footnote{This should not be confused with the constancy of the
   curvature tensor (or of the sectional curvature) under
   {\em parallel translation}\/.
   Indeed, these latter conditions provide two alternative definitions
   of the {\em Riemannian locally symmetric spaces}\/
   \cite[pp.\ Theorem 8.1.1]{Wolf_77}.
   At least one book \cite{Doubrovine_82} makes the unfortunate decision
   to use the term ``space of constant curvature'' for what everyone else calls
   ``Riemannian locally symmetric space''.}
Much is known about spaces of constant curvature:
see \cite{Wolf_77}, and especially Corollary 2.4.10, Theorem 2.5.1,
Theorem 2.7.1 and Corollary 2.7.2.
For our purposes, the following results suffice:

\begin{theorem}
\label{thm3.1}
Let $M$ be a connected homogeneous Riemannian manifold
of dimension $n$ and constant curvature $k$.  Then either
\begin{itemize}
  \item[(a)]  $k<0$ and $M$ is isometric to the hyperbolic space $H^n$
     of ``radius'' $k^{-1/2}$ in $\R^{n+1}$
\end{itemize}
or
\begin{itemize}
  \item[(b${}_m$)]  [$0 \le m \le n$]
    $k=0$ and $M$ is isometric to the product $\R^m \times T^{n-m}$
    of a Euclidean space with a flat Riemannian torus
    $T^{n-m} \equiv \R^{n-m}/\Gamma$
    (here $\Gamma$ is a discrete subgroup of the translation group $\R^{n-m}$,
     generated by $n-m$ linearly independent vectors)
\end{itemize}
or else
\begin{itemize}
  \item[(c)]  $k>0$ and $M$ is isometric to one of the following:
\begin{itemize}
  \item[(i)]  the sphere $S^n$ of radius $k^{-1/2}$ in $\R^{n+1}$
  \item[(ii)]  the real projective space $RP^n \equiv S^n / \{ \pm I \}$
  \item[(iii)]  [only for $n$ odd]  the quotient space $S^n/Z_m$ with $m>2$;
     here $S^n$ is considered as the unit sphere of $\C^{(n+1)/2}$,
     and the group $Z_m \subset U(1)$ acts by scalar multiplication
  \item[(iv)]  [only for $n=4l+3$, $l$ integer]  one of the quotient spaces
     $S^n/D^*_m$ with $m>2$, $S^n/T^*$, $S^n/O^*$ or $S^n/I^*$;
     here $S^n$ is considered as the unit sphere of $\Q^{(n+1)/4}$
     ($\Q$ = quaternions),
     and $D^*_m$, $T^*$, $O^*$ and $I^*$ are the dihedral, tetrahedral,
     octahedral and icosahedral groups lifted to $SU(2) \simeq U(1,\Q)$
     and acting by scalar multiplication
\end{itemize}
\end{itemize}
In particular, $M$ is compact in cases (b${}_0$) and (c), and only these.
\end{theorem}

\proof
This is an immediate consequence of
\cite[Theorem 2.7.1 and Corollary 2.7.2]{Wolf_77}.
\qed

\begin{corollary}
\label{cor3.2}
Let $M$ be a connected Riemannian symmetric space
of dimension $n$ and constant curvature $k$.  Then either
\begin{itemize}
  \item[(a)]  $k<0$ and $M$ is isometric to the hyperbolic space $H^n$
\end{itemize}
or
\begin{itemize}
  \item[(b${}_m$)]  [$0 \le m \le n$]
    $k=0$ and $M$ is isometric to a product $\R^m \times T^{n-m}$
\end{itemize}
or else
\begin{itemize}
  \item[(c)]  $k>0$ and $M$ is isometric to either the sphere $S^n$
       or the real projective space $RP^n \equiv S^n / \{ \pm I \}$.
\end{itemize}
In particular, $M$ is compact in cases (b${}_0$) and (c), and only these.
\end{corollary}

\proof
Since every Riemannian symmetric space is homogeneous,
it suffices to analyze the cases in Theorem \ref{thm3.1}.
By \cite[Theorem 8.3.12]{Wolf_77},
the quotient $S^n/\Gamma$ is a symmetric space if and only if
$\Gamma$ is a discrete subgroup of
$\Delta \equiv$ centralizer of $SO(n+1)$ in $O(n+1)$
= $\{ \pm I \}$.
\qed

A connected Riemannian manifold is said to be {\em irreducible}\/
if it is not locally isometric to a product of lower-dimensional manifolds.
Equivalently, $M$ is irreducible if for some (or all) $p \in M$
the linear holonomy group $\Psi_p$ acts irreducibly on the
tangent space $T_p M$.

The irreducible Riemannian symmetric spaces admitting a mirror have been
completely classified by Iwahori \cite{Iwahori_66}:

\begin{theorem}
\label{thm3.3}
Let $M$ be an irreducible Riemannian symmetric space of dimension $n$.
Then the following are equivalent:
\begin{itemize}
   \item[(a)] $M$ admits an involutive isometry whose fixed-point manifold
      has at least one connected component of codimension 1.
   \item[(b)] $M$ is a space of constant curvature.
   \item[(c)] $M$ is isometric to either the sphere $S^n$,
       the real projective space $RP^n \equiv S^n / \{ \pm I \}$,
       or the hyperbolic space $H^n$.
\end{itemize}
\end{theorem}
[Here (b) $\Longleftrightarrow$ (c) follows from Corollary \ref{cor3.2},
and (c) $\Longrightarrow$ (a) is easy.
The hard part of the proof is (a) $\Longrightarrow$ (b).]

Thus, the only irreducible compact Riemannian symmetric spaces
admitting a mirror are the examples we already know:
the sphere $S^n$ and the real projective space $RP^n$.

{\em Remarks.}\/
1. It is, as far as we know, an open question whether (a) $\Longrightarrow$ (b)
in Theorem \ref{thm3.3} holds also for irreducible homogeneous Riemannian
manifolds that are not symmetric spaces.

2.  Some relevant related work can be found in \cite{Sabinin_70,Oniscik_80};
unfortunately, we have not yet been able to obtain a copy of these articles.
For related work on reflections in {\em complex}\/ manifolds,
see \cite{Gottschling_69,Meschiari_72}.

\medskip

Next we would like to classify the {\em reducible}\/
Riemannian symmetric spaces admitting a mirror
(or more interestingly, admitting a transitive group of mirrors).
However, since $\sigma$-models taking values in a reducible
Riemannian symmetric space are of lesser interest
for physical applications,
we leave this classification for the mathematicians.
But it is not difficult to show that a transitive group of mirrors
can exist only if $M$ is a sphere, a product of spheres,
or the quotient of such a space by a discrete group.\footnote{One
   amusing application involves a $\sigma$-model taking values in
   the space $S^1 \times S^1$ (i.e.\ two uncoupled $XY$ models):
   in addition to the standard Wolff algorithm based on
   the codimension-1 reflections
   $\theta \to -\theta$ and $\phi \to -\phi$,
   there is another algorithm based on the codimension-1 reflections
   $\theta \leftrightarrow \phi$ and
   $\theta \leftrightarrow -\phi$.
   Physically, this algorithm takes chunks from one system and pastes them
   into the other system (possibly reflected).
   We conjecture that this algorithm will also have $z \approx 0$.}

\section{Numerical Study of Embedding Algorithms for the $O(4)$ Model}
   \label{s4}

In Section \ref{s2} we conjectured that the idealized embedding algorithm
can perform well (i.e.\ have $z \ll 2$) {\em only}\/
if it is based on a involutive isometry $T$
whose fixed-point manifold has codimension 1.
In order to test this conjecture, we have studied the
two-dimensional $N$-vector model for $N=4$ using the
embedding algorithm based on a codimension-2 reflection
[i.e.\ \reff{eq2.3} with $r=2$].
For comparison, we have also generated additional data
(extending \cite{Edwards_89}) on this same model
using the codimension-1 reflection.

\subsection{Quantities to be Measured}   \label{s4.1}

For any observable $A$, define the unnormalized autocorrelation function
\be
  C_{AA}(t)  \;=\;   \< A_s A_{s+t} \>   -  \< A \> ^2  \,,
\ee
where expectations are taken {\em in equilibrium\/}.
The corresponding normalized autocorrelation function is
\be
  \rho_{AA}(t)  \;=\;  C_{AA}(t) / C_{AA}(0) \,.
\ee
We then define the {\em integrated autocorrelation time}
\begin{eqnarray}
\tau_{int,A}  & =&
 \half \sum_{{t} \,=\, - \infty}^{\infty} \rho_{AA} (t)\nonumber\\
 &=&  \half \ +\  \sum_{{t} \,=\, 1}^{\infty} \  \rho_{AA} (t)
\end{eqnarray}
[The factor of $\half$ is purely a matter of convention;  it is
inserted so that $\tau_{int,A} \approx\ \tau$ if
$\rho_{AA}(t) \approx e^{-|t|/ \tau}$ with $\tau \gg 1$.]
The integrated autocorrelation time controls the statistical error
in Monte Carlo measurements of $\< A \>$.  More precisely,
the sample mean
\begin{equation}
\bar A \ \ \equiv\ \ {1 \over n }\  \sum_{t=1}^n \ A_t
\end{equation}
has variance
\begin{eqnarray}
\var( \bar A )  &= &
  {1 \over n^2} \ \sum_{r,s=1}^n \ C_{AA} (r-s) \nonumber \\
 &=& {1 \over n }\ \sum_{{t} \,=\, -(n-1)}^{n-1}
  (1 -  {{|t| \over n }} ) C_{AA} (t) \label{var_observa}  \\
 &\approx&  {1 \over n }\ (2 \tau_{int,A} ) \ C_{AA} (0)
   \qquad {\rm for}\ n\gg \tau \label{var_observb}
\end{eqnarray}
Thus, the variance of $\bar{A}$ is a factor $2 \tau_{int,A}$
larger than it would be if the $\{ A_t \}$ were
statistically independent.
Stated differently, the number of ``effectively independent samples''
in a run of length $n$ is roughly $n/2 \tau_{int,A}$.
The autocorrelation time $\tau_{int,A}$ (for interesting observables $A$)
is therefore a ``figure of (de)merit'' of a Monte Carlo algorithm.

We shall measure static quantities (expectations) and dynamic quantities
(autocorrelation times) for the following observables:
\begin{eqnarray}
  \scrm^2    &=&  \left( \sum_x \bsigma_x \right)^2       \\[0.2cm]
  \scrf      &=&  \half \left[
                           \left| \sum_x e^{2\pi i x_1/L} \bsigma_x \right| ^2
                           \,+\,
                           \left| \sum_x e^{2\pi i x_2/L} \bsigma_x \right| ^2
                        \right]                           \\[0.2cm]
  \scre      &=&  \sum_{\< xx' \>}  \bsigma_x \cdot \bsigma_{x'}
\end{eqnarray}
The mean values of these observables give information on different aspects
of the 2-point function
\begin{subeqnarray}
   G(x)         & = &   \< \bsigma_0 \cdot \bsigma_x \>        \\[0.1cm]
   \gtilde(p)   & = &   \sum_x e^{ip\cdot x} \< \bsigma_0 \cdot \bsigma_x \>
\end{subeqnarray}
In particular, we are interested in the {\em susceptibility}\/
\be
  \chi  \;=\;  \gtilde(0)   \;=\;  V^{-1} \< \scrm^2 \>
\ee
and the analogous quantity at the smallest nonzero momentum
\be
  F  \;=\;  \gtilde(p) | _{|p| = 2\pi/L}  \;=\;   V^{-1} \< \scrf \>    \;.
\ee
By combining these we can obtain the {\em (second-moment) correlation length}\/
\be
 \label{def_xi}
  \xi   \;=\;     \left( { \displaystyle  (\chi/F) - 1
                           \over
                           \displaystyle  4 \sin^2 (\pi/L)
                         }
                  \right) ^{1/2}
\ee
(see the Remark below).
Finally, we have the (negative) {\em energy}\/
\be
  E   \;=\;  2G(x) | _{|x| = 1}   \;=\;   V^{-1}  \< \scre \>   \;.
\ee
Here $V=L^2$ is the number of sites in the lattice.

\medskip

{\em Remark.}
The definition \reff{def_xi} is sometimes summarized \cite{Edwards_89}
by saying that we are fitting $\gtilde(p)$ to the Ansatz
\be
 \label{bad-form}
  \gtilde(p) = Z \left[ \xi^{-2} + 4 \sum_{i=1}^d \sin^2 (p_i/2) \right] ^{-1}
\ee
at $p=0$ and $|p| = 2\pi/L$.
This is of course true; but it is important to emphasize
that we are {\em not}\/ assuming that $\gtilde(p)$
really has the free-field form \reff{bad-form} at general $p$
(of course it doesn't).
Rather, we simply use this form to motivate {\em one}\/ reasonable definition
of the second-moment correlation length $\xi$ on a finite lattice.
Another definition --- slightly different but equally reasonable --- would be
\begin{eqnarray}
 \label{def_xiprime}
  \xi'     &=&    \left( {1 \over 2d} \,
                         { \displaystyle \sum_x
                              \left( \sum_{i=1}^d (L/\pi)^2 \sin^2 (\pi x_i/L)
                              \right)  G(x)
                           \over
                           \displaystyle  \sum_x G(x)
                         }
                  \right) ^{1/2}                            \nonumber \\[0.2cm]
   &=&  {L^2 \over 4\pi^2} \left( 1 - {F \over \chi} \right)   \;.
\end{eqnarray}
The two definitions coincide in the infinite-volume limit $L \to\infty$.
Finally, let us emphasize that {\em neither}\/ of these quantities is equal
to the exponential correlation length (= 1/mass gap)
\be
   \xi_{exp}   \;=\;   \lim\limits_{|x| \to \infty}  {-|x|  \over  \log G(x)}
   \;.
\ee
However, $\xi$ (or $\xi'$) and $\xi_{exp}$ are believed to scale
in the same way as $\beta \to\infty$.

\medskip

The integrated autocorrelation time $\tau_{int,A}$ can be estimated
by standard procedures of statistical time-series analysis
\cite{Priestley_81,Anderson_71}.
These procedures also give statistically valid {\em error bars}\/
on $\< A \>$ and $\tau_{int,A}$.
For more details, see \cite[Appendix C]{Madras_88}.
In this paper we have used in all cases a
self-consistent truncation window of width $6 \tau_{int,A}$.
In setting error bars on $\xi$ we have used the triangle inequality;
such error bars are overly conservative, but we did not feel it was worth
the trouble to measure the cross-correlations between $\scrm^2$ and $\scrf$.

\subsection{Numerical Results}\label{s4.2}

In Table \ref{table_codim1} we present results using the codimension-1
embedding algorithm (i.e.\ the usual Wolff embedding);
the induced Ising model is simulated by $N_{hit}$ hits of the
standard SW algorithm.
Additional data on larger lattices (but only for $N_{hit} = 1$)
can be found in \cite{Edwards_89}.
In Tables \ref{table_codim2_32} and \ref{table_codim2_64}
we present the analogous results for the codimension-2 algorithm;
here the induced Ising model is simulated by $N_{hit}$ hits of the
standard SW algorithm generalized in the obvious way to a mixed
ferromagnetic/antiferromagnetic Ising model.\protect\footnote{The
   data points at $\beta = 2.20$, $N_{hit} = 1$ for $L=32,64$
   may be unreliable, because the data discarded at the beginning of the run
   may have been insufficient to ensure equilibrium.
   (Indeed, the values for $\chi$, $\xi$ and $E$ seem systematically low.)
   Unfortunately, the raw data files for these runs were lost,
   so we are unable to reanalyze them with a larger discard interval.
   We include these data points in Tables
   \ref{table_codim2_32} and \ref{table_codim2_64}
   simply to show the order of magnitude of $\taux$ and $\taue$.
   We do {\em not}\/ include them in the table of merged static data
   (Table \ref{o4_merged_table1}).}
\vspace{.1cm}

\protect\begin{table}[p]
{\small
\vspace*{-1cm}
\addtolength{\tabcolsep}{-1.0mm}
\hspace*{-0.2cm}
 \protect\footnotesize
\begin{tabular}{|r|c|r|r|r@{\ (}r r@{\ (}r r@{\ (}r|r@{\ (}r r@{\ (}r|} \hline
\multicolumn{14}{|c|}{Codimension $= 1$} \\ \hline
$L$ & $\beta$ & $N_{hit}$ & Sweeps
  & \multicolumn{2}{|c}{$\chi$}
  & \multicolumn{2}{c}{$\xi$}
  & \multicolumn{2}{c|}{$E$}
  & \multicolumn{2}{c}{$\taux$}
  & \multicolumn{2}{c|}{$\taue$} \\
\hline\hline
 32 & 1.70 &   1 & 100000 &   27.4 &  0.2) &  3.54 & 0.01) & 0.98438 & 0.00037)
     &   4.53 &   0.16)& 11.39 & 0.61)\\
 32 & 1.70 &  10 &  50000 &   28.0 &  0.2) &  3.71 & 0.02) & 0.98400 & 0.00037)
     &   3.61 &   0.15)&  5.80 & 0.31)\\
 32 & 1.70 &  20 &  50000 &   27.4 &  0.2) &  3.58 & 0.02) & 0.98399 & 0.00039)
     &   3.76 &   0.16)&  6.11 & 0.33)\\
 32 & 1.70 & $\infty$ & & \multicolumn{6}{|c|}{\ }
                                & 3.68 &   0.11)&  5.96 & 0.23)\\
\hline
 32 & 1.80 &   1 &  50000 &   41.4 &  0.4) &  4.70 & 0.03) & 1.04361 & 0.00057)
     &   5.77 &   0.31)& 13.57 & 1.11)\\
 32 & 1.80 &  10 &  50000 &   40.9 &  0.3) &  4.61 & 0.03) & 1.04445 & 0.00039)
     &   4.04 &   0.18)&  6.34 & 0.36)\\
 32 & 1.80 &  20 &  50000 &   41.2 &  0.3) &  4.65 & 0.03) & 1.04422 & 0.00039)
     &   4.44 &   0.21)&  6.76 & 0.39)\\
 32 & 1.80 & $\infty$ & & \multicolumn{6}{|c|}{\ }
                                & 4.24 &   0.14)&  6.55 & 0.27)\\
\hline
 32 & 1.90 &   1 &  50000 &   62.8 &  0.6) &  6.03 & 0.04) & 1.10225 & 0.00054)
     &   6.80 &   0.40)& 13.00 & 1.05)\\
 32 & 1.90 &  10 &  50000 &   61.6 &  0.5) &  5.92 & 0.03) & 1.10092 & 0.00037)
     &   4.37 &   0.21)&  6.35 & 0.36)\\
 32 & 1.90 &  20 &  50000 &   63.3 &  0.5) &  6.08 & 0.04) & 1.10168 & 0.00037)
     &   4.83 &   0.23)&  6.21 & 0.34)\\
 32 & 1.90 & $\infty$ & & \multicolumn{6}{|c|}{\ }
                                & 4.60 &   0.16)&  6.28 & 0.25)\\
\hline
 32 & 2.00 &   1 &  50000 &   94.4 &  0.8) &  7.76 & 0.05) & 1.15555 & 0.00053)
     &   7.77 &   0.48)& 13.46 & 1.10)\\
 32 & 2.00 &  10 &  52000 &   93.4 &  0.7) &  7.64 & 0.04) & 1.15518 & 0.00039)
     &   5.59 &   0.29)&  7.60 & 0.46)\\
 32 & 2.00 &  20 &  50000 &   94.0 &  0.7) &  7.71 & 0.04) & 1.15491 & 0.00038)
     &   5.53 &   0.29)&  7.16 & 0.42)\\
 32 & 2.00 & $\infty$ & & \multicolumn{6}{|c|}{\ }
                                & 5.56 &   0.21)&  7.38 & 0.31)\\
\hline
 32 & 2.10 &   1 &  60000 &  134.9 &  1.0) &  9.64 & 0.06) & 1.20484 & 0.00045)
     &   9.51 &   0.60)& 13.57 & 1.01)\\
 32 & 2.10 &  10 &  50000 &  136.3 &  0.8) &  9.78 & 0.05) & 1.20390 & 0.00035)
     &   5.92 &   0.32)&  6.89 & 0.40)\\
 32 & 2.10 &  20 &  50000 &  134.1 &  0.9) &  9.58 & 0.05) & 1.20396 & 0.00036)
     &   6.18 &   0.34)&  6.90 & 0.40)\\
 32 & 2.10 & $\infty$ & & \multicolumn{6}{|c|}{\ }
                                & 6.05 &   0.23)&  6.89 & 0.28)\\
\hline
 32 & 2.20 &   1 &  60000 &  178.2 &  1.1) & 11.59 & 0.07) & 1.24760 & 0.00047)
     &  10.61 &   0.70)& 15.73 & 1.27)\\
 32 & 2.20 &  10 &  52000 &  177.4 &  0.9) & 11.52 & 0.06) & 1.24765 & 0.00033)
     &   6.23 &   0.34)&  7.02 & 0.41)\\
 32 & 2.20 &  20 &  50000 &  180.1 &  0.9) & 11.64 & 0.06) & 1.24872 & 0.00035)
     &   6.39 &   0.36)&  7.42 & 0.45)\\
 32 & 2.20 & $\infty$ & & \multicolumn{6}{|c|}{\ }
                                & 6.31 &   0.25)&  7.22 & 0.30)\\
\hline
\hline
 64 & 2.00 &   1 &  60000 &   98.6 &  0.8) &  7.72 & 0.05) & 1.15384 & 0.00023)
     &   4.76 &   0.21)& 12.39 & 0.89)\\
 64 & 2.00 &  10 &  50000 &   99.7 &  0.8) &  7.87 & 0.05) & 1.15400 & 0.00018)
     &   3.79 &   0.17)&  6.48 & 0.37)\\
 64 & 2.00 &  20 &  50000 &   99.4 &  0.8) &  7.91 & 0.05) & 1.15353 & 0.00019)
     &   3.94 &   0.17)&  6.92 & 0.40)\\
 64 & 2.00 & $\infty$ & & \multicolumn{6}{|c|}{\ }
                                & 3.87 &   0.12)&  6.70 & 0.27)\\
\hline
 64 & 2.10 &   1 &  60000 &  160.2 &  1.4) & 10.38 & 0.06) & 1.20229 & 0.00023)
     &   6.24 &   0.32)& 13.73 & 1.03)\\
 64 & 2.10 &  10 &  50000 &  163.1 &  1.4) & 10.61 & 0.06) & 1.20145 & 0.00018)
     &   4.76 &   0.23)&  7.23 & 0.43)\\
 64 & 2.10 &  20 &  50000 &  159.1 &  1.3) & 10.28 & 0.06) & 1.20191 & 0.00018)
     &   4.31 &   0.20)&  6.90 & 0.40)\\
 64 & 2.10 & $\infty$ & & \multicolumn{6}{|c|}{\ }
                                & 4.54 &   0.15)&  7.07 & 0.29)\\
\hline
 64 & 2.20 &   1 &  60000 &  256.4 &  2.2) & 13.70 & 0.08) & 1.24507 & 0.00022)
     &   7.49 &   0.42)& 14.17 & 1.08)\\
 64 & 2.20 &  10 &  50000 &  258.2 &  1.9) & 13.75 & 0.08) & 1.24552 & 0.00017)
     &   4.86 &   0.24)&  7.24 & 0.44)\\
 64 & 2.20 &  20 &  50000 &  257.1 &  2.0) & 13.70 & 0.08) & 1.24535 & 0.00017)
     &   5.32 &   0.27)&  6.90 & 0.40)\\
 64 & 2.20 & $\infty$ & & \multicolumn{6}{|c|}{\ }
                                & 5.09 &   0.18)&  7.07 & 0.30)\\
\hline
 64 & 2.30 &   1 &  60000 &  393.4 &  2.8) & 17.75 & 0.10) & 1.28437 & 0.00021)
     &   8.06 &   0.47)& 13.75 & 1.03)\\
 64 & 2.30 &  10 &  50000 &  397.7 &  2.7) & 17.88 & 0.10) & 1.28444 & 0.00016)
     &   6.04 &   0.33)&  7.38 & 0.45)\\
 64 & 2.30 &  20 &  50000 &  395.1 &  2.6) & 17.82 & 0.10) & 1.28444 & 0.00017)
     &   5.74 &   0.31)&  7.22 & 0.43)\\
 64 & 2.30 & $\infty$ & & \multicolumn{6}{|c|}{\ }
                                & 5.89 &   0.23)&  7.30 & 0.31)\\
\hline
 64 & 2.40 &   1 &  60000 &  549.6 &  3.3) & 21.76 & 0.12) & 1.31948 & 0.00020)
     &   8.93 &   0.54)& 13.87 & 1.05)\\
 64 & 2.40 &  10 &  50000 &  554.5 &  3.0) & 22.04 & 0.12) & 1.31945 & 0.00016)
     &   6.54 &   0.37)&  7.66 & 0.47)\\
 64 & 2.40 &  20 &  50000 &  556.8 &  3.1) & 22.09 & 0.12) & 1.31952 & 0.00016)
     &   6.67 &   0.39)&  7.57 & 0.47)\\
 64 & 2.40 & $\infty$ & & \multicolumn{6}{|c|}{\ }
                                & 6.61 &   0.27)&  7.62 & 0.33)\\
\hline
 64 & 2.50 &   1 &  60000 &  701.1 &  3.4) & 25.54 & 0.13) & 1.35087 & 0.00019)
     &   9.56 &   0.60)& 14.43 & 1.11)\\
 64 & 2.50 &  10 &  50000 &  708.8 &  3.0) & 25.85 & 0.12) & 1.35091 & 0.00014)
     &   6.18 &   0.34)&  7.11 & 0.42)\\
 64 & 2.50 &  20 &  50250 &  708.1 &  3.0) & 25.91 & 0.12) & 1.35080 & 0.00015)
     &   6.22 &   0.34)&  7.14 & 0.42)\\
 64 & 2.50 & $\infty$ & & \multicolumn{6}{|c|}{\ }
                                & 6.20 &   0.24)&  7.12 & 0.30)\\
\hline
\end{tabular}
\caption{
   Results at $L=32$ and $L=64$ from the codimension-1 algorithm.
   The runs were either started from an equilibrium configuration
   or else had $1000$ sweeps discarded for equilibration.
   Rows labelled $N_{hit} = \infty$ are averages of the $N_{hit} = 10,20$
   values.
   Standard error is shown in parentheses.
}
\label{table_codim1}
}
\end{table}

%
%
\protect\begin{table}[p]
\vspace*{-1.3cm}
{\small
\addtolength{\tabcolsep}{-1.0mm}
\hspace*{-0.3cm}
 \protect\footnotesize
\begin{tabular}{|r|c|r|r|r@{\ (}r r@{\ (}r r@{\ (}r|r@{\ (}r r@{\ (}r|} \hline
\multicolumn{14}{|c|}{Codimension $= 2$} \\ \hline
$L$ & $\beta$ & $N_{hit}$ & Sweeps
  & \multicolumn{2}{|c}{$\chi$}
  & \multicolumn{2}{c}{$\xi$}
  & \multicolumn{2}{c|}{$E$}
  & \multicolumn{2}{c}{$\taux$}
  & \multicolumn{2}{c|}{$\taue$} \\   \hline
\hline
 32 & 1.70 &   1 & 100000 &   28.0 &  0.2) &  3.68 & 0.02) & 0.98453 & 0.00042)
       &   8.92 &   0.42)& 14.06 & 0.82)\\
 32 & 1.70 &   5 & 100000 &   27.8 &  0.2) &  3.68 & 0.01) & 0.98398 & 0.00028)
       &   3.57 &   0.11)&  6.75 & 0.28)\\
 32 & 1.70 &  10 &  50000 &   28.3 &  0.2) &  3.75 & 0.02) & 0.98460 & 0.00036)
       &   2.80 &   0.10)&  5.63 & 0.30)\\
 32 & 1.70 &  20 &  50000 &   28.0 &  0.2) &  3.69 & 0.02) & 0.98465 & 0.00037)
       &   2.51 &   0.09)&  5.72 & 0.31)\\
 32 & 1.70 &  40 &  50000 &   27.6 &  0.2) &  3.57 & 0.02) & 0.98422 & 0.00037)
       &   2.30 &   0.08)&  5.86 & 0.32)\\
 32 & 1.70 &  80 &  80000 &   27.7 &  0.1) &  3.65 & 0.01) & 0.98350 & 0.00028)
       &   2.24 &   0.06)&  5.45 & 0.22)\\
 32 & 1.70 & 160 & 100000 &   28.0 &  0.1) &  3.70 & 0.01) & 0.98440 & 0.00026)
       &   2.23 &   0.05)&  5.76 & 0.22)\\
\hline
 32 & 1.80 &   1 & 200000 &   42.5 &  0.4) &  4.83 & 0.03) & 1.04459 & 0.00031)
       &  20.63 &   1.03)& 16.62 & 0.75)\\
 32 & 1.80 &   5 & 300000 &   41.1 &  0.2) &  4.63 & 0.01) & 1.04460 & 0.00018)
       &   7.33 &   0.18)&  8.06 & 0.21)\\
 32 & 1.80 &  10 & 250000 &   41.3 &  0.2) &  4.67 & 0.01) & 1.04479 & 0.00018)
       &   5.25 &   0.12)&  7.25 & 0.19)\\
 32 & 1.80 &  20 & 250000 &   41.3 &  0.1) &  4.68 & 0.01) & 1.04447 & 0.00018)
       &   4.17 &   0.09)&  6.72 & 0.17)\\
 32 & 1.80 &  40 & 250000 &   41.4 &  0.1) &  4.68 & 0.01) & 1.04477 & 0.00018)
       &   3.69 &   0.07)&  6.95 & 0.18)\\
 32 & 1.80 &  80 & 100000 &   41.1 &  0.2) &  4.65 & 0.02) & 1.04444 & 0.00028)
       &   3.37 &   0.10)&  6.76 & 0.28)\\
 32 & 1.80 & 160 &  80000 &   41.6 &  0.2) &  4.71 & 0.02) & 1.04429 & 0.00031)
       &   3.15 &   0.10)&  6.61 & 0.30)\\
 32 & 1.80 & 320 & 100000 &   41.5 &  0.2) &  4.69 & 0.02) & 1.04440 & 0.00027)
       &   3.19 &   0.09)&  6.38 & 0.25)\\
\hline
 32 & 1.90 &   1 & 200000 &   61.7 &  0.7) &  5.93 & 0.05) & 1.10152 & 0.00033)
       &  43.76 &   3.18)& 20.17 & 1.00)\\
 32 & 1.90 &   5 & 300000 &   62.0 &  0.4) &  5.94 & 0.02) & 1.10167 & 0.00019)
       &  14.80 &   0.51)&  9.74 & 0.27)\\
 32 & 1.90 &  10 & 300000 &   62.5 &  0.3) &  6.00 & 0.02) & 1.10175 & 0.00018)
       &  10.79 &   0.32)&  8.65 & 0.23)\\
 32 & 1.90 &  20 & 250000 &   62.2 &  0.3) &  5.98 & 0.02) & 1.10143 & 0.00019)
       &   8.58 &   0.25)&  7.97 & 0.22)\\
 32 & 1.90 &  40 & 250000 &   61.8 &  0.3) &  5.93 & 0.02) & 1.10167 & 0.00019)
       &   6.95 &   0.18)&  7.98 & 0.22)\\
 32 & 1.90 &  80 & 100000 &   62.0 &  0.4) &  5.97 & 0.03) & 1.10147 & 0.00030)
       &   6.24 &   0.25)&  8.30 & 0.37)\\
 32 & 1.90 & 160 & 100000 &   62.3 &  0.4) &  5.96 & 0.03) & 1.10212 & 0.00029)
       &   5.54 &   0.21)&  7.69 & 0.34)\\
 32 & 1.90 & 320 & 100000 &   61.9 &  0.4) &  5.95 & 0.02) & 1.10156 & 0.00029)
       &   5.39 &   0.20)&  7.80 & 0.34)\\
 32 & 1.90 & 640 &  50000 &   61.7 &  0.5) &  5.92 & 0.03) & 1.10120 & 0.00040)
       &   4.95 &   0.24)&  7.33 & 0.44)\\
\hline
 32 & 2.00 &   1 & 200000 &   93.2 &  1.4) &  7.63 & 0.09) & 1.15507 & 0.00037)
       &  92.52 &   9.76)& 26.83 & 1.52)\\
 32 & 2.00 &   5 & 300000 &   93.8 &  0.7) &  7.67 & 0.04) & 1.15510 & 0.00020)
       &  29.77 &   1.46)& 12.41 & 0.39)\\
 32 & 2.00 &  10 & 300000 &   92.7 &  0.6) &  7.61 & 0.03) & 1.15493 & 0.00018)
       &  22.17 &   0.94)& 10.20 & 0.29)\\
 32 & 2.00 &  20 & 250000 &   94.3 &  0.5) &  7.74 & 0.03) & 1.15522 & 0.00019)
       &  16.42 &   0.66)&  9.58 & 0.29)\\
 32 & 2.00 &  40 & 250000 &   93.4 &  0.5) &  7.64 & 0.03) & 1.15517 & 0.00020)
       &  12.52 &   0.44)&  9.45 & 0.29)\\
 32 & 2.00 &  80 & 100000 &   94.8 &  0.7) &  7.76 & 0.04) & 1.15542 & 0.00030)
       &  10.85 &   0.56)&  9.17 & 0.44)\\
 32 & 2.00 & 160 & 100000 &   94.5 &  0.7) &  7.73 & 0.04) & 1.15532 & 0.00031)
       &  10.16 &   0.50)&  9.64 & 0.47)\\
 32 & 2.00 & 320 & 100000 &   93.4 &  0.7) &  7.64 & 0.04) & 1.15552 & 0.00032)
       &   9.64 &   0.47)&  9.97 & 0.49)\\
 32 & 2.00 & 640 & 100000 &   94.5 &  0.6) &  7.76 & 0.04) & 1.15494 & 0.00032)
       &   9.24 &   0.44)&  9.73 & 0.47)\\
\hline
 32 & 2.10 &   5 & 105000 &  137.4 &  1.6) &  9.90 & 0.10) & 1.20400 & 0.00034)
       &  45.50 &   4.68)& 13.66 & 0.77)\\
 32 & 2.10 &  10 & 300000 &  134.5 &  0.8) &  9.60 & 0.10) & 1.20388 & 0.00020)
       &  35.91 &   1.93)& 12.78 & 0.41)\\
 32 & 2.10 &  20 & 300000 &  134.5 &  0.7) &  9.61 & 0.04) & 1.20391 & 0.00019)
       &  26.97 &   1.26)& 11.74 & 0.36)\\
 32 & 2.10 &  40 & 300000 &  134.4 &  0.7) &  9.62 & 0.04) & 1.20394 & 0.00018)
       &  21.30 &   0.88)& 10.76 & 0.32)\\
 32 & 2.10 &  80 & 400000 &  134.3 &  0.5) &  9.60 & 0.03) & 1.20391 & 0.00016)
       &  17.66 &   0.58)& 10.80 & 0.28)\\
 32 & 2.10 & 160 & 400000 &  134.8 &  0.4) &  9.66 & 0.03) & 1.20386 & 0.00016)
       &  14.88 &   0.45)& 10.77 & 0.28)\\
 32 & 2.10 & 320 & 105000 &  135.4 &  0.9) &  9.70 & 0.10) & 1.20409 & 0.00031)
       &  14.76 &   0.87)& 11.17 & 0.57)\\
 32 & 2.10 & 640 & 100000 &  134.1 &  0.9) &  9.59 & 0.05) & 1.20390 & 0.00031)
       &  13.16 &   0.74)& 10.28 & 0.51)\\
\hline
 32 & 2.20 &   1 & 250000 &  172.4 &  2.2) & 11.13 & 0.14) & 1.24682 & 0.00033)
       & 177.34 &  23.39)& 34.40 & 2.00)\\
 32 & 2.20 &   5 & 301000 &  180.0 &  1.1) & 11.68 & 0.07) & 1.24783 & 0.00020)
       &  55.67 &   3.72)& 15.41 & 0.54)\\
 32 & 2.20 &  10 & 300000 &  178.7 &  1.0) & 11.59 & 0.06) & 1.24791 & 0.00019)
       &  44.06 &   2.62)& 13.19 & 0.43)\\
 32 & 2.20 &  20 & 300000 &  179.7 &  0.9) & 11.68 & 0.06) & 1.24783 & 0.00018)
       &  34.04 &   1.78)& 12.42 & 0.39)\\
 32 & 2.20 &  40 & 100000 &  179.7 &  1.3) & 11.64 & 0.08) & 1.24786 & 0.00032)
       &  28.45 &   2.36)& 12.74 & 0.71)\\
 32 & 2.20 &  80 & 100000 &  179.0 &  1.2) & 11.59 & 0.08) & 1.24791 & 0.00031)
       &  23.63 &   1.78)& 11.38 & 0.60)\\
 32 & 2.20 & 160 & 100100 &  178.8 &  1.2) & 11.59 & 0.08) & 1.24801 & 0.00031)
       &  21.67 &   1.56)& 12.14 & 0.66)\\
 32 & 2.20 & 320 & 100000 &  179.7 &  1.1) & 11.66 & 0.07) & 1.24805 & 0.00031)
       &  19.23 &   1.31)& 11.61 & 0.62)\\
\hline
\end{tabular}
\caption{
   Results at $L=32$ from the codimension-2 algorithm.
   All runs were started in equilibrium,
   except for $\beta=2.20$, $N_{hit}=1$,
   where 5000 sweeps (probably not enough) were discarded for equilibration.
   Standard error is shown in parentheses.
}
\label{table_codim2_32}
}
\end{table}

%
%
%

%
%

\protect\begin{table}
\vspace*{-1.2cm}
{\small
\addtolength{\tabcolsep}{-1.0mm}
\hspace*{-0.5cm}
 \protect\footnotesize
\begin{tabular}{|r|c|r|r|r@{\ (}r r@{\ (}r r@{\ (}r|r@{\ (}r r@{\ (}r|} \hline
\multicolumn{14}{|c|}{Codimension $= 2$} \\ \hline
$L$ & $\beta$ & $N_{hit}$ & Sweeps
  & \multicolumn{2}{|c}{$\chi$}
  & \multicolumn{2}{c}{$\xi$}
  & \multicolumn{2}{c|}{$E$}
  & \multicolumn{2}{c}{$\taux$}
  & \multicolumn{2}{c|}{$\taue$} \\   \hline
\hline
 64 & 2.00 &   1 & 200000 &   99.7 &  1.9) &  7.86 & 0.10) & 1.15404 & 0.00017)
       &  89.48 &   9.28)& 24.59 & 1.34)\\
 64 & 2.00 &   5 & 300500 &   98.8 &  0.9) &  7.79 & 0.05) & 1.15368 & 0.00010)
       &  31.71 &   1.60)& 11.73 & 0.36)\\
 64 & 2.00 &  10 & 250000 &  100.3 &  0.8) &  7.93 & 0.05) & 1.15382 & 0.00010)
       &  21.06 &   0.95)&  9.98 & 0.31)\\
 64 & 2.00 &  20 & 200000 &   98.6 &  0.8) &  7.80 & 0.04) & 1.15368 & 0.00011)
       &  15.99 &   0.70)&  9.56 & 0.33)\\
 64 & 2.00 &  40 & 100200 &  100.1 &  1.0) &  7.88 & 0.06) & 1.15392 & 0.00015)
       &  12.32 &   0.67)&  9.19 & 0.44)\\
 64 & 2.00 &  80 & 100000 &  100.3 &  0.9) &  7.97 & 0.05) & 1.15371 & 0.00015)
       &   9.59 &   0.46)&  9.06 & 0.43)\\
 64 & 2.00 & 160 & 100000 &   98.9 &  0.8) &  7.75 & 0.04) & 1.15374 & 0.00014)
       &   7.43 &   0.32)&  8.40 & 0.38)\\
 64 & 2.00 & 320 & 100000 &   99.1 &  0.7) &  7.80 & 0.04) & 1.15381 & 0.00014)
       &   6.53 &   0.26)&  8.61 & 0.39)\\
\hline
 64 & 2.10 &   1 &  50000 &  152.8 &  9.9) & 10.00 & 0.40) & 1.20202 & 0.00036)
       & 292.77 & 110.90)& 30.05 & 3.66)\\
 64 & 2.10 &   5 & 950000 &  159.9 &  1.3) & 10.36 & 0.05) & 1.20184 & 0.00006)
       &  81.32 &   3.70)& 13.58 & 0.25)\\
 64 & 2.10 &  10 & 700000 &  158.9 &  1.2) & 10.26 & 0.05) & 1.20180 & 0.00006)
       &  56.19 &   2.47)& 12.06 & 0.25)\\
 64 & 2.10 &  20 & 750000 &  159.7 &  1.0) & 10.35 & 0.04) & 1.20189 & 0.00005)
       &  40.65 &   1.48)& 11.16 & 0.21)\\
 64 & 2.10 &  40 & 400000 &  161.7 &  1.2) & 10.51 & 0.05) & 1.20181 & 0.00008)
       &  29.75 &   1.26)& 10.37 & 0.26)\\
 64 & 2.10 &  80 & 400000 &  161.1 &  1.1) & 10.47 & 0.05) & 1.20179 & 0.00008)
       &  24.17 &   0.92)& 10.19 & 0.26)\\
 64 & 2.10 & 160 & 100000 &  159.0 &  2.0) & 10.29 & 0.09) & 1.20197 & 0.00015)
       &  20.22 &   1.42)& 10.26 & 0.52)\\
 64 & 2.10 & 320 & 100250 &  162.2 &  1.8) & 10.50 & 0.10) & 1.20184 & 0.00015)
       &  16.39 &   1.04)& 10.55 & 0.54)\\
 64 & 2.10 & 640 &  25000 &  162.5 &  3.3) & 10.48 & 0.15) & 1.20178 & 0.00028)
       &  13.96 &   1.62)&  9.06 & 0.85)\\
\hline
 64 & 2.20 &   1 & 200000 &  244.3 & 10.3) & 12.88 & 0.39) & 1.24504 & 0.00020)
       & 540.92 & 141.41)& 36.60 & 2.49)\\
 64 & 2.20 &  5 & 1000000 &  256.5 &  2.5) & 13.69 & 0.10) & 1.24521 & 0.00006)
       & 165.44 &  10.43)& 15.61 & 0.30)\\
 64 & 2.20 & 10 & 1000000 &  258.1 &  2.0) & 13.75 & 0.08) & 1.24532 & 0.00005)
       & 109.43 &   5.61)& 13.98 & 0.26)\\
 64 & 2.20 &  20 & 800000 &  261.9 &  2.0) & 14.01 & 0.08) & 1.24540 & 0.00006)
       &  84.23 &   4.24)& 12.99 & 0.26)\\
 64 & 2.20 &  40 & 600000 &  257.0 &  1.9) & 13.71 & 0.08) & 1.24534 & 0.00006)
       &  60.90 &   3.01)& 12.15 & 0.27)\\
 64 & 2.20 &  80 & 400100 &  260.0 &  2.2) & 13.94 & 0.08) & 1.24541 & 0.00008)
       &  50.95 &   2.82)& 11.79 & 0.32)\\
 64 & 2.20 & 160 & 400000 &  255.6 &  2.0) & 13.65 & 0.08) & 1.24536 & 0.00008)
       &  45.13 &   2.35)& 11.50 & 0.30)\\
 64 & 2.20 & 320 & 200199 &  254.3 &  2.5) & 13.54 & 0.10) & 1.24527 & 0.00011)
       &  35.28 &   2.30)& 11.26 & 0.42)\\
 64 & 2.20 & 640 & 102840 &  251.1 &  3.2) & 13.36 & 0.12) & 1.24524 & 0.00015)
       &  29.52 &   2.46)& 11.74 & 0.62)\\
\hline
\end{tabular}
\caption{
   Results at $L=64$ from the codimension-2 algorithm.
   All runs were started in equilibrium,
   except for $\beta=2.20$, $N_{hit}=1$,
   where 10000 sweeps (probably not enough) were discarded for equilibration.
   Standard error is shown in parentheses.
}
\label{table_codim2_64}
}
\end{table}

\protect\begin{table}[p]
{\small
 \protect\footnotesize
\begin{center}
\begin{tabular}{|r|c|r@{\ (}r r@{\ (}r r@{\ (}c|} \hline
\multicolumn{8}{|c|}{Merged $O(4)$-Model Static Data} \\ \hline
$L$&$\beta$
  &\multicolumn{2}{c}{$\chi$}
  &\multicolumn{2}{c}{$\xi$}
  &\multicolumn{2}{c|}{$E$}  \\   \hline
 32 & 1.62 &   21.0 &  0.1) &  3.09 & 0.02) & 0.93417 & 0.00054)\\
 32 & 1.70 &   27.8 &  0.0) &  3.65 & 0.01) & 0.98418 & 0.00011)\\
 32 & 1.80 &   41.3 &  0.0) &  4.67 & 0.01) & 1.04456 & 0.00007)\\
 32 & 1.90 &   62.1 &  0.1) &  5.96 & 0.01) & 1.10161 & 0.00007)\\
 32 & 2.00 &   93.7 &  0.1) &  7.68 & 0.01) & 1.15509 & 0.00004)\\
 32 & 2.10 &  134.6 &  0.2) &  9.63 & 0.01) & 1.20398 & 0.00004)\\
 32 & 2.20 &  178.8 &  0.2) & 11.60 & 0.02) & 1.24787 & 0.00004)\\
 32 & 2.30 &  221.1 &  0.3) & 13.42 & 0.02) & 1.28685 & 0.00004)\\
 32 & 2.40 &  258.5 &  0.3) & 14.94 & 0.03) & 1.32155 & 0.00007)\\
\hline
 64 & 1.70 &   27.9 &  0.2) &  3.65 & 0.02) & \multicolumn{2}{c|}{\ }\\
 64 & 1.80 &   41.2 &  0.3) &  4.60 & 0.10) & \multicolumn{2}{c|}{\ }\\
 64 & 1.90 &   64.1 &  0.5) &  6.10 & 0.10) & \multicolumn{2}{c|}{\ }\\
 64 & 2.00 &   99.4 &  0.2) &  7.83 & 0.01) & 1.15378 & 0.00004)\\
 64 & 2.10 &  160.5 &  0.3) & 10.40 & 0.01) & 1.20184 & 0.00002)\\
 64 & 2.15 &  202.9 &  0.7) & 11.93 & 0.03) & 1.22410 & 0.00003)\\
 64 & 2.20 &  257.9 &  0.4) & 13.77 & 0.02) & 1.24532 & 0.00001)\\
 64 & 2.22 &  283.3 &  1.0) & 14.58 & 0.04) & 1.25349 & 0.00003)\\
 64 & 2.24 &  309.5 &  1.0) & 15.38 & 0.04) & 1.26149 & 0.00003)\\
 64 & 2.26 &  337.4 &  1.1) & 16.20 & 0.04) & 1.26932 & 0.00003)\\
 64 & 2.28 &  364.7 &  1.2) & 16.93 & 0.04) & 1.27692 & 0.00003)\\
 64 & 2.30 &  395.0 &  0.7) & 17.81 & 0.03) & 1.28442 & 0.00001)\\
 64 & 2.35 &  471.5 &  1.3) & 19.87 & 0.05) & 1.30244 & 0.00003)\\
 64 & 2.40 &  551.6 &  0.8) & 21.92 & 0.03) & 1.31950 & 0.00001)\\
 64 & 2.43 &  601.5 &  1.3) & 23.22 & 0.05) & 1.32931 & 0.00002)\\
 64 & 2.45 &  632.5 &  1.2) & 23.97 & 0.05) & 1.33563 & 0.00002)\\
 64 & 2.50 &  706.7 &  0.9) & 25.77 & 0.04) & 1.35092 & 0.00001)\\
 64 & 2.55 &  779.0 &  1.1) & 27.49 & 0.05) & 1.36545 & 0.00002)\\
 64 & 2.60 &  847.9 &  1.0) & 29.06 & 0.04) & 1.37928 & 0.00002)\\
\hline
\end{tabular}
\end{center}
\caption{
   Best estimates of susceptibility, correlation length and energy
   for the $O(4)$ model on $32 \times 32$ and $64 \times 64$ lattices,
   from Tables \protect\ref{table_codim1}--\protect\ref{table_codim2_64}
   and \protect\cite{Edwards_89,MGMC_O4}.
   Standard error is shown in parentheses.
}
\label{o4_merged_table1}
}
\end{table}

\vspace{.1cm}

In Table \ref{o4_merged_table1}
we summarize, for the convenience of the reader,
our best estimates of the static quantities $\chi$, $\xi$ and $E$
on lattices of size $L = 32, 64$.
These estimates come from merging all the data in
Tables \ref{table_codim1}--\ref{table_codim2_64}
together with the data from our earlier work
on the codimension-1 algorithm \cite{Edwards_89}
and from a separate work using the
multi-grid Monte Carlo algorithm \cite{MGMC_O4}.\footnote{The
   $\beta = 2.20$, $N_{hit} = 1$ points from
   Tables \ref{table_codim2_32} and \ref{table_codim2_64}
   are {\em not}\/ included in these means.
   In all other cases, $\chi$, $\xi$ and $E$ from the different algorithms
   agree within error bars;  this is strong evidence that all the programs
   are correct!}
We use these merged data in our finite-size-scaling analyses
whenever $\xi$ is required.
(An analogous table for $L=128,256$ can be found in \cite{MGMC_O4}.)

In the codimension-1 algorithm, the induced Ising model is ferromagnetic,
and 10 SW hits are already enough to produce an ``almost independent''
sample from the induced Ising model.  As a result, the autocorrelation times
for $N_{hit} = 10, 20$ are equal within error bars,
and their average can be taken to represent the idealized ($N_{hit} = \infty$)
algorithm, as shown in Table \ref{table_codim1}.

%
%
\protect\begin{table}
\begin{center}
\begin{tabular}{|r|c|r@{\ (}c|} \hline
\multicolumn{4}{|c|}{Codimension = 2} \\ \hline
$L$&$\beta$ &\multicolumn{2}{c|}{Frustration} \\   \hline
\hline
 32 & 1.70 &    0.2571 & 0.0001)  \\
 32 & 1.80 &    0.2452 & 0.0001)  \\
 32 & 1.90 &    0.2329 & 0.0001)  \\
 32 & 2.00 &    0.2205 & 0.0001)  \\
 32 & 2.10 &    0.2086 & 0.0001)  \\
 32 & 2.20 &    0.1973 & 0.0001)  \\
\hline
 64 & 2.00 &    0.2209 & 0.0001)  \\
 64 & 2.10 &    0.2092 & 0.0001)  \\
 64 & 2.20 &    0.1980 & 0.0001)  \\
\hline
\end{tabular}
\end{center}
\caption{
  Average fraction of frustrated plaquettes in the induced Ising model
  arising from the codimension-2 algorithm.
  Standard error is shown in parentheses.
}
\label{table_frust}
\end{table}

In the codimension-2 algorithm, by contrast, the induced Ising model is
highly frustrated (see Table \ref{table_frust}),
and all known algorithms for updating this model
are extremely inefficient.  Both the standard SW algorithm
(as generalized to mixed ferromagnetic/antiferromagnetic Ising models)
and the single-site Metropolis algorithm have {\em very slow}\/
convergence to equilibrium, which gets worse as $\beta$ and $L$ are increased.
A combination of SW and Metropolis appears to be no better (in rate of
convergence per unit CPU time) than SW alone.
Therefore, we had no choice but to simulate the induced Ising model
with the best algorithm available to us --- namely, standard SW ---
and to use an {\em enormous}\/ number of hits in an effort to approximate
the behavior of the idealized embedding algorithm.

(After our runs were completed, Kandel, Ben-Av and Domany
\cite{Kandel_frustrated} reported very encouraging results for simulating
a frustrated Ising model using an ingenious new algorithm
of Swendsen-Wang type.
Their algorithm is apparently successful in cases of
{\em full}\/ frustration, but not in cases of partial frustration.
In the codimension-2 $O(4)$ application at the $\beta$ and $L$ values
studied here, the induced Ising model has approximately
20--25\% of the plaquettes frustrated.
We therefore do not expect the Kandel {\em et al.}\/ algorithm to work
miracles, but even a factor-of-2 speedup over standard SW would be
highly desirable.  We have not yet had an opportunity to try the Kandel
{\em et al.}\/ algorithm in our induced Ising model, but we hope to do
so in the near future.)

\begin{figure}
\epsfxsize=\columnwidth
\centerline{\epsffile{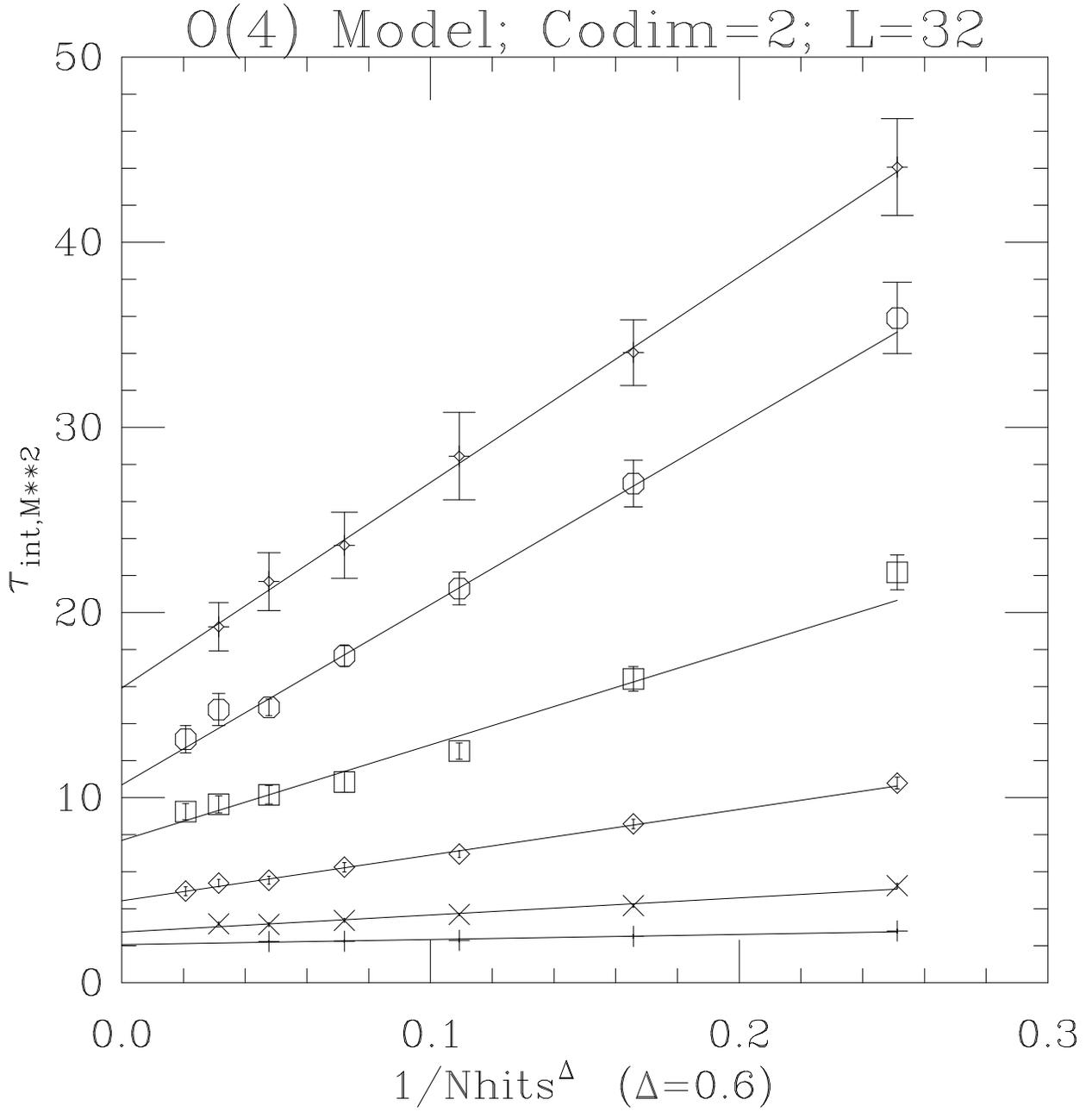}}
\caption{
  $\taux$ versus $1/N_{hit}^\Delta$ with $\Delta = 0.6$,
  for $L=32$ and $\beta =$ 1.7 ($+$), 1.8 ($\times$),
  1.9 ($\Diamond$), 2.0 ($\Box$), 2.1 ($\bigcirc$),
  2.2 ($\circ$).
}
\label{fig_extrap_32_0.6}
\end{figure}

\begin{figure}
\epsfxsize=\columnwidth
\epsffile{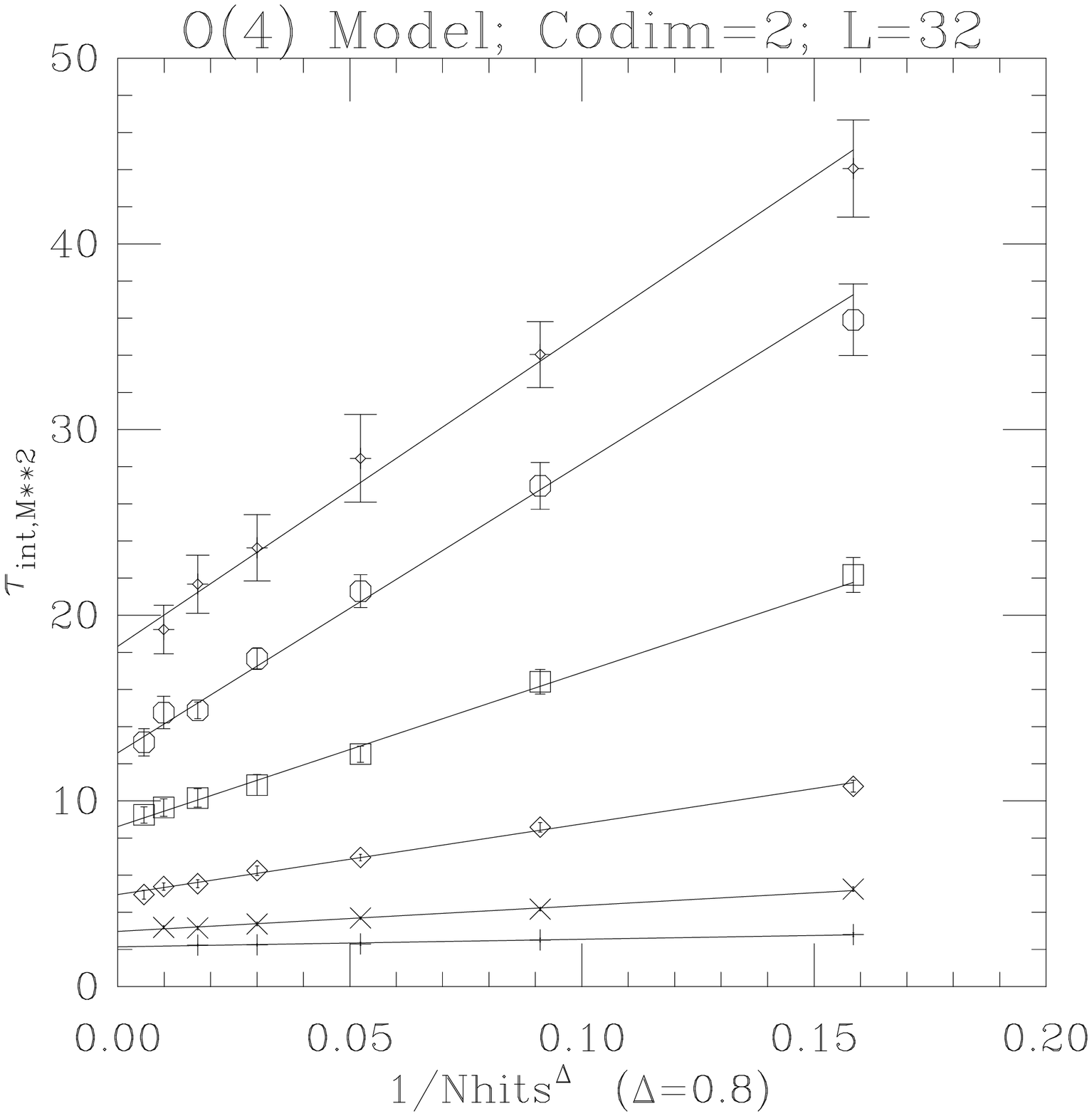}
\caption{
  $\taux$ versus $1/N_{hit}^\Delta$ with $\Delta = 0.8$,
  for $L=32$ and $\beta =$ 1.7 ($+$), 1.8 ($\times$),
  1.9 ($\Diamond$), 2.0 ($\Box$), 2.1 ($\bigcirc$),
  2.2 ($\circ$).
}
\label{fig_extrap_32_0.8}
\end{figure}

\begin{figure}
\epsfxsize=\columnwidth
\epsffile{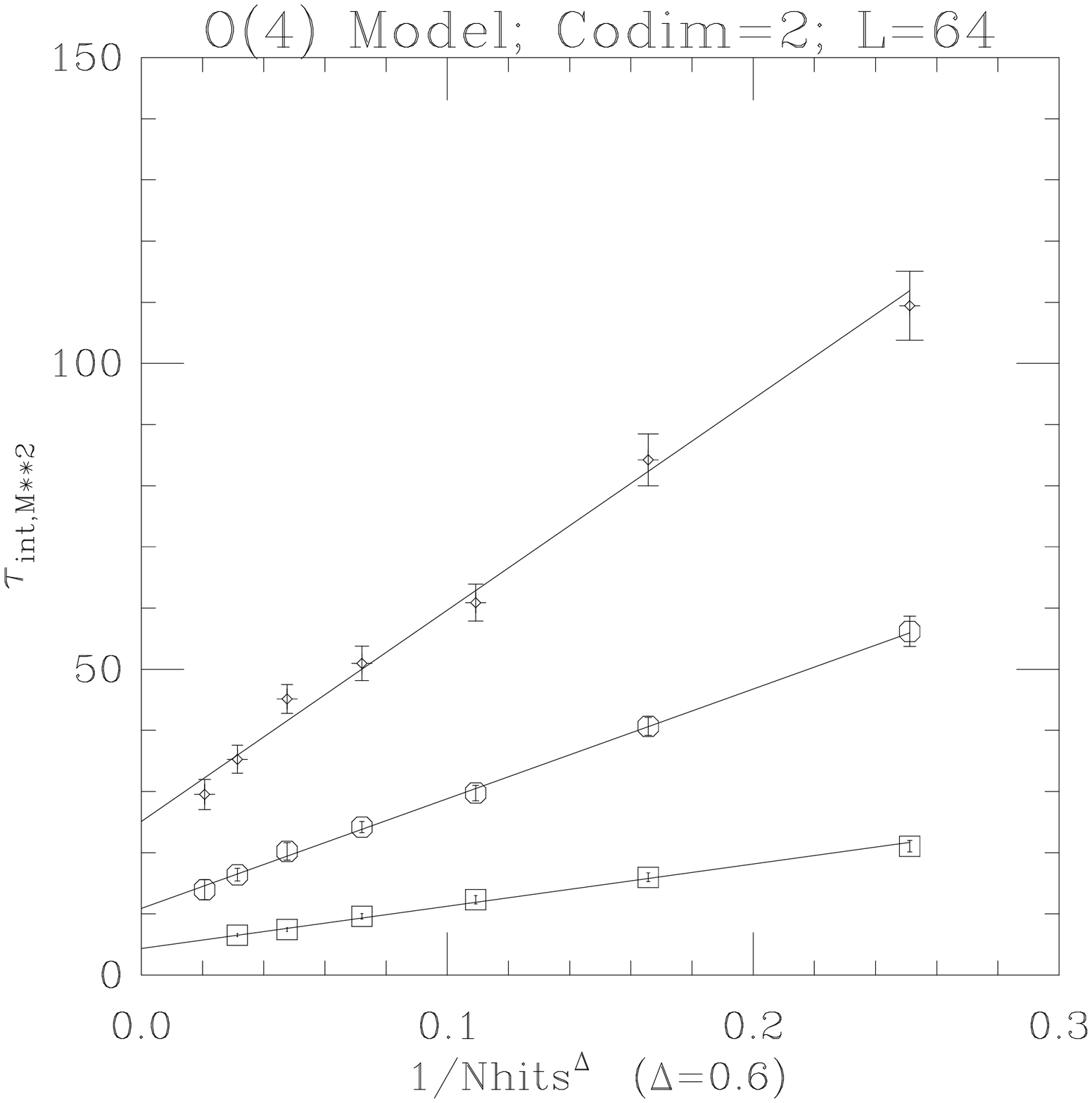}
\caption{
  $\taux$ versus $1/N_{hit}^\Delta$ with $\Delta = 0.6$,
  for $L=64$ and $\beta =$ 2.0 ($\Box$), 2.1 ($\bigcirc$),
  2.2 ($\circ$).
}
\label{fig_extrap_64_0.6}
\end{figure}

{}From Tables \ref{table_codim2_32} and \ref{table_codim2_64}, we see
that even at $N_{hit} = 640$, the autocorrelation time $\taux$ for the
codimension-2 algorithm has not stabilized (except for $L=32$ and
$\beta \ltapprox 1.8$).  Therefore, we are obliged to attempt an {\em
extrapolation}\/ of our data to $N_{hit} = \infty$.  This turns out to
be a very tricky business, as the behavior of $\taux$ as a function of
$N_{hit}$ is extremely complicated (and we have no {\em theoretical}\/
understanding of it).  We tried empirical fits of the form
\be
   \taux(N_{hit})   \;=\;   a + {b \over N_{hit}^\Delta}
 \label{extrap_ansatz}
\ee
where $a \equiv \taux(N_{hit} = \infty)$ and $b$ are variable,
and $\Delta$ is some fixed exponent.  (We also tried fits with $\Delta$
variable, but these fits were quite unstable, and the error bars on
$\Delta$ were large.) Reasonable fits can be obtained provided that one
discards the data from the lowest values of $N_{hit}$ (namely,
$N_{hit} = 1,5$); but the optimal exponent $\Delta$ is not well
determined, and it moreover shows a clear (and more-or-less but not
quite monotonic) variation with $\beta$ and $L$.  Of course, we have no
explanation whatsoever for these empirical observations.  (That is, we
have no {\em theoretical}\/ understanding of the dynamic behavior of
the SW algorithm on a highly frustrated Ising model.) Therefore, the
best we can do is to report the results of our extrapolations ---
allowing for a wide range of values of $\Delta$ --- and let the reader
judge their reasonableness.  In Figures \ref{fig_extrap_32_0.6},
\ref{fig_extrap_32_0.8} and \ref{fig_extrap_64_0.6} we plot $\taux$
versus $1/N_{hit}^\Delta$ for a few selected values of $\Delta$.  In
Table \ref{codim2_extrap_M2} we show the extrapolated values of
$\taux(N_{hit} = \infty)$ as a function of the exponent $\Delta$, based
on a least-squares fit to the data with $N_{hit} \ge 10$.  Our
subjective selection of the ``best'' fit is marked in boldface.  The
$\Delta$ values in a range $\pm 0.1$ around this ``best'' value yield
also reasonable fits;  we take these three extrapolants as a 68\%
subjective confidence interval on $\taux(N_{hit} = \infty)$.  Clearly
there is a very wide range of ``reasonable'' extrapolants, especially
for $L=64$ at the higher values of $\beta$.  Our final results will
therefore have a very large systematic uncertainty.
%
%
%

%
%
%
%
%
%

%
%
%
{\small
\begin{table}
\addtolength{\tabcolsep}{-1.5mm}
\setlength{\doublerulesep}{2.0mm}
\hspace*{-2.1cm}
 \protect\footnotesize
\begin{tabular}{|c|ccccccccc|}
\hline
$\beta\;\backslash\;\Delta$ &0.4& 0.5& 0.6& 0.7& 0.8& 0.9& 1.0& 1.1& 1.2 \\
\hline\hline
1.70  & 1.93(0.08)  & 2.01(0.07)  & 2.07(0.06)  & 2.11(0.05)  &
  2.14(0.05)  & 2.16(0.04)  & {\bf 2.18(0.04)}  & 2.19(0.04)  & 2.20(0.04)  \\
{[3 DF]}& 2.71  & 2.05  & 1.50  & 1.06  & 0.73  & 0.50  & 0.36 & 0.32  & 0.34\\
\hline
1.80  & 2.32(0.10)  & 2.57(0.08)  & 2.74(0.07)  & 2.87(0.06)  &
  2.96(0.06)  & {\bf 3.04(0.06)}  & 3.10(0.05)  & 3.16(0.05)  & 3.20(0.05)  \\
{[4 DF]}& 19.17  & 12.80  & 7.86  & 4.34  & 2.16  & 1.20 & 1.31 & 2.30 & 4.03\\
\hline
1.90  & 3.45(0.19)  & 4.04(0.16)  & 4.44(0.14)  & {\bf 4.73(0.13)}  &
  {\bf 4.95(0.12)}  & 5.12(0.12)  & 5.26(0.11)  & 5.38(0.11)  & 5.48(0.10)  \\
{[5 DF]}& 11.00  & 5.48  & 2.25  & 1.22  & 2.16  & 4.81  & 8.86  &
                                                            14.01  & 19.94  \\
\hline
2.00  & 5.91(0.43)  & 6.98(0.36)  & 7.70(0.32)  & 8.22(0.29)  &
   {\bf 8.62(0.27)}  & 8.94(0.26)  & 9.20(0.25)  & 9.42(0.24)  & 9.60(0.24)  \\
{[5 DF]}& 22.16  & 14.70  & 8.78  & 4.51  & 1.87  & 0.77  & 1.04  &
                                                               2.50  & 4.93  \\
\hline
2.10  & 7.10(0.70)  & 9.23(0.57)  & 10.69(0.49)  & {\bf 11.77(0.43)}  &
   12.59(0.39)  & 13.23(0.36)  & 13.75(0.34)  & 14.17(0.33)  & 14.51(0.32)  \\
{[5 DF]}& 11.16  & 5.92  & 2.88  & 1.94  & 2.88  & 5.42  & 9.25  &
                                                            14.08  & 19.60  \\
\hline
2.20  & 10.99(1.60)  & 13.96(1.34)  & {\bf 15.92(1.18)}  & 17.31(1.08)
  & 18.33(1.01)   & 19.12(0.96)  & 19.74(0.92)  & 20.24(0.89)  & 20.66(0.87)\\
{[4 DF]}& 1.10  & 0.43  & 0.20  & 0.38  & 0.93  & 1.82 & 2.99 & 4.39  & 5.98\\
\hline \multicolumn{10}{c}{\quad} \\ \hline
2.00  & 1.52(0.40)  & 3.23(0.32)  & {\bf 4.36(0.27)}  & 5.15(0.24)
  & 5.73(0.22)   & 6.16(0.21)  & 6.50(0.20)  & 6.76(0.20)  & 6.97(0.19)  \\
{[4 DF]}& 2.80  & 1.09  & 1.65  & 4.32  & 8.85  & 14.92  & 22.21  &
                                                         30.40  & 39.19  \\
\hline
2.10  & 3.67(1.16)  & 7.98(0.97)  & {\bf 10.90(0.85)}  & 13.03(0.76)
  & 14.65(0.70)   & 15.93(0.66)  & 16.97(0.63)  & 17.82(0.61)  & 18.53(0.59)\\
{[5 DF]}& 6.03  & 2.11  & 0.99  & 2.48  & 6.28  & 12.01  & 19.30  &
                                                              27.76  & 37.05\\
\hline
2.20  & 12.01(2.34)  & {\bf 19.84(1.94)}  & 25.09(1.70)  & 28.85(1.54)
  & 31.66(1.44)   & 33.84(1.36)  & 35.56(1.30)  & 36.95(1.26)  & 38.08(1.23)\\
{[5 DF]}& 4.43  & 3.48  & 4.58  & 7.53  & 12.07  & 17.88  & 24.63  &
                                                              32.03  & 39.79\\
\hline
\end{tabular}
\caption{
   Extrapolated values of $\taux(N_{hit} = \infty)$ for $L=32$
   (upper half of table) and $L=64$ (lower half),
   based on a least-squares fit to (\protect\ref{extrap_ansatz})
   using the data for $N_{hit} \ge 10$;
   error bars are one standard deviation, {\em statistical error only}\/.
   The second line is the value of $\chi^2$;
   number of degrees of freedom is indicated in the first column.
   Our subjective selection of the ``best'' fit is marked in boldface.
}
\label{codim2_extrap_M2}
\end{table}
}

For the autocorrelation time $\taue$,
the extrapolation to $N_{hit} = \infty$ is fortunately less problematic:
in most cases $\taue$ is constant within error bars for
$N_{hit} \gtapprox 40$, or at worst it shows in this region
a very slow decrease as a function of $N_{hit}$.
In Table \ref{codim2_extrap_E}
we show the extrapolated values of $\taue(N_{hit} = \infty)$
as a function of the exponent $\Delta$,
based on a least-squares fit to the data with $N_{hit} \ge 10$.
Here the extrapolation is in most cases insensitive to the choice of $\Delta$,
and there is no discernible systematic preference for one or another
value of $\Delta$.  Therefore we have chosen (rather arbitrarily)
$\Delta = 1$ as the ``best'' extrapolant,
and the range $0.8 \le \Delta \le 1.2$ as defining a
68\% subjective confidence interval on $\taue(N_{hit} = \infty)$.

%
%
%
{\small
\begin{table}
\addtolength{\tabcolsep}{-1.5mm}
\setlength{\doublerulesep}{2.0mm}
\hspace*{-1.9cm}
 \protect\footnotesize
\begin{tabular}{|c|ccccccccc|}
\hline
$\beta\;\backslash\;\Delta$ &0.4& 0.5& 0.6& 0.7& 0.8& 0.9& 1.0& 1.1& 1.2 \\
\hline\hline
1.70  & 5.65(0.30)  & 5.65(0.25)  & 5.65(0.22)  & 5.65(0.20)  &
  5.66(0.18)   & 5.66(0.17)  & {\bf 5.66(0.16)}  & 5.66(0.16)  & 5.66(0.15) \\
{[3 DF]}& 1.55  & 1.55  & 1.55  & 1.55  & 1.55  & 1.56  & 1.56 & 1.56 & 1.56\\
\hline
1.80  & 6.28(0.24)  & 6.37(0.20)  & 6.44(0.18)  & 6.49(0.16)  &
  6.52(0.15)   & 6.55(0.14)  & {\bf 6.57(0.14)}  & 6.59(0.13)  & 6.61(0.13) \\
{[4 DF]}& 3.21  & 3.23  & 3.25  & 3.29  & 3.32  & 3.36  & 3.40 & 3.43 & 3.47\\
\hline
1.90  & 7.36(0.27)  & 7.48(0.23)  & 7.55(0.21)  & 7.61(0.19)  &
  7.65(0.18)   & 7.68(0.17)  & {\bf 7.71(0.16)}  & 7.73(0.16)  & 7.75(0.15) \\
{[5 DF]}& 3.76  & 3.65  & 3.56  & 3.49  & 3.43  & 3.39  & 3.35 & 3.33 & 3.31\\
\hline
2.00  & 9.33(0.34)  & 9.37(0.29)  & 9.39(0.26)  & 9.41(0.24)  &
  9.42(0.23)   & 9.43(0.22)  & {\bf 9.44(0.21)}  & 9.45(0.20)  & 9.46(0.19) \\
{[5 DF]}& 4.31  & 4.07  & 3.83  & 3.60  & 3.39  & 3.19  & 3.01 & 2.85 & 2.70\\
\hline
2.10  & 9.77(0.32)  & 10.02(0.27)  & 10.19(0.24)  & 10.31(0.22)  &
  10.40(0.20) & 10.47(0.19) & {\bf 10.52(0.18)} & 10.57(0.17)  & 10.61(0.17)\\
{[5 DF]}& 5.59 & 4.79 & 4.12  & 3.57  & 3.13  & 2.79  & 2.56  & 2.40  & 2.32\\
\hline
2.20  & 11.06(0.57)  & 11.28(0.49)  & 11.42(0.44)  & 11.52(0.40)  &
  11.60(0.38) & 11.66(0.36) & {\bf 11.71(0.34)} & 11.75(0.33)  & 11.78(0.32)\\
{[4 DF]}& 2.06  & 1.98  & 1.93  & 1.90  & 1.89  & 1.90 & 1.92 & 1.96  & 2.00\\
\hline \multicolumn{10}{c}{\quad} \\ \hline
2.00 & 8.00(0.37) & 8.22(0.31) & 8.36(0.28) & 8.46(0.26) & 8.54(0.24)
  &  8.60(0.23) & {\bf 8.65(0.22)} & 8.69(0.22) & 8.72(0.21)  \\
{[4 DF]}& 0.77 & 0.84 & 0.97 & 1.13 & 1.32 & 1.53 & 1.76 & 2.00 & 2.24  \\
\hline
2.10  & 8.96(0.34)  & 9.29(0.29)  & 9.51(0.25)  & 9.67(0.23)
   & 9.80(0.21) & 9.90(0.20) & {\bf 9.98(0.19)} & 10.04(0.18)  & 10.10(0.17)\\
{[5 DF]}& 4.81  & 4.06  & 3.52  & 3.15  & 2.94  & 2.86 & 2.89 & 3.02  & 3.23\\
\hline
2.20 & 10.34(0.30) & 10.70(0.25) & 10.95(0.22) & 11.13(0.21) &
  11.26(0.19) &  11.37(0.18) & {\bf 11.45(0.17)} & 11.52(0.17) & 11.58(0.16)\\
{[5 DF]}& 2.31 & 1.49 & 1.01 & 0.84 & 0.92 & 1.20 & 1.65 & 2.22 & 2.88  \\
\hline
\end{tabular}
\caption{
   Extrapolated values of $\taue(N_{hit} = \infty)$ for $L=32$
   (upper half of table) and $L=64$ (lower half),
   based on a least-squares fit to (\protect\ref{extrap_ansatz})
   using the data for $N_{hit} \ge 10$;
   error bars are one standard deviation, {\em statistical error only}\/.
   The second line is the value of $\chi^2$;
   number of degrees of freedom is indicated in the first column.
   Our subjective selection of the ``best'' fit is marked in boldface.
}
\label{codim2_extrap_E}
\end{table}
}

We can now make a finite-size-scaling analysis of the dynamic critical
behavior, using the Ansatz
\be
  \tau_{int,A}(\beta,L)  \;\sim\;
     \xi(\beta,L)^{z_{int,A}} \, g_A \Bigl( \xi(\beta,L)/L  \Bigr)
\ee
for $A = \scrm^2, \scre$.
Here $g_A$ is an unknown scaling function, and
$g_A(0) = \lim_{x \downarrow 0} g_A(x)$
is supposed to be finite and nonzero.\footnote{
   It is of course equivalent to use the Ansatz
   $$\tau_{int,A}(\beta,L)  \;\sim\;
     L^{z_{int,A}} \, h_A \Bigl( \xi(\beta,L)/L  \Bigr)  \;,$$
   and indeed the two Ans\"atze are related by $h_A(x) = x^{z_{int,A}} g_A(x)$.
   However, to determine whether
   $\lim_{x \downarrow 0} g_A(x) = \lim_{x \downarrow 0} x^{-z_{int,A}} h_A(x)$
   is nonzero, it is more convenient to inspect a graph of $g_A$
   than one of $h_A$.
}
We emphasize that the dynamic critical exponent $z_{int,A}$
is in general {\em different}\/ from the exponent $z_{exp}$
associated with the exponential autocorrelation time $\tau_{exp}$
\cite{Sokal_Lausanne,Caracciolo_90,Sokal_LAT90}.

\begin{figure}
\epsfxsize=\columnwidth
\epsffile{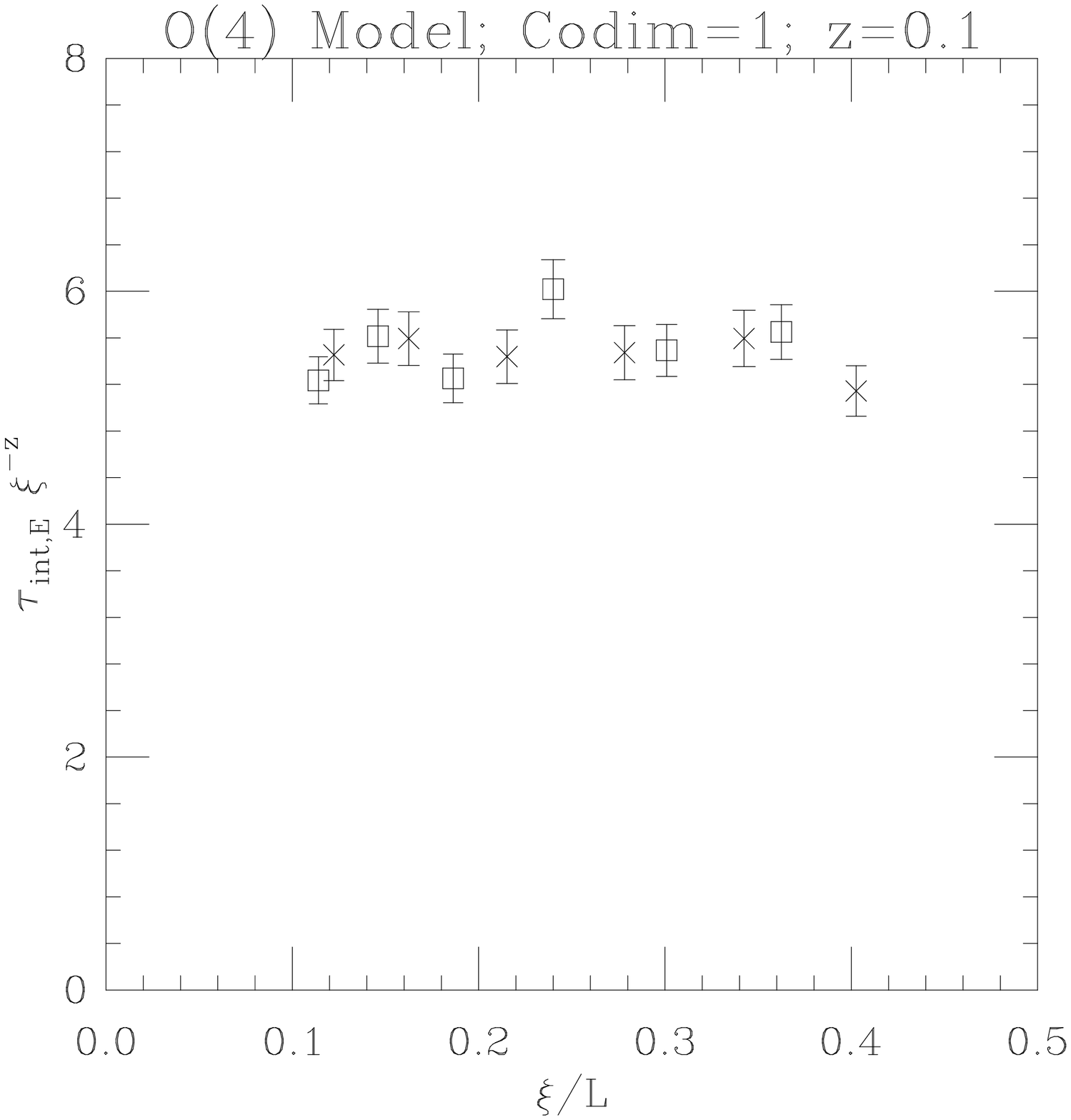}
\caption{
  Dynamic finite-size-scaling plot of $\taue \xi^{-z_{int,\scre}}$
  versus $\xi/L$ for $z_{int,\scre} = 0.1$,
  for the idealized ($N_{hit} = \infty$) codimension-1 algorithm
  on $L=32$ ($\Box$) and $L=64$ ($\times$).
}
\label{swwo4c1.extrap.tauE.0p1}
\end{figure}

\begin{figure}
\epsfxsize=\columnwidth
\epsffile{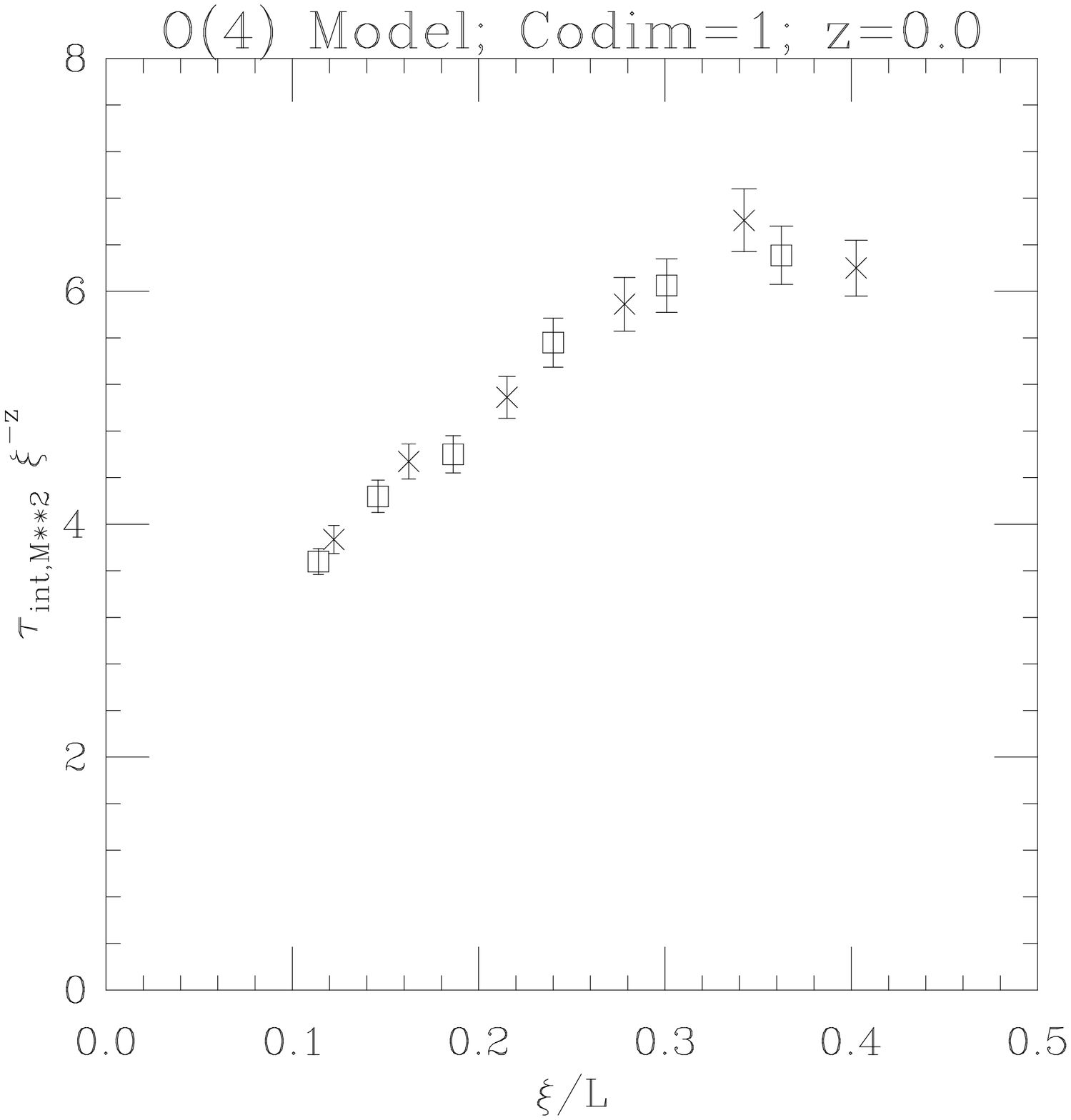}
\caption{
  Dynamic finite-size-scaling plot of $\taux \xi^{-z_{int,\scrm^2}}$
  versus $\xi/L$ for $z_{int,\scrm^2} = 0$,
  for the idealized ($N_{hit} = \infty$) codimension-1 algorithm
  on $L=32$ ($\Box$) and $L=64$ ($\times$).
}
\label{swwo4c1.extrap.tauM2.0p0}
\end{figure}

For the codimension-1 idealized algorithm ($N_{hit} = \infty$),
the dynamic critical exponents are very close to zero:
we estimate $z_{int,\scre} = 0.1 \pm 0.1$ and
$z_{int,\scrm^2} = 0.0 \pm 0.1$ (subjective 68\% confidence intervals).
Finite-size-scaling plots of $\tau_{int,A} \xi^{-z_{int,A}}$
versus $\xi/L$ are shown in Figures
\ref{swwo4c1.extrap.tauE.0p1} and \ref{swwo4c1.extrap.tauM2.0p0}.
These estimates are of course in agreement with our previous results
\cite{Edwards_89}
showing that even for $N_{hit} = 1$ we have $z$ very close to zero.

%
%
%
%
%
%

\begin{figure}
\epsfxsize=\columnwidth
\epsffile{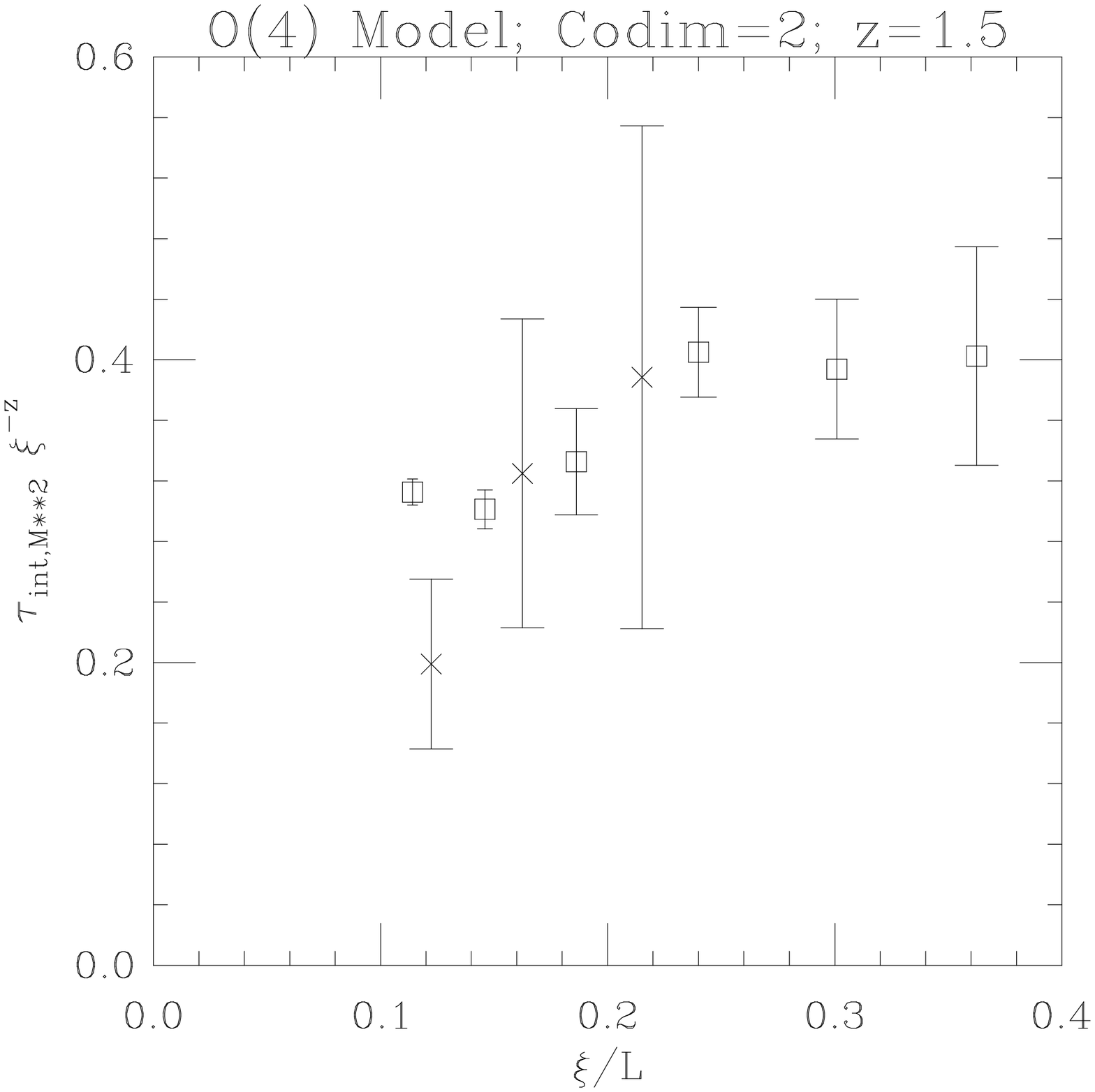}
\caption{
  Dynamic finite-size-scaling plot of $\taux \xi^{-z_{int,\scrm^2}}$
  versus $\xi/L$ for $z_{int,\scrm^2} = 1.5$,
  for the idealized ($N_{hit} = \infty$) codimension-2 algorithm.
  Data points are from the extrapolations in Table
  \protect\ref{codim2_extrap_M2}
  with $L=32$ ($\Box$) and $L=64$ ($\times$).
}
\label{swwo4c2.extrap.tauM2.1p5}
\end{figure}

\begin{figure}
\epsfxsize=\columnwidth
\epsffile{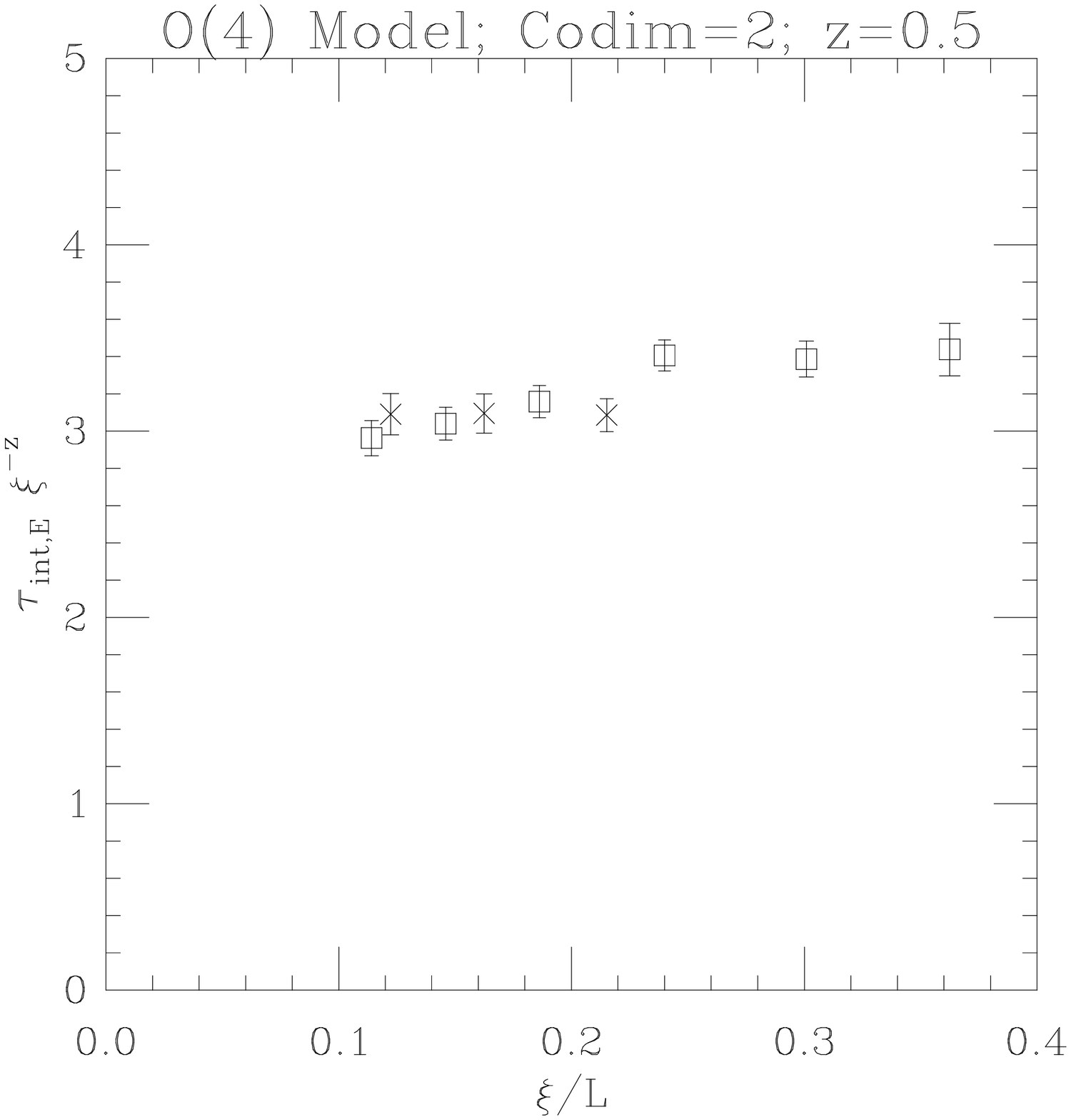}
\caption{
  Dynamic finite-size-scaling plot of $\taue \xi^{-z_{int,\scre}}$
  versus $\xi/L$ for $z_{int,\scre} = 0.5$,
  for the idealized ($N_{hit} = \infty$) codimension-2 algorithm.
  Data points are from the extrapolations in Table
  \protect\ref{codim2_extrap_E}
  with $L=32$ ($\Box$) and $L=64$ ($\times$).
}
\label{swwo4c2.extrap.tauE.0p5}
\end{figure}

For the codimension-2 idealized algorithm,
a similar analysis using the extrapolated data from
Tables \ref{codim2_extrap_M2} and \ref{codim2_extrap_E}
yields the estimates $z_{int,\scrm^2} = 1.5 \pm 0.5$ and
$z_{int,\scre} = 0.5 \pm 0.2$ (subjective 68\% confidence intervals).
The finite-size-scaling plots corresponding to these
values of $z$ are shown in Figures
\ref{swwo4c2.extrap.tauM2.1p5} and \ref{swwo4c2.extrap.tauE.0p5}.
Clearly the estimate for $z_{int,\scrm^2}$ is
{\em very}\/ imprecise, as a result of the large systematic error bars
in the extrapolation of $\taux$ for $L=64$.
All we can say with much certainty is that $z_{int,\scrm^2}$
is far from zero;  it is consistent with our data,
but not at all guaranteed, that $z_{int,\scrm^2} \approx 2$.

%
%
%
%
%
%

It is interesting to note the relative sizes of $\taux$ as compared to $\taue$
(and also of $z_{int,\scrm^2}$ as compared to $z_{int,\scre}$).
For the traditional local algorithms (e.g.\ single-site heat-bath),
$\scrm^2$ is a much slower mode than $\scre$, because it is much more
strongly coupled to the long-wavelength spin waves that evolve slowly
in the local dynamics.\footnote{
   See e.g.\ \cite{MGMC_O4} for some quantitative measurements.
}
For the codimension-1 idealized Wolff algorithm, by contrast,
the two observables have roughly equal autocorrelation times
(and in fact $\scre$ is a little slower):
the Wolff collective moves are apparently equally effective at equilibrating
fluctuations on all length scales.
For the codimension-2 idealized algorithm, however, $\scrm^2$ is again
much slower than $\scre$:  this confirms our view that the codimension-2
reflection is ineffective at equilibrating long-wavelength spin waves,
and that its effect is ``primarily local''.

We wish to emphasize once again the importance of studying the
{\em idealized}\/ embedding algorithm,
and the misleading conclusions that can be caused by the failure to do so.
Indeed, suppose that we had studied the codimension-2 algorithm
only for $N_{hit} = 1$.
Then, by the usual dynamic finite-size-scaling analysis,
we would have concluded that $z_{int,\scrm^2} \approx 3$
and $z_{int,\scre} \approx 0.75$ ({\em very}\/ roughly),
i.e.\ that the codimension-2 embedding algorithm is disastrously bad.
Now, this conclusion happens to be true, but not for the reason claimed!
In fact, the enormous autocorrelation times $\taux$
in the $N_{hit} = 1$ algorithm reflect primarily the inability of the
SW algorithm to update efficiently the highly frustrated induced
Ising model (as one can verify by comparing to larger values of $N_{hit}$),
and {\em not}\/ any intrinsic defect of the codimension-2 embedding.
The intrinsic properties of the embedding are found
{\em only}\/ by considering $N_{hit} \to \infty$.

The CPU time for this program is approximately
$5.35 \times N_{hit} L^2$ $\mu$s/sweep on a Cray Y-MP 8/432.
Thus, the total CPU time for the the runs reported here is approximately
4000 Cray hours.\footnote{
   This is only an ``equivalent'' figure, as the runs were actually
   performed on a variety of supercomputers and RISC workstations:
   see the Acknowledgments.
}

\section{Discussion}  \label{s5}

Let us now place our analysis of Wolff-type embedding algorithms
into the more general context of {\em arbitrary}\/ embedding algorithms.
The general idea \cite{Sokal_LAT90} is to
``foliate'' the configuration space of the original model into ``leaves''
isomorphic to the configuration space of some ``embedded'' model.
One then moves around the current leaf, using any legitimate Monte Carlo
algorithm for simulating the conditional probability distribution restricted
to that leaf (i.e.\ the induced Hamiltonian for the embedded model).
Of course, one must combine this move with other moves, or with a different
foliation, in order to make the algorithm ergodic.
(This same structure arises also in multi-grid Monte Carlo,
where it is termed ``partial resampling''
\cite{Sokal_Lausanne,MGMC_PRL,MGMC_1}.)

As noted in the Introduction, the performance of an embedding algorithm
is determined by the combined effect of two {\em completely distinct}\/ issues:
\begin{itemize}
 \item[i)]  How well the embedding
     captures the important large-scale collective modes
     of the original model.  The point is that these modes
     must be realizable by motions {\em within}\/ a leaf.
 \item[ii)]  How well some particular algorithm
     (e.g.\ standard SW or single-cluster SW)
     succeeds in updating the embedded model.
\end{itemize}
It is crucial to employ a test procedure that {\em disentangles}\/ these
two issues, if one wants to obtain {\em physical insight}\/
into why a particular embedding algorithm does or does not work well.

In all of the recently-invented embedding algorithms ---
Wolff-type algorithms for nonlinear $\sigma$-models
and $N_t = 1$ $SU(2)$ gauge theories \cite{Evertz_lat90},
spin-flip algorithms for one-component
scalar-field models \cite{Brower-Tamayo},
and reflection algorithms for solid-on-solid (SOS)
and anharmonic-crystal models
\cite{SOS_embedding} ---
the embedded model happens to be an Ising model.
But the principle is much more general:
for example, one might consider embeddings
of $U(1)$ spins in a higher $\sigma$-model,
$U(1)$ or $Z_N$ spins in an $SU(N)$ gauge theory, etc.
Of course, the nontrivial problem is to find an embedding
that captures at least some of the important collective modes of the
original model, and once this has been done,
to find an efficient algorithm for updating the embedded model.

In Section \ref{s2} we explained why
an idealized embedding algorithm will do a good job
of handling long-wavelength spin waves {\em if}\/,
in the induced Hamiltonian for the embedded model,
$x$-space is divided
into large disconnected regions which are almost decoupled from each other.
We furthermore conjectured that in a Wolff-type embedding of
Ising spins into a nonlinear $\sigma$-model,
this is the {\em only}\/ mechanism by which
long-wavelength spin waves can be handled well;
and our numerical results in Section \ref{s4}
gave modest support for this conjecture in at least one case.
However, we wish to point out that whatever the status of this conjecture
for embeddings of {\em Ising}\/ spins, it certainly {\em cannot}\/ be true
as a general proposition about embedding algorithms.
To see this, consider the extreme case in which there is only one leaf
(namely, the whole configuration space):
then the {\em idealized}\/ embedding algorithm performs perfectly
(it is just independent sampling from the Gibbs measure of the original model),
but the induced Hamiltonian for the embedded model does not decouple anywhere
(it is just the original Hamiltonian).
This rather trivial extreme example shows that decoupling cannot be the
{\em only}\/ mechanism by which an {\em idealized}\/ embedding algorithm
can work well.
More generally, consider any two foliations $\scrf_1$ and $\scrf_2$
of the same configuration space,
such that $\scrf_1$ is a refinement of $\scrf_2$
(i.e.\ the leaves of $\scrf_2$ are unions of leaves from $\scrf_1$).
It is intuitively clear that the {\em idealized}\/ embedding algorithm
based on the foliation $\scrf_2$ (plus possibly other moves)
will perform better than the one based on foliation $\scrf_1$
(plus the same other moves):
in $\scrf_2$ there is more freedom to move around within a leaf.
On the other hand, the induced Hamiltonian based on $\scrf_2$
will have fewer (if any) surfaces in $x$-space along which it decouples,
compared to the induced Hamiltonian based on $\scrf_1$:
the larger the leaves, the more stringent the requirement
of a complete decoupling.
So ``performance'' and ``decoupling'' have opposite monotonicities
as a function of the ``size'' of the leaves.
It follows that decoupling {\em cannot}\/ be the general principle that
explains the good or bad performance of embedding algorithms.

In particular, it is not justified to insist
on the codimension-1 property when considering non-Ising embeddings ---
and this is fortunate, since by Theorem \ref{thmA.9}
(see also the remarks following it),
{\em only}\/ Ising embeddings can have the codimension-1 property!

There are many possibilities for embedding algorithms:
each such algorithm can be interpreted as
``reducing'' one simulation problem to another (hopefully simpler) one.
For example, consider a (possibly frustrated) $SU(N)$ principal chiral model
defined by the Hamiltonian
\be
   H( \{ U \} )   \;=\;   -\sum_{\< xy \>}
     \hbox{Re tr}  (A_{xy} U_x^\dagger U_y)
     \;,
\ee
where the $A_{xy}$ are $N \times N$ complex matrices.
Then one can embed $U(1)$ (= $XY$) spins as follows:
let $T = {\rm diag}(i,-i,0,\ldots,0) \in {\germansu}(N)$,
let $R$ be a random element of $SU(N)$, and embed the field $\{\theta_x\}$
of $XY$ spins according to the rule
\be
   U_x^{new}   \;=\;   R \exp(\theta_x T) R^{-1} U_x^{old}   \;.
\ee
It is easy to see that the induced $XY$ Hamiltonian is of the form
\be
   H(\{\theta\})   \;=\;   -\sum_{\< xy \>}
     \left[  \alpha_{xy} \cos(\theta_x - \theta_y)
           + \beta_{xy} \sin(\theta_x - \theta_y)
     \right]  \;,
\ee
i.e.\ it is a nearest-neighbor $XY$ model, with couplings that are in general
{\em frustrated}\/ (even if the original $SU(N)$ model is ferromagnetic).
We conjecture that the {\em idealized}\/ embedding algorithm
corresponding to this embedding has dynamic critical exponent $z \approx 0$,
at least if the $SU(N)$ model is ferromagnetic:
the idea is that the spin waves in $SU(N)$,
which are approximately Gaussian, can be obtained by superposing
spin waves in the various $U(1)$ subgroups generated by the $R T R^{-1}$.
If this is the case, then the problem of simulating an
$SU(N)$ principal chiral model (possibly even a frustrated one)
has been reduced to the problem of simulating a frustrated $XY$ model.
Unfortunately, we have no idea how to carry out efficiently the latter
simulation.\footnote{
   The multi-grid method \cite{MGMC_1,MGMC_XY} is applicable to the latter
   problem, and its performance might not be totally disastrous.
   But the multi-grid method is also applicable to the original $SU(N)$ model
   \cite{MGMC_1,MGMC_O4}, and it is easy to see that multi-grid updates on the
   induced $XY$ model are simply a subset of the multi-grid updates on the
   $SU(N)$ model.  Thus, at least for the idealized two-grid cycle and
   presumably also for the other cycles, nothing is gained by first embedding
   $U(1)$ variables.  An embedding is useful only if there exists an efficient
   algorithm for simulating the embedded model that for some reason
   does {\em not}\/ generalize to the original model.
}
But if, some day in the future, the latter problem should be solved,
then it is useful to know that the former problem would also be solved.

It is also interesting to note that the Wolff and multi-grid algorithms
can be understood from a unified perspective.  In both cases one exploits
an exact symmetry of the model (global reflection in the case of Wolff,
global rotation in the case of multi-grid) and applies it in an
inhomogeneous way (constant on clusters in the case of Wolff,
constant on cubical blocks in the case of multi-grid).
In both cases the energy cost is a surface term.
This perspective may be useful in suggesting generalizations of the
Wolff and/or multi-grid algorithms to broader classes of models,
especially lattice gauge theories.

For example, consider a $U(1)$ gauge theory with Hamiltonian
\be
   H  \;=\;  - \sum\limits_{P}  \beta_P
               \cos \left( \sum\limits_{\ell \in P} \theta_\ell \right)   \;,
\ee
where $\theta_\ell \in [0,2\pi]$ is a gauge potential on the oriented
link $\ell$, $\sum_{\ell \in P} \theta_\ell$ is a properly oriented sum
over the links bounding the plaquette $P$, and $\beta_P \ge 0$ for all $P$.
(Usually all $\beta_P$ will be equal.)
Next let $\{ \theta_\ell^\circ \}$ be an arbitrary curvature-free gauge field,
i.e.\ a ground state for the Hamiltonian $H$.\footnote{
   In free boundary conditions, such a field $\{ \theta_\ell^\circ \}$
   would be a gauge transform of the identity.
   However, in periodic boundary conditions (i.e.\ on the torus $T^d$)
   there are $d$ additional linearly independent possibilities for
   $\{ \theta_\ell^\circ \}$, which give arbitrary values to the
   $d$ independent Polyakov loops.  [In fancy language, these solutions
   are representatives of a basis for the first cohomology group
   $H^1(T^d; \R/\zed) \simeq (\R/\zed)^d$.]
}
Now it is easy to see that a global reflection of $\{ \theta_\ell \}$
around $\{ \theta_\ell^\circ \}$ is a symmetry of $H$.
Let us therefore consider applying this reflection in an
inhomogeneous way, i.e.\ let us consider \cite{Brower_private}
the embedding of Ising spins $\{ \varepsilon_\ell \}$ defined by
\be
 \label{u1_star1}
   \theta_\ell \;\longto\;
   \theta_\ell^{new} \,\equiv\,
     \theta_\ell^\circ + \varepsilon_\ell (\theta_\ell - \theta_\ell^\circ)
   \;.
\ee
Under this updating, the energy changes only on those plaquettes $P$
for which the $\varepsilon_\ell$ ($\ell \in P$) are not all equal;
heuristically, the energy cost is a ``surface term''.
Therefore, we conjecture that the {\em idealized}\/ embedding algorithm
based on \reff{u1_star1}
--- with, say, a random choice of $\{ \theta_\ell^\circ \}$ ---
has dynamic critical exponent $z \approx 0$.
Now, the induced Ising Hamiltonian corresponding to \reff{u1_star1}
is of the form
\be
 \label{u1_star2}
   H(\{\varepsilon_\ell\})   \;=\;
   - \sum_{P} \beta_P
   \left[ A_P \varepsilon_{\ell_1} \varepsilon_{\ell_2} \varepsilon_{\ell_3}
              \varepsilon_{\ell_4}
          + (B_{P,\ell_1 \ell_2} \varepsilon_{\ell_1} \varepsilon_{\ell_2}
          + \hbox{5 similar terms})
   \right]
   \,+\, {\rm const}
\ee
where $P = \{\ell_1,\ell_2,\ell_3,\ell_4\}$ and
\begin{subeqnarray}
   A_P   & = &   \sin(\theta_{\ell_1} - \theta_{\ell_1}^\circ)
                 \sin(\theta_{\ell_2} - \theta_{\ell_2}^\circ)
                 \sin(\theta_{\ell_3} - \theta_{\ell_3}^\circ)
                 \sin(\theta_{\ell_4} - \theta_{\ell_4}^\circ)      \\[2mm]
   B_{P,\ell_1 \ell_2}   & = &
             -   \sin(\theta_{\ell_1} - \theta_{\ell_1}^\circ)
                 \sin(\theta_{\ell_2} - \theta_{\ell_2}^\circ)
                 \cos(\theta_{\ell_3} - \theta_{\ell_3}^\circ)
                 \cos(\theta_{\ell_4} - \theta_{\ell_4}^\circ)      \\[2mm]
   {\rm etc.}  & &  \nonumber
\end{subeqnarray}
This is {\em not}\/ a $Z_2$ gauge model;
rather, it is a $Z_2$ spin model with a curious mixture of 2-spin and 4-spin
couplings.  Unfortunately, we have no idea how to simulate efficiently
such a model.\footnote{
   It is worth noting that the 4-spin couplings
   $A_P \varepsilon_{\ell_1} \varepsilon_{\ell_2}
        \varepsilon_{\ell_3} \varepsilon_{\ell_4}$
   can be eliminated in favor of 2-spin couplings, by introducing
   a new spin $\varepsilon_P$ at the center of each plaquette
   and coupling this spin individually to
   $\varepsilon_{\ell_1}, \varepsilon_{\ell_2},
    \varepsilon_{\ell_3}, \varepsilon_{\ell_4}$
   (analogously to the well-known ``star-triangle transformation'').
   The problem is then to find an efficient simulation algorithm
   for the resulting pair-interacting Ising model,
   which is in general {\em frustrated}\/.
}
But it would be worth investigating our conjecture
that the idealized embedding algorithm works well;
and if this conjecture is found to be true,
then it would be worth investigating algorithms for simulating \reff{u1_star2}.

One might try generalizing this algorithm to an $SU(2)$ gauge theory,
using the codimension-1 reflection
\be
 \label{su2_star3}
   U_\ell  \;\longto\;   - U_\ell^\circ U_\ell^\dagger U_\ell^\circ  \;,
\ee
where $\{ U_\ell^\circ \}$ is a random curvature-free gauge field
(see Example 4 of Section \ref{s3.1}).
Unfortunately, a global application of \reff{su2_star3}
is {\em not}\/ a symmetry of the Wilson Hamiltonian,
because
\be
 \label{su2_star4}
   \real \tr (U_1 U_2 U_3 U_4)   \;\neq\;
   \real \tr (U_1^\dagger U_2^\dagger U_3^\dagger U_4^\dagger)
\ee
for a non-Abelian group.  Therefore, the induced Ising Hamiltonian
would contain magnetic-field terms that contribute a {\em bulk}\/ energy,
and the idealized embedding algorithm would probably {\em not}\/ work well.
On the other hand, equality in \reff{su2_star4} {\em almost}\/ holds
if the fields $U_1,U_2,U_3,U_4$ are close to the identity.
Therefore, one might hope that for $\beta \gg 1$
(i.e.\ near the continuum limit), the idealized embedding algorithm
corresponding to \reff{su2_star3} would work well
{\em in Landau gauge}\/ with the choice $U_\ell^\circ \equiv I$.
Of course, to implement this idea in practice one would have to find
an efficient algorithm for Landau gauge-fixing ---
a nontrivial problem \cite{Davies_88,Mandula_90,Hulsebos_91}
for which the conventional algorithms also suffer from
critical slowing-down.
But the situation may not be completely hopeless.
(We remark that other collective-mode algorithms,
such as Fourier acceleration and multi-grid, may also need
Landau gauge-fixing in order to perform well in the
non-Abelian case \cite{FA_MG_gauge}.)

Let us conclude by mentioning several recent studies
of Wolff-type embedding algorithms
\cite{Hasenbusch_89b,Janke_90,Jansen-Wiese_91,Wolff_review_91}
which complement our own.
Hasenbusch and Meyer \cite{Hasenbusch_89b}
studied the three-dimensional $XY$ model at $\beta \le \beta_c$ using
the codimension-1 Wolff embedding with standard SW updates
($N_{hit}=1$); they found $z_{int,\scre} \approx 0.46$ and
$z_{int,\scrm^2} \approx 0.31$.  Janke \cite{Janke_90} studied the same
model using single-cluster SW updates, and found $z_{int,\scre,CPU}
\approx 0.25$ and $z_{int,\scrm^2,CPU} \approx 0$.  However, it should
be noted that these exponents may well be due to critical slowing-down
in the inner SW or 1CSW subroutine; indeed, they are roughly of the
same order of magnitude as the dynamic critical exponents of the SW and
1CSW algorithms for the ordinary three-dimensional
Ising model
\cite{Swendsen_87,Wolff_89d,Tamayo_90,Wang_90,Coddington_private}.\footnote{
   Since the induced Ising model in the codimension-1 Wolff algorithm is
   ferromagnetic and short-range (though slightly disordered),
   one might expect it to have a dynamic critical exponent under the SW or 1CSW
   algorithm that is approximately (though probably not exactly) equal to
   that of the ordinary ferromagnetic nearest-neighbor Ising model.
   We say ``not exactly'' because the disorder --- which has long-range
   correlations --- probably does change the dynamic universality class.
}
Therefore, it remains possible that the {\em idealized}\/
Wolff algorithm for the three-dimensional $XY$ model could have $z \approx 0$.
In the near future we hope to study the idealized algorithm for this model.
(Since in this case the SW algorithm simulates the induced Ising model
reasonably well, albeit not perfectly, it will probably not be necessary
to go beyond $N_{hit} \approx 50$.)

Jansen and Wiese \cite{Jansen-Wiese_91} have very recently studied the
two-dimensional $\CP^3$ and $\CP^4$ models,
using the codimension-2 embedding $z \to I_1 z$
(see Example 3 in Section \ref{s3.1} above)
and single-cluster SW updates.
They also studied, for purposes of comparison,
a single-site Metropolis algorithm.
They found that $z_{exp,\scrm^2} \approx 2$ for both algorithms.
Unfortunately, this fact alone does {\em not}\/ constitute evidence
for our conjecture regarding the codimension-1 property:
the slowing-down observed by Jansen and Wiese may well be due to
the inability of the 1CSW algorithm to update efficiently the frustrated
Ising model that is induced by the codimension-2 embedding.
It would be necessary to study the {\em idealized}\/ embedding algorithm
to draw a definitive conclusion regarding the merit of this embedding.

Finally, a recent review talk of Wolff \cite{Wolff_review_91}
contains many interesting ideas, and gives some preliminary (negative) results
on embedding algorithms for the two-dimensional $SU(3)$ principal chiral model.
In particular, Wolff's results are consistent with our thesis that
the codimension-1 and isometry properties must hold if the
algorithm is to work well.
However, as in the work by Jansen and Wiese,
no definitive conclusion can be drawn
except from a study of the {\em idealized}\/ embedding algorithm.

\section*{Acknowledgments}

We wish to thank Richard Brower,
Ferenc Niedermayer, Claudio Parrinello and Ulli Wolff
for helpful discussions about embedding algorithms,
and Oliver Attie, Sylvain Cappell and Fabio Podest\`a
for helpful discussions about topology and geometry.
The computations reported here were carried out on
a loosely coupled MIMD parallel computer
(with local memory and message-passing communication
via Internet/Bitnet/Decnet and four neural networks)
consisting of the following processors:
the Cray X-MP at CRTN-ENEL (Pisa);
the Cray Y-MP at CINECA (Bologna);
the Cray Y-MP at the Pittsburgh Supercomputing Center;
the Cyber 205 at the John von Neumann Supercomputer Center$\dagger$;
and the ETA-10G, ETA-10Q, Cray Y-MP, Silicon Graphics 4D/240GTX
and numerous DECstation 5000, IBM RS-6000/320 and IBM RS-6000/530 workstations
at SCRI (Tallahassee).
We thank all these organizations for their generous contribution to
this research.
The authors' research was supported in part by
the Istituto Nazionale di Fisica Nucleare (S.C.\ and A.P.),
U.S.\ Department of Energy contract DE-FC05-85ER250000 (R.G.E.),
U.S.\ Department of Energy contract DE-FG02-90ER40581 (A.D.S.),
U.S.\ National Science Foundation grants DMS-8705599 and DMS-8911273 (A.D.S.),
and NATO Collaborative Research Grant CRG 910251 (S.C.\ and A.D.S.).

\appendix
\section{Some Topology and Geometry}
\label{appendix_A}

\subsection{Submanifolds and connectedness}  \label{secA.1}

Let $X$ be a connected finite-dimensional metrizable\footnote{
   For a connected finite-dimensional $C^\infty$ manifold $X$,
   the following conditions are equivalent:
   $X$ is metrizable;  $X$ is separable and metrizable;
   $X$ is second countable;  $X$ is Lindel\"of;  $X$ is paracompact;
   there exists a Riemannian metric on $X$.
   (For a proof, see \cite[vol.\ I, p.\ 271]{Kobayashi_69} and
   \cite[Theorem VIII.6.5]{Dugundji_66}.
   See also \cite[p.\ 207]{Royden_88}.)
   The purpose of imposing this condition is to exclude pathological examples
   such as the ``long line'' \cite[Example 45]{Steen_78}, which are
   locally Euclidean \cite[Problem 3.12.18(b)]{Engelking_77}
   and even admit a $C^\omega$ differential structure \cite[p.\ 15]{Hirsch_76},
   but are globally ``too big''.
}
$C^\infty$ manifold, and let $S$ be a subset of $X$
(usually a submanifold or union of submanifolds).
We wish to know whether or not $X \setminus S$ is connected.

Our first theorem asserts that a manifold cannot be disconnected by deleting
a subset of codimension $\ge 2$.  To make this statement meaningful,
we first have to define what we mean by ``dimension'' of an arbitrary
subset $S \subset X$.  The appropriate notion is provided by a branch of
topology called {\em dimension theory}\/ \cite{Hurewicz_48,Engelking_78},
which assigns to each separable metric space $S$ a dimension $\dim S$
($= -1, 0, 1, 2, \ldots$ or $\infty$) having the following properties:
\begin{itemize}
  \item[(a)]  Dimension is a topological invariant, i.e.\ $\dim S_1 = \dim S_2$
    if $S_1$ and $S_2$ are homeomorphic.
  \item[(b)]  The empty set has dimension $-1$.
  \item[(c)]  A nonempty open subset of an $n$-dimensional manifold ($n \ge 0$)
    has dimension $n$.
  \item[(d)]  If $S_1 \subset S_2$, then $\dim S_1 \le \dim S_2$.
  \item[(e)]  If $S$ is the union of countably many {\em closed}\/ subsets
    $S_i$, then $\dim S = \sup\limits_i \dim S_i$.  (It is crucial here that
    the sets $S_i$ be closed.
    Indeed, it can be proven that any $n$-dimensional
    space can be decomposed as the union of $n+1$ zero-dimensional subsets.)
\end{itemize}
We then have the following fundamental result:

\begin{theorem}
\label{thmA.1}
Let $X$ be a connected finite-dimensional metrizable $C^0$ manifold,
and let $S$ be a subset of $X$ satisfying $\dim S \le \dim X -2$.
(In particular, this holds if $S$ is the union of countably many
{\em closed}\/ submanifolds $S_i$ of codimension $\ge 2$.)
Then $X \setminus S$ is connected.
\end{theorem}

\proof
See \cite[p.\ 48, Corollary 1]{Hurewicz_48}
or \cite[p.\ 80, Theorem 1.8.19]{Engelking_78}.
\qed

\noindent
This can be rephrased as:  if $X \setminus S$ is disconnected, then $S$ must
have codimension 0 or 1.  In particular, if $X \setminus \bigcup_i S_i$
is disconnected and the $S_i$ are closed, then at least one of the $S_i$
must have codimension 0 or 1.

\bigskip

Next we wish to study theorems going in the opposite direction,
i.e.\ asserting that $X \setminus S$ is disconnected.  The naive converse
of Theorem \ref{thmA.1} is false:

{\em Example 1.}\/  Let $X$ be the torus $T^N$, and let $S$ be the ``slice''
$T^{N-1} \times \{a\}$ for some $a \in T^1$.
Then $S$ has codimension 1, but $X \setminus S$ is connected.

{\em Example 2.}\/  Let $X$ be the real projective space
$RP^{N-1} \equiv S^{N-1} / Z_2$, and let $S$ be the equator in $X$.
Then $S$ has codimension 1, but $X \setminus S$ is connected.

The key fact in both of these examples is that the manifold $X$ is not
simply connected.  Indeed, if we assume that $X$ is simply connected,
then we can prove a converse to Theorem \ref{thmA.1}:

\begin{theorem}
\label{thmA.2}
Let $X$ be a simply connected finite-dimensional metrizable $C^\infty$
manifold, and let $S$ be a closed codimension-1 submanifold (without
boundary) of $X$.  Then $X \setminus S$ is disconnected.
\end{theorem}

\medskip\par\noindent{\sc Proof\ }
(explained to us by Sylvain Cappell).
Fix a point $p \in S$, and let $U$ be a small open neighborhood of $p$.
Let $\gamma\colon\, [0,1] \to U$ be a smooth curve that intersects
$S$ exactly once, doing so transversally at $p$.
Then $q_0 \equiv \gamma(0)$ and $q_1 \equiv \gamma(1)$
are points in $U \setminus S$ ``on opposite sides of $S$''.
Now, if $X \setminus S$ is connected (and hence path-connected),
then there exists a smooth curve $\widetilde{\gamma}$ in $X \setminus S$
running from $q_1$ to $q_0$.
In that case $\alpha \equiv \gamma \circ \widetilde{\gamma}$ is a loop in $X$,
which intersects $S$ exactly once.
By a slight modification near $q_0$ and $q_1$, we can assume that
$\alpha$ is smooth.

Now recall the basic ideas of {\em intersection theory mod 2}\/
\cite[Section 2.4]{Guillemin_74}:
Let $M$ and $X$ be finite-dimensional metrizable $C^\infty$ manifolds,
with $M$ compact, and let $S$ be a closed submanifold of $X$ satisfying
$\dim M + \dim S = \dim X$.
If $f\colon\, M \to X$ is a smooth map that is transversal to $S$
(see Section \ref{secA.2} for the precise definition),
we define $I_2(f,S)$ to be the cardinality of $f^{-1}[S]$ mod 2.
A fundamental theorem states that if $f_0,f_1$ are homotopic and
are both transversal to $S$, then $I_2(f_0,S) = I_2(f_1,S)$.

To apply this theory, we let $M = S^1$ and $f=\alpha$.
By construction $I_2(\alpha,S) = 1$.
On the other hand, since $X$ is simply connected, $\alpha$ is homotopic
to a constant map $\beta\colon\, S^1 \to X$, where the constant can be chosen
to be $\notin S$;  so $I_2(\beta,S) = 0$.  But this is a contradiction.
\qed

\medskip

{\em Remarks.}\/   1.  For $S$ compact and connected,
this proof can be found in \cite[Theorem 4.4.6]{Hirsch_76}.

2.  A fancier way of phrasing this proof is to use the
language of homology theory.
There is a natural bilinear map
$H_1(X;Z_2) \times H_{n-1}(X;Z_2) \to Z_2$ (where $n = \dim X$),
called the ``Poincar\'e duality'' or ``intersection pairing''
\cite[Theorem 65.1]{Munkres_84}:
if $\alpha$ is a loop and $S$ is an ($n-1$)-dimensional submanifold,
then $[\alpha] \otimes [S]$ counts (mod 2)
the number of times that $\alpha$ intersects $S$.
For our loop $\alpha$, this intersection number is 1,
so $[\alpha]$ is a nontrivial element of $H_1(X;Z_2)$.
But this yields a contradiction if
the first homology group of $X$ mod 2 is trivial [i.e.\ $H_1(X;Z_2) = 0$].
Now, since $H_1(X;\zed)$ is the quotient of the first homotopy group
$\pi_1(X)$ by its commutator, simple connectedness implies the triviality
of $H_1(X;\zed)$ and hence of $H_1(X;Z_2)$; but the latter condition is weaker.
For example, the lens spaces $L(n,k)$ are 3-dimensional manifolds
having $H_1(X;\zed) = Z_n$ \cite[pp.\ 238--243]{Munkres_84};
so for $n$ odd, we have $H_1(X;Z_2) = 0$ but $\pi_1(X) \neq 0$.
For an even more extreme example, let $G$ be an arbitrary
finite simple group, and let $X$ be a compact polyhedron such that
$\pi_1(X) = G$ \cite[Theorem 6.4.6]{Hilton_65};
then $H_1(X;\zed) = H_1(X;Z_2) = 0$.
\bigskip

If $X$ has nontrivial first homology mod 2, then removing a single
codimension-1 submanifold may not disconnect $X$, as the two preceding
examples show.  Nevertheless, by removing {\em several}\/ codimension-1
submanifolds we can disconnect $X$.  For simplicity, we restrict attention
to the case where $X$ is compact.

\begin{theorem}
\label{thmA.3}
Let $X$ be a compact connected $n$-dimensional metrizable $C^\infty$ manifold,
and let $k = {\rm rank}\, H_1(X;Z_2)$.
Let $S_1,\ldots,S_l$ be disjoint closed codimension-1 submanifolds
(without boundary) of $X$.
Then $X \setminus \bigcup_{i=1}^{l} S_i$ has at least $l-k+1$ connected
components.  In particular, if $l \ge k+1$, then
$X \setminus \bigcup_{i=1}^{l} S_i$ is disconnected.
\end{theorem}

\proof
We imitate the proof of the Alexander duality theorem
\cite[Theorem 71.1]{Munkres_84}.
Set $A = \bigcup_{i=1}^{l} S_i$, and let $l' \ge l$ be the number of
connected components of $A$.  Then, by the Poincar\'e duality theorem
\cite[Theorem 65.1]{Munkres_84}, the $(n-1)^{st}$ cohomology group mod 2
of $A$ has rank $l'$, i.e. $H^{n-1}(A;Z_2) \simeq H_0(A;Z_2) \simeq Z_2^{l'}$.
On the other hand, by Poincar\'e duality we have
$H^{n-1}(X;Z_2) \simeq H_1(X;Z_2) \simeq Z_2^{k}$.
Now there is an exact sequence
\be
  H^n(X;Z_2)  \,\stackrel{j^*}{\longleftarrow}\,
  H^n(X,A;Z_2)  \,\stackrel{\delta^*}{\longleftarrow}\,
  H^{n-1}(A;Z_2)  \,\stackrel{i^*}{\longleftarrow}\,
  H^{n-1}(X;Z_2)
\ee
where $i\colon\; A \to X$ and $j\colon\; (X,\emptyset) \to (X,A)$
are inclusions, and $\delta^*$ is the cohomology coboundary homomorphism
\cite[Theorem 43.1, compare Theorem 23.3]{Munkres_84}.
So $\delta^*$ induces an isomorphism
\be
   \ker j^*  \;\simeq\;  H^{n-1}(A;Z_2) / i^*[H^{n-1}(X;Z_2)]
             \;\simeq\;  Z_2^m
\ee
where $m \ge l' -k \ge l-k$.

Now let $\Gamma_{(2)}$ be the unique nonzero element of $H^n(X;Z_2) \simeq Z_2$
(also called an ``orientation class for $X$ over $Z_2$'');
then $j_* \Gamma_{(2)}$ is a nonzero element of $H^n(X,A;Z_2)$.
Let $k\colon\; X \setminus A \to X$ be inclusion.
Then by Poincar\'e duality \cite[Theorems 67.1 and 67.2]{Munkres_84}
and Lefschetz duality \cite[Theorem 70.6]{Munkres_84} we have the diagram
%
\def\mapdown#1{\Big\downarrow\rlap{$\vcenter{\hbox{$\scriptstyle#1$}}$}}
\newlength{\arrow}
\settowidth{\arrow}{$\displaystyle\searrow$}
\setlength{\arrow}{0.45\arrow}
\def\mapse#1{\displaystyle\searrow\hspace{-\arrow}
                              \rlap{$\vcenter{\hbox{$\scriptstyle#1$}}$}}
\be
\begin{array}{ccc}
   H^n(X,A;Z_2)     &  \stackrel{j^*}{\longrightarrow}   &   H^n(X;Z_2)   \\
   \mapdown{\phi_*} &  \mapse{\cap j_* \Gamma_{(2)}}     &
                                          \mapdown{\cap \Gamma_{(2)}}     \\
   H_0(X \setminus A;Z_2)  &  \stackrel{k_*}{\longrightarrow}   & H_0(X;Z_2)
\end{array}
\ee
where $\phi_*$ is the Lefschetz duality isomorphism,
$\cap \Gamma_{(2)}$ is the Poincar\'e duality isomorphism,
and the diagram commutes up to sign.
Therefore,
\be
   \ker j^*  \;\simeq\;  \ker k_*  \;.
\ee
On the other hand, by exactness of the sequence
\be
  0 \,\longrightarrow\,  \widetilde{H}_0(X \setminus A)
    \,\longrightarrow\,  H_0(X \setminus A)
    \,\stackrel{k_*}{\longrightarrow}\,  H_0(X)
    \,\longrightarrow\,  0
\ee
\cite[exercise 71.1]{Munkres_84}, we have
\be
   \ker k_*  \;\simeq\;  \widetilde{H}_0(X \setminus A)  \;.
\ee
Combining these isomorphisms, we conclude that
\be
   \widetilde{H}_0(X \setminus A)  \;\simeq\;  Z_2^m  \;,
\ee
i.e.\ $X \setminus A$ has $m+1 \ge l-k+1$ connected components.
\qed

\medskip

{\em Remarks.}\/  1.  In the foregoing proof it is not necessary that
$S_1,\ldots,S_l$ be disjoint;
it suffices that they be ``linearly independent mod 2'', in the sense that
${\rm rank}\, H^{n-1}$$(\bigcup_{i=1}^{l}$ $S_i; Z_2) \ge l$.

2. Is the bound $l \ge k+1$ best possible?
We suspect that it {\em is}\/ best possible under the hypothesis
${\rm rank}\, H^{n-1}(\bigcup_{i=1}^{l} S_i; Z_2) \ge l$,
but that it is {\em not}\/ best possible under the stronger hypothesis that
$S_1,\ldots,S_l$ are disjoint.
Indeed, if $X$ is an orientable surface (= 2-dimensional manifold)
of genus $g$, then $k = {\rm rank}\, H_1(X;Z_2) = 2g$;
and while $2g+1$ {\em independent}\/ circles (= codimension-1 submanifolds)
may be needed to disconnect $X$, only $g+1$ {\em disjoint}\/ circles
are needed to disconnect $X$
\cite[Exercise 9.3.17 and Theorem 9.3.6]{Hirsch_76}.

3.  If $X$ is a connected manifold and $S$ is a {\em connected}\/ closed
codimension-1 submanifold, then $X \setminus S$ is either connected or else
has exactly two connected components \cite[Lemma 4.4.4]{Hirsch_76}.
Thus, each disjoint ``cut'' creates {\em at most one}\/ new connected
component.

\subsection{Transversality and genericity}  \label{secA.2}

Let $X$ and $Y$ be finite-dimensional metrizable $C^r$ manifolds
($1 \le r \le \infty$), let $f\colon\; X \to Y$ be a $C^r$ map,
and let $Z$ be a codimension-$k$ $C^r$ submanifold of $Y$.
We wish to show that ``generically'' the set
$f^{-1}[Z] \equiv \{x\colon\; f(x) \in Z \}$
is a submanifold of codimension $k$ in $X$.
The appropriate tool is a branch of differential topology called
{\em transversality theory}\/ \cite{Hirsch_76,Guillemin_74,Abraham_67}.
We say that {\em $f$ is transversal to $Z$}\/
(denoted $f \transversal Z$)
if, for every $x \in X$, either
\begin{itemize}
  \item[(a)]  $f(x) \notin Z$;  or
  \item[(b)]  $(D_x f)(T_x X) + T_{f(x)} Z = T_{f(x)} Y$,
    i.e.\ the image of the tangent space $T_x X$ under the linear map $D_x f$
    contains a subspace of $T_{f(x)} Y$ that is complementary to $T_{f(x)} Z$.
\end{itemize}
(For a nice intuitive discussion, with pictures, see
\cite[pp.\ 27 ff.]{Guillemin_74}.)
Transversality is just the right condition we need to control the
inverse image $f^{-1}[Z]$:

\begin{theorem}
\label{thmA.4}
Let $X,Y$ be finite-dimensional $C^r$ manifolds ($1 \le r \le \infty$),
let $Z$ be a codimension-$k$ $C^r$ submanifold of $Y$,
and let $f\colon\; X \to Y$ be a $C^r$ map that is transversal to $Z$.
Then $f^{-1}[Z]$ is either empty or else a $C^r$ submanifold
(not necessarily connected) of codimension $k$.
Moreover, if $X$ is Lindel\"of (resp.\ compact) and $Z$ is closed,
then $f^{-1}[Z]$ has only countably many (resp.\ finitely many)
connected components.
\end{theorem}

\proof
See \cite[Theorem 1.3.3]{Hirsch_76}, \cite[p.\ 28]{Guillemin_74},
\cite[p.\ 24]{Palis_82} or \cite[pp.\ 45--46]{Abraham_67}.
\qed

Now the point is that {\em transversality is generic}\/:

\begin{theorem}
\label{thmA.5}
Let $X,Y$ be finite-dimensional metrizable $C^r$ manifolds
($1 \le r \le \infty$),
and let $Z$ be a $C^r$ submanifold of $Y$.
Then the set of maps
\be
   \scra_Z  \;\equiv\;  \{ f \in C^r(X,Y)\colon\;
                           f \hbox{ is transversal to } Z  \}
\ee
contains a dense $G_\delta$ subset of $C^r(X,Y)$
in the $C^r$ compact-open topology (also called the ``weak topology'').
In fact, if $X$ is compact and $Z$ is closed, then $\scra_Z$
is a dense open subset of $C^r(X,Y)$.
\end{theorem}

\proof
See \cite[Theorem 3.2.1]{Hirsch_76};  other references are
\cite[pp.\ 25--26]{Palis_82} and \cite[pp.\ 46--50]{Abraham_67}.
\qed

Putting together Theorems \ref{thmA.4} and \ref{thmA.5}, we conclude that
for ``almost all'' maps $f$, the inverse image $f^{-1}[Z]$ is either empty
or else a submanifold of codimension $k$.

We cannot exclude the possibility that $f^{-1}[Z]$ is empty:
indeed, it is perfectly possible for the image $f[X]$ to avoid completely
the submanifold $Z$.
However, the point is that in physical applications there will be a
{\em nonzero probability}\/ for $f[X]$ to intersect $Z$.
We could go on to formalize this idea:
we would assume that a compact Lie group $G$ acts transitively on $Y$,
and we would seek to prove that under appropriate conditions the set
\be
  \{ f \in C^r(X,Y)\colon\;
      \mu_{Haar}( \{g \in G\colon\, (g \circ f)^{-1}[Z] \neq \emptyset \} ) > 0
  \}
\ee
contains a dense $G_\delta$ subset of $C^r(X,Y)$.
But we are physicists, not mathematicians, and enough is enough.

\medskip

Let us summarize the upshot of all this topology
for our physical application.
Consider a $\sigma$-model with $x$-space $X$ and target space $M$,
and let $Z$ be a closed codimension-1 submanifold of $M$.
We reason as follows:

(a) If $M$ is simply connected,
then Theorem \ref{thmA.2} implies that $M \setminus Z$ is disconnected.
With ``high probability'' one expects the field $\bsigma$
to intersect more than one connected component of $M \setminus Z$.
In this case the set $X \setminus \bsigma^{-1}[Z]$ is disconnected.

(b) For arbitrary $M$, Theorems \ref{thmA.4} and \ref{thmA.5}
and the subsequent remarks imply that with ``high probability''
$\bsigma^{-1}[Z]$ is a nonempty closed codimension-1 submanifold of $X$.
If $X$ is simply connected, Theorem \ref{thmA.2} then implies that
the set $X \setminus \bsigma^{-1}[Z]$ is disconnected.

(c) If neither $M$ nor $X$ is simply connected, then one uses
Theorem \ref{thmA.3} in place of Theorem \ref{thmA.2}.
One expects the probability of the desired event to be smaller,
but still nonzero (uniformly for $1 \ll \xi \ltapprox L$).

\subsection{Fixed points of isometries}   \label{secA.3}

Next we discuss the fixed-point set for isometries of a Riemannian manifold.

\begin{theorem}
\label{thmA.6}
Let $M$ be a finite-dimensional Riemannian manifold,
and let $\scrs$ be any set of isometries of $M$.
Let $F$ be the set of points of $M$ which are left fixed by all elements of
$\scrs$.
Then:
\begin{itemize}
  \item[(a)]  $F$ is a closed set.
  \item[(b)]  Each connected component $F_i$ of $F$ is a closed
     totally geodesic\footnote{
  A submanifold $N \subset M$ is called
  {\em totally geodesic at a point $x\in N$}\/
  if each $M$-geodesic which is tangent to $N$ at $x$ lies in $N$.
  The submanifold $N$ is called {\em totally geodesic}\/
  if it is totally geodesic at each of its points.
}
     submanifold of $M$.
  \item[(c)]  For each $F_i$, there exists an open set $U_i \supset F_i$
     that intersects none of the other connected components of $F$.
  \item[(d)]  If $M$ is Lindel\"of (resp.\ compact), there are at most
     countably many (resp.\ finitely many) $F_i$.
\end{itemize}
\end{theorem}

\proof
(a) is trivial, since an isometry is necessarily continuous.
(b) is \cite[p.\ 59, Theorem II.5.1]{Kobayashi_72},
and (c) is implicit in the proof given there.
To prove (d), consider the open cover of $M$ consisting of the $\{ U_i \}$
together with $V \equiv M \setminus F$.
Since $M$ is Lindel\"of (resp.\ compact), there exists
a countable (resp.\ finite) subcover;  but by definition of $U_i$,
this subcover must include all of the $\{ U_i \}$, since otherwise
it couldn't cover all of $F$.
\qed

Some partial converses to Theorem \ref{thmA.6}(b) are mentioned in
Further Remarks 1 and 2 at the end of this section.

The following lemma shows that isometries are completely determined locally
(it is analogous to analytic continuation of holomorphic functions,
but even stronger):

\begin{lemma} \cite[p.\ 62, Lemma I.11.2]{Helgason_78}
\label{lemmaA.7}
Let $M$ be a connected finite-dimensional Riemannian manifold,
and let $\varphi$ and $\psi$ be isometries of $M$.
Suppose that there exists a point $p \in M$
for which $\varphi(p) = \psi(p)$ and $(D\varphi)_p = (D\psi)_p$.
Then $\varphi = \psi$.
\end{lemma}

It follows immediately that a fixed-point manifold of codimension 0
can occur only in the case of the identity map:

\begin{theorem}
\label{thmA.8}
Let $M$ be a connected finite-dimensional Riemannian manifold,
and let $\varphi$ be an isometry of $M$.
If the set of fixed points of $\varphi$ contains a nonempty open set
(i.e.\ a submanifold of codimension 0),
then $\varphi$ is the identity map
(and hence the fixed-point set is all of $M$).
\end{theorem}

By a slightly more subtle argument, we can show that if the fixed-point
manifold has codimension 1, then the map must be involutive:

\begin{theorem}
\label{thmA.9}
Let $M$ be a connected finite-dimensional Riemannian manifold,
and let $\varphi$ be an isometry of $M$.
If the set of fixed points of $\varphi$ contains
a submanifold of codimension 1,
then $\varphi$ is involutive (i.e.\ $\varphi^2$ is the identity map).
\end{theorem}

\proof
Let $N \subset {\rm Fix}(\varphi)$ be a submanifold of $M$ of codimension 1,
and let $p \in N$.  Then in some small neighborhood $U \ni p$ we can choose
local coordinates $(x_1,\ldots,x_n)$ such that $U \cap N$ is given by
$x_1 = 0$.  In this basis, $(D\varphi)_p$ has the form
\be
   (D\varphi)_p   \;=\;
   \left(
   \begin{tabular}{c|c}
      $v_1$   &   $v_2 \quad\; \cdots \quad\; v_n$   \\
      \hline
              &                                    \\
      {\large 0}  &  {\large\it I}                 \\[2mm]
   \end{tabular}
   \right)   \;.
\ee
Since $(D\varphi)_p$ leaves invariant the metric tensor $G_p$
[i.e.\ $(D\varphi)_p^T G_p (D\varphi)_p = G_p$]
and $G_p$ is nondegenerate, it follows that
$\det (D\varphi)_p = \pm 1$, i.e.\ $v_1 = \pm 1$.
We now claim that $(D\varphi)_p^2 = I$:

(a)  If $v_1 = +1$, then we must have $v_2 = \ldots = v_n = 0$,
because a matrix that leaves invariant a positive-definite quadratic form
must be diagonalizable (over $\C$), i.e.\ it must not have a nontrivial
Jordan block.
Then $(D\varphi)_p = I$.

(b)  If $v_1 = -1$, then
\be
   (D\varphi)_p^2   \;=\;
   \left(
   \begin{tabular}{c|c}
      $v_1^2$   &   $(v_1+1)v_2 \quad\; \cdots \quad\; (v_1+1)v_n$   \\
      \hline
              &                                    \\
      {\large 0}  &  {\large\it I}                 \\[2mm]
   \end{tabular}
   \right)   \;=\;   I   \;.
\ee
(Alternatively, we can argue that there exists a change of basis setting
$v_2 = \ldots = v_n = 0$.)

Thus, in either case, $\varphi^2$ is an isometry satisfying
$\varphi^2(p) = p$ and $(D\varphi^2)_p = (D\varphi)_p^2 = I$.
Lemma \ref{lemmaA.7} then implies that $\varphi^2$ is the identity map.
[In case (a), $\varphi$ is itself the identity map, while in case (b)
it is not.]
\qed

{\em Remark.}\/
This theorem seems to be well known to differential geometers,
but we have been unable to find a published reference.
Some vaguely related theorems, which may conceivably be of interest
in future generalizations of the embedding method, are:

a)  Let $\scrs$ be a {\em one-parameter group}\/ of isometries
(or more generally, a {\em connected abelian}\/ Lie group of isometries)
of a finite-dimensional Riemannian manifold $M$.
Then each connected component of ${\rm Fix}(\scrs)$ has {\em even}\/
codimension \cite{Kobayashi_58} \cite[p.\ 60, Theorem II.5.3]{Kobayashi_72}.

b)  Let $T\colon\; M \to M$ be a smooth map which is periodic of {\em odd}\/
period $p$ (here $p \equiv \min\{q\colon\, T^q = \hbox{identity} \}$).
Then each connected component of ${\rm Fix}(\scrs)$ has {\em even}\/
codimension \cite{Smith_45}.  Note that $T$ need not be an isometry;
this theorem is purely algebraic and topological.
For related material, see also \cite{Borel_55}.

\medskip

{\em Further Remarks.}
1.  Vanhecke and collaborators
\cite{Vanhecke_83,Tondeur_88,Chen-Vanhecke_88,Chen-Vanhecke_89,%
Tondeur_89a,Tondeur_89b}
have recently investigated reflections in Riemannian manifolds from a
point of view opposite (but complementary) to ours.
We start from an involutive isometry and seek to study its fixed-point
manifold.  They start, by contrast, from a submanifold $N \subset M$,
and define the local reflection $\varphi_N$ about $N$;
this local reflection is automatically involutive, and they ask under
what conditions it is a local isometry.
In view of Theorem \ref{thmA.6}(b) [or more precisely its local analogue],
a necessary condition for $\varphi_N$ to be a local isometry is that
$N$ be totally geodesic.  It is natural to ask whether this condition is
sufficient.  Vanhecke {\em et al.}\/ prove the following interesting theorem
(\cite[Theorem 5.7]{Vanhecke_83} and
 \cite[Corollaries 4(a) and 5]{Chen-Vanhecke_89}),
which is somewhat reminiscent of Theorem \ref{thm3.3}:
Let $M$ be a Riemannian manifold.  Then the following are equivalent:

(a)  For each geodesic curve (= totally geodesic 1-dimensional submanifold)
$N \subset M$, the local reflection $\varphi_N$ is a local isometry.

(b)  For each totally geodesic submanifold
$N \subset M$, the local reflection $\varphi_N$ is a local isometry.

(c)  $M$ is a space of constant curvature.

\smallskip

2.  Another partial converse to Theorem \ref{thmA.6}(b) is due to
Chen and Nagano \cite[Theorem 3.1]{Chen_77}:
In the manifold $Q_m = SO(m+2) / (SO(m) \times SO(2))$ for $m \ge 2$,
a complete connected submanifold $N \subset Q_m$ is totally geodesic
if {\em and only if}\/ it is a connected component of the fixed-point set
of some {\em finite}\/ set of {\em involutive}\/ isometries of $Q_m$.
It would be interesting to know to which other symmetric spaces, if any,
this result extends.
Some totally geodesic submanifolds $N \subset Q_m$ of codimension $m$
have been found by Niki\'c \cite{Nikic_80}.

3.  In Example 3 of Section \ref{s3.1}, we proved that in $\CP^{N-1}$
($N \ge 3$)
there are no isometries having a fixed-point manifold of codimension
1.  Wolf \cite{Wolf_63} has shown much more:  in $\CP^{N-1}$ ($N \ge
3$) there are no closed totally geodesic submanifolds of codimension
1.  Moreover, the same holds for quaternionic projective space
$\QP^{N-1}$.  In fact, Wolf \cite[Theorem 1]{Wolf_63} obtains a
complete list of the totally geodesic submanifolds of $S^{N-1}$,
$RP^{N-1}$, $\CP^{N-1}$, $\QP^{N-1}$ and {\em CayleyP}${}^2$.

\subsection{Codimension and frustration}   \label{secA.4}

Here we prove some theorems mentioned in Section \ref{s2.3},
regarding the relations between codimension and frustration.
Theorem \ref{thmA.10} and Corollary \ref{corA.11}
state that non-frustration implies codimension 1.
Corollary \ref{corA.12} shows further that in the case of an irreducible
symmetric space, non-frustration occurs {\em only}\/ in the
codimension-1 algorithm for the $N$-vector model ---
i.e.\ the original Wolff algorithm.

\begin{theorem}
\label{thmA.10}
Let $M$ be a finite-dimensional Riemannian manifold with metric tensor $g$,
and let $T$ be an involutive isometry of $M$.
Let $E \colon\; M \times M \to \R$ be a function satisfying
$E(\bsigma,\bsigma') = a + b d(\bsigma,\bsigma')^2 + o(d(\bsigma,\bsigma')^2)$
as $\bsigma' \to \bsigma$,
where $a \in \R$, $b > 0$, and $d$ is the geodesic distance on $M$.
Define
\be
  J(\bsigma,\bsigma')   \;=\;   E(\bsigma,T\bsigma') - E(\bsigma,\bsigma')  \;.
\ee

Now let $\bsigma^* \in {\rm Fix}(T)$.  Suppose that there exists an integer
$m \ge 3$ and a neighborhood $U \ni \bsigma^*$ such that
for all $\bsigma_1,\ldots,\bsigma_m,\bsigma_{m+1} \equiv \bsigma_1 \in U$
we have
\be
   \prod\limits_{j=1}^m  J(\bsigma_j,\bsigma_{j+1})   \;\ge\;   0   \;.
\ee
Then the connected component of ${\rm Fix}(T)$ containing $\bsigma^*$
has codimension 1.
\end{theorem}

\proof
Use a chart on $U$ such that $\bsigma^* = 0$
(by abuse of language we identify a point in $U$ with its coordinates
given by the chart) and $g_{\mu\nu}(\bsigma^*) = \delta_{\mu\nu}$.
Then the linear map $(DT)_{\sigma^*}$ is represented in these coordinates
by an orthogonal matrix ${\bf T}$ satisfying ${\bf T}^2 = I$.
Thus, ${\bf T}$ is symmetric with eigenvalues $\pm 1$,
and the number of negative eigenvalues equals the codimension
of the connected component of ${\rm Fix}(T)$ containing $\bsigma^*$.
Finally, $J$ is given in these coordinates by
\begin{subeqnarray}
 J(\bsigma,\bsigma')
  & = &  b \left[ d(\bsigma,T\bsigma')^2 - d(\bsigma,\bsigma')^2 \right]
      \,+\, o\!\left(d(\bsigma,\bsigma')^2, d(\bsigma,T\bsigma')^2\right) \\
  & = &  b \left[ (\bsigma - {\bf T} \bsigma')^2 - (\bsigma-\bsigma')^2 \right]
      \,+\, o(\bsigma^2,\bsigma'^2)                                      \\
  & = &  2b \bsigma \cdot (I - {\bf T}) \bsigma'  \,+\, o(\bsigma^2,\bsigma'^2)
\end{subeqnarray}

Now, if ${\rm rank}(I - {\bf T}) \ge 2$ and $m \ge 3$,
it is easy to choose $\bsigma_1,\ldots,\bsigma_m$ with arbitrarily small
magnitudes such that $\bsigma_i \cdot (I - {\bf T}) \bsigma_{i+1} > 0$
for $i = 1,\ldots,m-1$
and $\bsigma_m \cdot (I - {\bf T}) \bsigma_1 < 0$.
[Using polar coordinates in some fixed two-dimensional subspace of
${\rm Ran}(I - {\bf T})$, let $\bsigma_j$ point at angle
$\theta_j = (j-1) \Theta/(m-1)$ for $j=1,\ldots,m$,
where $\pi < \Theta < 3\pi/2$.]
This proves the theorem.
\qed

\begin{corollary}
\label{corA.11}
Let $M$, $T$, $E$ and $J$ be as in Theorem \ref{thmA.10}.
Suppose that for each $\bsigma^* \in {\rm Fix}(T)$ there exists
an integer $m_{\sigma^*} \ge 3$ and a neighborhood
$U_{\sigma^*} \ni \bsigma^*$ such that for all
$\bsigma_1,\ldots,\bsigma_m,\bsigma_{m+1} \equiv \bsigma_1 \in U_{\sigma^*}$
we have
\be
   \prod\limits_{j=1}^m  J(\bsigma_j,\bsigma_{j+1})   \;\ge\;   0   \;.
\ee
Then either
\begin{itemize}
  \item[(a)]  ${\rm Fix}(T) = \emptyset$
\end{itemize}
or else
\begin{itemize}
  \item[(b)]  every connected component of ${\rm Fix}(T)$ has codimension 1.
\end{itemize}
\end{corollary}

\proof
Immediate.
\qed

\begin{corollary}
\label{corA.12}
Let $M$, $T$, $E$ and $J$ be as in Theorem \ref{thmA.10},
and assume further that $M$ is an irreducible compact Riemannian
symmetric space of dimension $n$.
Suppose that for each $\bsigma^* \in {\rm Fix}(T)$ there exists
an integer $m_{\sigma^*} \ge 3$ and a neighborhood
$U_{\sigma^*} \ni \bsigma^*$ such that for all
$\bsigma_1,\ldots,\bsigma_m,\bsigma_{m+1} \equiv \bsigma_1 \in U_{\sigma^*}$
we have
\be
   \prod\limits_{j=1}^m  J(\bsigma_j,\bsigma_{j+1})   \;\ge\;   0   \;.
\ee
Then either
\begin{itemize}
  \item[(a)]  ${\rm Fix}(T) = \emptyset$
\end{itemize}
or else
\begin{itemize}
  \item[(b)]  $M$ is isometric to $S^n$, and under this isometry
       $T$ is the codimension-1 reflection $\bsigma \to I_1 \bsigma$.
\end{itemize}
\end{corollary}

\proof
If ${\rm Fix}(T) \neq \emptyset$, then by Corollary \ref{corA.11}
each connected component of ${\rm Fix}(T)$ must have codimension 1.
By Theorem \ref{thm3.3}, $M$ must be isometric to either $S^n$ or $RP^n$.
But in Section \ref{s3.1} (Example 2), we classified the involutive isometries
of $RP^n$:  in particular, we found that for $n \ge 2$,
every involutive isometry with ${\rm Fix}(T) \neq \emptyset$
has at least one connected component of ${\rm Fix}(T)$ of codimension $\ge 2$.
So $M$ must be isometric to $S^n$, and the theorem follows from our
classification of the involutive isometries of $S^n$
(Section \ref{s3.1}, Example 1).
\qed

\section{Involutive Isometries of $SU(N)$}
\label{appendix_B}

Let $G$ be a compact simple Lie group\footnote{
   We make the usual abuse of language,
   and call a Lie group ``simple'' if its Lie {\em algebra}\/
   is simple (i.e.\ has no nontrivial ideals).  Of course $G$ need {\em not}\/
   be simple in the group-theoretic sense:  for example, $SU(N)$ has a
   nontrivial (but discrete) center, isomorphic to $Z_N$.
}%
, with Lie algebra ${\germang}$.
Then there is a {\em unique}\/ (up to constant multiples)
Riemannian metric on $G$ that is invariant under both left and right
translations;  at the identity element $e \in G$ this metric is given by
the negative of the Killing-Cartan form
\be
   B(X,Y)  \;=\;  \tr ({\rm ad} X \, {\rm ad} Y)
   \qquad\qquad \hbox{for }  X,Y \in {\germang}
\ee
(see \cite[pp.\ 57--58, Lemme 5]{Doubrovine_82}).
We shall always consider $G$ to be equipped with this unique bi-invariant
Riemannian metric.

For $G=SU(N)$, the Lie algebra ${\germang} = {\germansu}(N)$ is the
space of all traceless antihermitian $N\times N$ matrices, and the
Killing-Cartan form is
\be
   B(X,Y)  \;=\;  - {\rm const}  \times  \real\tr (X^\dagger Y)
   \qquad\qquad \hbox{for }  X,Y \in {\germansu}(N)   \;.
\ee
The isometry group of $SU(N)$ has been determined by Cartan \cite{Cartan_27}
(see also Wolf \cite[secs.\ 2.4 and 4.1.2]{Wolf_62}):

\begin{theorem} \label{thmB.1}  \cite{Cartan_27} \quad
 The isometries of $SU(N)$ are the following:
 \begin{itemize}
   \item[(a)]  $A \longto UAV$
   \item[(b)]  $A \longto UA^\dagger V$
   \item[(c)]  $A \longto U \bar{A} V$
   \item[(d)]  $A \longto UA^T V$
 \end{itemize}
 for $U,V \in SU(N)$.
 For $N \ge 3$, two isometries in this list are equal if and only if
 they belong to the same class (a), (b), (c) or (d) and in addition
 their determining matrices $(U,V)$ and $(U',V')$ satisfy
 $U'=CU$, $V' = C^{-1}V$ for some $C \in \scrc \equiv \hbox{center of } SU(N)$.
 For $N=2$ the same statement holds provided that we consider only the classes
 (a) and (b).  [For $N=2$ the classes (c) and (d) are redundant because
 $\bar{A} = JA(-J)$ and $A^T = J A^\dagger (-J)$, where
 $J = i \btau^2 = \left(\begin{array}{cc} 0 & 1 \\ -1 & 0 \end{array}\right)$
 and $\pm J \in SU(2)$.]
\end{theorem}

\bigskip

Our next task is to classify these isometries modulo conjugacy:

\begin{theorem} \label{thmB.2}
 Two isometries of $SU(N)$ are conjugate if and only if
 they belong to the same class (a), (b), (c) or (d)
 [for $N=2$, (a) or (b)] and in addition
 their determining matrices $(U,V)$ and $(U',V')$ are related in
 one of the ways (i)--(iv) listed below:
 \medskip\par\noindent
 Class (a):
 \begin{itemize}
  \item[(i)]  $U' = CXUX^{-1}$,          \quad $V' = C^{-1} Y^{-1} V Y$
  \item[(ii)] $U' = CXV^\dagger X^{-1}$, \quad $V' = C^{-1} Y^{-1} U^\dagger Y$
  \item[(iii)]$U' = CX \bar{U} X^{-1}$,  \quad $V' = C^{-1} Y^{-1} \bar{V} Y$
  \item[(iv)] $U' = CXV^T X^{-1}$,       \quad $V' = C^{-1} Y^{-1} U^T Y$
 \end{itemize}
 \medskip\par\noindent
 Class (b):
 \begin{itemize}
  \item[(i)]   $U' = CXUY$,          \quad $V' = C^{-1} X V Y$
  \item[(ii)]  $U' = CXV^\dagger Y$, \quad $V' = C^{-1} X U^\dagger Y$
  \item[(iii)] $U' = CX \bar{U} Y$,  \quad $V' = C^{-1} X \bar{V} Y$
  \item[(iv)]  $U' = CXV^T Y$,       \quad $V' = C^{-1} X U^T Y$
 \end{itemize}
 \medskip\par\noindent
 Class (c):
 \begin{itemize}
  \item[(i)]   $U' = CXUX^T$,          \quad $V' = C^{-1} Y^T V Y$
  \item[(ii)]  $U' = CXV^\dagger X^T$, \quad $V' = C^{-1} Y^T U^\dagger Y$
  \item[(iii)] $U' = CX \bar{U} X^T$,  \quad $V' = C^{-1} Y^T \bar{V} Y$
  \item[(iv)]  $U' = CXV^T X^T$,       \quad $V' = C^{-1} Y^T U^T Y$
 \end{itemize}
 \medskip\par\noindent
 Class (d):
 \begin{itemize}
  \item[(i)]  $U' = CXU \bar{Y}$,         \quad $V'=C^{-1} \bar{X} V Y$
  \item[(ii)]$U' = CXV^\dagger \bar{Y}$, \quad $V'=C^{-1} \bar{X} U^\dagger Y$
  \item[(iii)]$U' = CX \bar{U}  \bar{Y}$, \quad $V'=C^{-1}  \bar{X} \bar{V} Y$
  \item[(iv)] $U' = CXV^T  \bar{Y}$,      \quad $V' = C^{-1}  \bar{X} U^T Y$
 \end{itemize}
In all cases $X,Y \in SU(N)$ and $C \in \scrc \equiv \hbox{center of } SU(N)$.
\end{theorem}

\proof
Consider an isometry $f$ of class (a), say $f(A) = UAV$,
and let us compute $g \circ f \circ g^{-1}$ with all possible isometries $g$:
\begin{itemize}
  \item[(i)]    $g(A) = XAY  \quad\Longrightarrow\quad
                   (g \circ f \circ g^{-1})(A) = (XUX^{-1}) A (Y^{-1}VY)$
  \item[(ii)]   $g(A) = XA^\dagger Y  \quad\Longrightarrow\quad
      (g \circ f \circ g^{-1})(A) = (XV^\dagger X^{-1}) A (Y^{-1}U^\dagger Y)$
  \item[(iii)]  $g(A) = X \bar{A} Y  \quad\Longrightarrow\quad
        (g \circ f \circ g^{-1})(A) = (X \bar{U} X^{-1}) A (Y^{-1} \bar{V} Y)$
  \item[(iv)]   $g(A) = XA^T Y  \quad\Longrightarrow\quad
                  (g \circ f \circ g^{-1})(A) = (XV^T X^{-1}) A (Y^{-1}U^T Y)$
\end{itemize}
The claim then follows from Theorem \ref{thmB.1}.
Analogous computations handle the cases when $f$ is an isometry of
class (b), (c) or (d).
\qed

\bigskip

Next we determine which isometries are involutive:

\begin{theorem}  \label{thmB.3}
An isometry of $SU(N)$ is involutive if and only if:
\begin{itemize}
  \item[Class (a):]  $U^2 = V^{-2} \in \scrc \equiv \hbox{center of } SU(N)$.
  \item[Class (b):]  $U V^{-1} \in \scrc$.
  \item[Class (c):]  \begin{tabbing}
                    Either \=  (2) \=     \kill
                    Either \>  (1)  \>  $U=U^T$ and $V=V^T$   \\
                    or     \>  (2) \>  $U=-U^T$ and $V=-V^T$ and $N$ is even.
                     \end{tabbing}
  \item[Class (d):]  \begin{tabbing}
                    Either \=  (2) \=     \kill
                    Either \>  (1)  \>  $U=\bar{V}$           \\
                    or     \>  (2) \>  $U=-\bar{V}$ and $N$ is even.
                     \end{tabbing}
\end{itemize}
\end{theorem}

\proof
We compute two successive applications of the given isometry,
and demand (using the last part of Theorem \ref{thmB.1})
that it be equal to the identity map.

{\em Class (a):}\/  $A \longto UAV \longto U(UAV)V$.
So we need $U^2 =C$, $V^2 = C^{-1}$ with $C \in \scrc$.

{\em Class (b):}\/  $A \longto UA^\dagger V \longto U(UA^\dagger V)^\dagger V
  = (UV^\dagger) A (U^\dagger V)$.
So we need $UV^\dagger = C$, $U^\dagger V = C^{-1}$ with $C \in \scrc$.
The two conditions are equivalent; they amount to $U=CV=VC$,
i.e.\ $UV^{-1} \in \scrc$.

{\em Class (c):}\/  $A \longto U \bar{A} V \longto U \overline{U \bar{A} V} V
  = (U \bar{U}) A (\bar{V} V)$.
So we need $U \bar{U} = C$, $\bar{V} V = C^{-1}$ with $C \in \scrc$.
Equivalently we need $U = CU^T$, $V = C^{-1} V^T$.
But then $U = CU^T = C(CU^T)^T = CC^T U$;
and since the central elements of $SU(N)$ are multiples of the identity matrix,
we have $C = C^T$;  so we need $C^2 = I$.
But this means that $C=I$ if $N$ is odd, or $C=\pm I$ if $N$ is even.

{\em Class (d):}\/  $A \longto U A^T V \longto U (U A^T V)^T V
  = (U V^T) A (U^T V)$.
So we need $U V^T = C$, $U^T V = C^{-1}$ with $C \in \scrc$.
Equivalently we need $U = C \bar{V}$, $V = C^{-1} \bar{U}$.
But for a central element of $SU(N)$, $C^{-1} = \bar{C}$,
so we can write the second equation as $V = \bar{C} \bar{U}$,
hence $\bar{V} = CU$.
But then we have $U = C^2 U$, so $C^2 = I$.
It follows that $C=I$ if $N$ is odd, or $C=\pm I$ if $N$ is even.
\qed

\bigskip

Next we classify the involutive isometries modulo conjugacy:

\begin{theorem} \label{thmB.4}
Every involutive isometry of $SU(N)$ is conjugate to one of the following:
\begin{itemize}
 \item[(a${}_{r,s}$)]  $A \longto I_r A I_s$ with $r+s$ even, $r \le s$
    and $r \le N/2$.  \\
   Here
  $I_r   \;=\;
   {\rm diag}( \underbrace{-1,\ldots,-1}_{\displaystyle{r} \hbox{ times}},
               \underbrace{+1,\ldots,+1}_{\displaystyle{N-r} \hbox{ times}} )$.
 \item[(b1)]  $A \longto A^\dagger$.
 \item[(b2)]  [Only for $N$ even]  $A \longto e^{2\pi i/N} A^\dagger \;$.
 \item[(c1)]  $A \longto \bar{A}$.
 \item[(c2a)] [Only for $N$ even]  $A \longto -J \bar{A} J$.
 \item[(c2b)] [Only for $N = 4k$, $k$ integer]
                                       $A\longto e^{2\pi i/ N} J\bar A J$.
 \item[(c2c)] [Only for $N = 4k+2$, $k$ integer]  $A \longto J \bar{A} J$.
 \item[(d1)]  $A \longto A^T$.
 \item[(d2)]  [Only for $N$ even]  $A \longto -A^T$.
\end{itemize}
Moreover, none of the above isometries
[for $N=2$, none of the above isometries of classes (a) and (b)]
are conjugate one to another.
\end{theorem}

\proof
{\em Class (a):}\/  The map $A \longto UAV$ is involutive iff
$U^2 = V^{-2} = C = e^{2\pi im/N} I$.  It then follows that the eigenvalues
of $U$ (resp.\ $V$) are all $\pm e^{\pi im/N}$ (resp.\ $\pm e^{-\pi im/N}$),
so the matrices $U$ and $V$ can be diagonalized as
$U = X(e^{\pi im/N} I_r) X^{-1}$,  $V = Y^{-1} (e^{-\pi im/N} I_s) Y$
with $X,Y \in SU(N)$ and $0 \le r,s \le N$.
Note that $m$, $r$ and $s$ must all have the same
parity modulo 2 (since $\det U = \det V = +1$), so in particular $r+s$
must be even.  By Theorem \ref{thmB.2}(a)(i) it follows that the map
$A \longto UAV$ is conjugate to
$A \longto (e^{\pi im/N} I_r) A (e^{-\pi im/N} I_s) = I_r A I_s$.
By Theorem \ref{thmB.2}(a)(ii) [or (iv)] we can take $r \le s$.
Trivially $I_r A I_s = I_{N-r} A I_{N-s}$, so we can take $r \le N/2$.
Finally, it follows from Theorem \ref{thmB.2}(a) that there are no further
conjugacies between members of this class.
\medskip

{\em Class (b):}\/  The map $A \longto UA^\dagger V$ is involutive iff
$U = e^{2\pi im/N} V$.  By Theorem \ref{thmB.2}(b)(i) with
$X = e^{-2\pi ik/N} V^{-1}$ and $Y=I$,
this map $A \longto e^{2 \pi im/N} VA^\dagger V$
is conjugate to $A \longto e^{2\pi i (m-2k)/N} A^\dagger$.
If $N$ is odd, the integer $k$ can be chosen so that $m-2k = 0 \pmod{N}$;
if $N$ is even, $k$ can be chosen so that $m-2k = 0 \hbox{ or } 1 \pmod{N}$.
Finally, it follows from Theorem \ref{thmB.2}(b) that if $N$ is even,
these two alternatives are not conjugate one to another.
\medskip

{\em Class (c1):}\/  Consider the map $A \longto U \bar{A} V$ with
$U = U^T$, $V = V^T$.  We can write $U = R+iS$ where $R$ and $S$ are
real symmetric matrices.  Expanding out $U U^\dagger = U^\dagger U = I$,
we conclude that $RS=SR$ and $R^2 + S^2 = I$.
It follows that we can simultaneously diagonalize $R$ and $S$
using a real rotation matrix $L \in SO(N)$:
\begin{subeqnarray}
   R   & = &   L \, {\rm diag}(\lambda_1,\ldots,\lambda_N) \, L^T      \\
   S   & = &   L \, {\rm diag}(\mu_1,\ldots,\mu_N) \, L^T
\end{subeqnarray}
with $\lambda_i,\mu_i$ real and $\lambda_i^2 + \mu_i^2 = 1$.
Hence
\begin{eqnarray}
   U   & = &   L \, {\rm diag}(e^{i\theta_1},\ldots,e^{i\theta_N}) \, L^T
                                                                  \nonumber \\
       & \equiv &   LDL^T
\end{eqnarray}
with $L \in SO(N)$ and $D$ a diagonal matrix in $SU(N)$.
Now let
$E = {\rm diag}(\pm e^{i\theta_1 /2},$ $e^{i\theta_2 /2}, \ldots,
                e^{i\theta_N /2})$,
with the sign chosen so that $E \in SU(N)$;  we have shown that
$U = XX^T$ where $X \equiv LE \in SU(N)$.  Similarly, $V=Y^T Y$ with
$Y \in SU(N)$.  It follows from Theorem \ref{thmB.2}(c)(i) that the map
$A \longto U \bar{A} V$ is conjugate to $A \longto \bar{A}$.
\medskip

{\em Class (c2):}\/  Consider the map $A \longto U \bar{A} V$ with
$U = -U^T$, $V = -V^T$ and $N$ even.  We can write $U = R+iS$ where $R$ and $S$
are real antisymmetric matrices.
Expanding out $U U^\dagger = U^\dagger U = I$,
we conclude that $RS=SR$ and $R^2 + S^2 = -I$.
It follows from Lemma \ref{lemma_RS} below
that $R$ and $S$ can be simultaneously brought into real Schur form
by a real rotation matrix $L \in SO(N)$:
\begin{subeqnarray}
   R   & = &   L \, {\rm blockdiag}
                        (\lambda_1 {\bf J}, \ldots, \lambda_{N/2} {\bf J})
                 \, L^T    \\
   S   & = &   L \, {\rm blockdiag}
                        (\mu_1 {\bf J}, \ldots, \mu_{N/2} {\bf J})   \, L^T
\end{subeqnarray}
with ${\bf J} = \left(\! \begin{array}{cc}
                           0  & 1  \\
                           -1 & 0
                         \end{array}
                \!\right)$,
$\lambda_i,\mu_i$ real and $\lambda_i^2 + \mu_i^2 = 1$.
Hence
\begin{eqnarray}
   U   & = &   L \, {\rm blockdiag}
            (e^{i\theta_1} {\bf J},\ldots,e^{i\theta_{N/2}} {\bf J}) \, L^T
                                                                  \nonumber \\
       & \equiv &   LDJL^T
\end{eqnarray}
with $L \in SO(N)$,
$D = {\rm diag}(e^{i\theta_1},e^{i\theta_1},\ldots,
                e^{i\theta_{N/2}},e^{i\theta_{N/2}})   \in SU(N)$,
and $J = {\rm blockdiag}({\bf J},\ldots,{\bf J})$.
Now let
$E = {\rm diag}(e^{i\theta_1 /2},e^{i\theta_1 /2},\ldots,
                e^{i\theta_{N/2} /2},e^{i\theta_{N/2} /2})$;
either $E \in SU(N)$ or else $e^{-\pi i/N} E \in SU(N)$,
according as $\det E = \pm 1$.  We have then shown that either
\be
    U = XJX^T  \quad\hbox{with}\quad X = LE \in SU(N)
\ee
or else
\be
  U = e^{2\pi i/N} XJX^T  \quad\hbox{with}\quad
                                    X = e^{-\pi i/N} LE \in SU(N)   \;.
\ee
And obviously the same holds for $V$.  Thus, by Theorem \ref{thmB.2}(c)(i)
we have shown that every isometry of class (c2) is
conjugate to either $A\to J\bar A J$ or $A\to e^{2\pi i/ N} J\bar A J$.

We now show that these two isometries are not conjugate one to the other.
We firstly prove that there do not exist $X, Y \in SU(N)$
and $C = cI \in \hbox{center of } SU(N)$ such that
$J = C X J X^T e^{2\pi i/ N}$ and $J = C^{-1} Y J Y^T$.
Indeed, if such matrices exist, then $C^{1/2} e^{\pi i/ N} X$ and
$C^{-1/2} Y$ are symplectic matrices and as such must have determinant $+1$
\cite[pp.\ 347--349]{Miller_72}.
But $\det (C^{1/2} e^{\pi i/ N} X) = - c^{N/2}$ and
$\det (C^{-1/2} Y) = c^{-N/2} = c^{N/2}$, which is a contradiction.
It easily follows, by using Theorem \ref{thmB.2}(c),
that the two isometries are not conjugate one to another.

It is convenient to make now a slightly different choice of representatives
from the two conjugacy classes.  Note that when when $N = 4k$ ($k$ integer),
the isometry $A\to J\bar A J$ is conjugate to $A\to - J\bar A J$
[take $X = i I$ and $Y = I$ in Theorem \ref{thmB.2}(c)(i)],
while for $N = 4k + 2$ the isometry $A\to e^{2\pi i/ N} J\bar A J$
is conjugate to $A\to - J\bar A J$
[take $X = e^{2\pi i(N-2) / 4N} I$, $Y = I$].
Thus, in either case we can choose $A \to -J \bar{A} J$ as one representative,
while for $N=4k$ (resp.\ $N=4k+2$) we take $A\to e^{2\pi i/ N} J\bar A J$
(resp.\ $A\to J\bar A J$) as the other.

Finally, we prove that isometry (c1) is not conjugate to (c2a), (c2b) or (c2c).
By Theorem \ref{thmB.2}(c), such a conjugacy would imply the existence of
$X \in SU(N)$ and $C \in \scrc$ such that $X J X^T = C$.
But then $X' \equiv C^{-1/2} X \in U(N)$ would satisfy
$X' J {X'}^T = I$, or equivalently $X' J = \overline{X'}$.
Taking complex conjugates, we obtain $\overline{X'} J = X'$
(since $J$ is real).  Then $X' J^2 = \overline{X'} J = X'$,
which implies that $J^2 = I$.
But in fact $J^2 = -I$, so we have a contradiction.

\medskip

{\em Class (d):}\/  The map $A \longto UA^T V$ is involutive iff
$V = \pm \bar{U}$ (with the $-$ sign allowed only if $N$ is even).
By Theorem \ref{thmB.2}(d)(i) with $X=U^{-1}$ and $Y=I$, this map
$A \longto \pm U A^T \bar{U}$ is conjugate to $A \longto \pm A^T$.
Finally, it follows from Theorem \ref{thmB.2}(d) that if $N$ is even,
these two alternatives are not conjugate one to another
[if $X \bar{Y} = C \in \scrc$,
then $\bar{X} Y = \bar{C} = C^{-1} \neq -C^{-1}$].
\qed

\medskip

\begin{lemma}
 \label{lemma_RS}
Let $A_1,\ldots,A_k$ be commuting real antisymmetric $N \times N$ matrices.
Then there exists a real rotation matrix $L \in SO(N)$ such that
for each $i = 1,\ldots,k$,
\be
   L^T A_i L   \;=\;
      \cases{ {\rm blockdiag} (\lambda_{i,1} {\bf J}, \ldots,
                               \lambda_{i,N/2} {\bf J}       )
              & if $N$ is even \cr
              {\rm blockdiag} (\lambda_{i,1} {\bf J}, \ldots,
                               \lambda_{i,(N-1)/2} {\bf J}, 0)
              & if $N$ is odd \cr
            }
\ee
where the $\lambda_{i,j}$ are real numbers.
\end{lemma}

\proof
Since $A_1,\ldots,A_k$ are commuting anti-hermitian matrices,
they have a common eigenvector $v \in \C^N$:
\be
   A_i v   \;=\;  i \lambda_i v  \qquad \hbox{for each } i  \;,
\ee
where the $\lambda_i$ are real.
Since the $A_i$ are real, we have also
\be
   A_i \bar{v}   \;=\;  -i \lambda_i \bar{v}  \qquad \hbox{for each } i   \;.
\ee
There are now two cases:

(a)  If the real and imaginary parts of $v$ are linearly independent
(i.e. $v$ is not a multiple of a real vector),
then $w_+  \equiv  (v + \bar{v})/ \| v + \bar{v} \|$
and $w_-  \equiv  (v - \bar{v})/ i\| v - \bar{v} \|$
are perpendicular unit vectors in $\R^N$, and
\begin{subeqnarray}
 \label{wplusminus}
   A_i w_+   & = &   -\lambda_i w_-   \\
   A_i w_-   & = &   \lambda_i w_+
\end{subeqnarray}
[If at least one of the $\lambda_i$ is nonzero, then $v \perp \bar{v}$
and hence $\| v + \bar{v} \| = \| v - \bar{v} \|$,
so the normalizations work out right.
If all of the $\lambda_i$ are zero, then \reff{wplusminus} is trivial.]
Moreover, the subspace of $\R^N$ orthogonal to $\{ w_+, w_- \}$
is invariant under each operator $A_i$.

(b)  If $v$ is a multiple of a real vector $w$
(which we take to be of unit norm),
then all the $\lambda_i$ must be zero.
Moreover, the subspace of $\R^N$ orthogonal to $w$
is invariant under each operator $A_i$.

We now continue the same process of reduction, working on the operators
$A_i$ restricted to $\{ w_+, w_- \} ^\perp$ or $\{w\} ^\perp$.
In this way we produce an orthonormal basis $\{ w_1,\ldots,w_N \} \in \R^N$
consisting of all the vectors $w_\pm$ or $w$ generated at each step;
we order this basis so as to put first all pairs $w_\pm$,
then all individual vectors $w$.
The columns of the desired matrix $L$ are given by this basis.
(If necessary, $w_1$ and $w_2$ can be interchanged so as to guarantee
$\det L = +1$.)
\qed

\bigskip

Finally we determine the fixed-point manifolds for each of the involutive
isometries listed in Theorem \ref{thmB.4}:

\begin{theorem} \label{thmB.5}
The fixed-point manifolds $F$ of the involutive isometries listed in
Theorem \ref{thmB.4} are:
\begin{itemize}
 \item[(a${}_{r,r}$)]  $A \longto I_r A I_r$:
   $F = $ matrices of the form
   $\left(\!\! \begin{array}{cc} B & 0 \\ 0 & C \end{array} \!\!\right)$
   with $B \in U(r)$, $C \in U(N-r)$ and $(\det B)(\det C) = 1$.
   This is the connected symmetric space $S( U(r) \times U(N-r) )$;
   it is a subgroup of $SU(N)$ of codimension $2r(N-r)$.
 \item[(a${}_{r,s}$)]  $A \longto I_r A I_s$ with $r \neq s$:
   $F = \emptyset$.
 \item[(b1)]  $A \longto A^\dagger$:
   $F$ is a disjoint union of components
   $F_r = \{ U I_r U^\dagger \colon\;  U \in SU(N) \}$
   for $r$ even ($0 \le r \le N$).
   The manifold $F_r$ is the symmetric space $SU(N)/S(U(r)\times U(N-r))$;
   it has codimension $r^2 + (N-r)^2 - 1$.
 \item[(b2)]  [$N$ even]  $A \longto e^{2\pi i/N} A^\dagger \;$:
   $F$ is a disjoint union of components
   $F_r = \{ e^{\pi i/N} U I_r U^\dagger \colon\;  U \in SU(N) \}$
   for $r$ odd ($0 \le r \le N$).
   The manifold $F_r$ is isometric to the symmetric space
   $SU(N)/S(U(r)\times U(N-r))$;
   it has codimension $r^2 + (N-r)^2 - 1$.
 \item[(c1)]  $A \longto \bar{A}$: $F$ consists of real matrices $A \in SO(N)$.
   This is a connected symmetric space;
   it is a subgroup of $SU(N)$ of codimension $\half (N^2 + N -2)$.
 \item[(c2a)] [$N$ even]  $A \longto -J \bar{A} J$:
   $F = \USp(N/2) \equiv Sp(N/2,\C) \cap U(N)$.
   This is a connected symmetric space;
   it is a subgroup of $SU(N)$ of codimension $\half (N^2-N-2)$.
 \item[(c2b)] [$N = 4k$, $k$ integer]  $A\longto e^{2\pi i/ N} J\bar A J$:
   $F = \emptyset$.
 \item[(c2c)] [$N = 4k+2$, $k$ integer]  $A \longto J \bar{A} J$:
   $F = \emptyset$.
 \item[(d1)]  $A \longto A^T$:
   $F = \{ U U^T \colon\;  U \in SU(N) \}$.
   The manifold $F$ is the connected symmetric space $SU(N)/SO(N)$;
   it has codimension $\half (N^2 + N)$.
 \item[(d2)]  [$N$ even]  $A \longto -A^T$:
   $F$ is a disjoint union of two components $F_1$ and $F_2$ with
   $F_1 = \{ U J U^T \colon\;  U \in SU(N) \}$ and
   $F_2 = \{ e^{2\pi i/N} U J U^T \colon\;  U \in SU(N) \}$.
   The manifold $F_1$ is the connected symmetric space
   $SU(N)/\USp(N/2)$; it has codimension $\half (N^2-N)$.
   The manifold $F_2$ is isometric to $F_1$.
\end{itemize}

\end{theorem}

\proof
{\em Class (a${}_{r,r}$):}\/  This is immediate.
\medskip

{\em Class (a${}_{r,s}$):}\/  If $A = I_r A I_s$ with $A$ invertible,
then $I_r = A I_s A^{-1}$, i.e.\ $I_r$ is similar to $I_s$.
But this is of course false if $r \neq s$.
\medskip

{\em Class (b1):}\/  If $A = A^\dagger$, then $A$ has eigenvalues $\pm 1$
and it can be diagonalized in the form
$A = UI_rU^\dagger$ with $U \in SU(N)$;
moreover, $r$ is even because $A \in SU(N)$.
Conversely, every matrix of this form is in $SU(N)$ and satisfies
$A = A^\dagger$.  Obviously these manifolds are disjoint,
as $I_r$ cannot be similar to $I_{s}$ if $r\not= s$.

Let us now consider $U,V \in SU(N)$ such that $U I_r U^\dagger =
V I_r V^\dagger$. Then $U^\dagger V I_r = I_r U^\dagger V$, i.e.\
$U^\dagger V$ commutes with $I_r$ and thus belongs to $S(U(r)\times U(N-r))$.
Thus $V = UK$ with $K\in S(U(r)\times U(N-r))$.
Conversely, if $V = UK$ with $K\in S(U(r)\times U(N-r))$,
then $U I_r U^\dagger = V I_r V^\dagger$.   Therefore, the manifold $F_r$
is in one-to-one correspondence with the cosets
$SU(N)/S(U(r)\times U(N-r))$.
\medskip

{\em Class (b2):}\/ If $A = e^{2\pi i/N} A^\dagger$, then $A e^{-\pi i/N}$
is a hermitian unitary matrix. One can then apply the same reasoning
used for class (b1), but $r$ must be odd rather than even.
\medskip

{\em Class (c1):}\/  This is immediate.
\medskip

{\em Class (c2a):}\/  For $A \in SU(N) \subset U(N)$,
clearly $A = - J \bar{A} J$ iff $A^T J A = A^T \bar{A} J = J$,
i.e.\ iff $A \in \USp(N/2)$.
In \cite[pp.\ 347--349]{Miller_72} it is proven that
$\USp(N/2) \subset SU(N)$ and that $\USp(N/2)$ is connected.
\medskip

{\em Classes (c2b) and (c2c):}\/  If $A = c J \bar{A} J$ and $A \in U(N)$,
then $A^T J A = -c A^T \bar{A} J = -cJ$.
Therefore $\sqrt{-c}A$ is a symplectic matrix, and thus
\cite[pp.\ 347--349]{Miller_72} $\det(\sqrt{-c}A) = +1$.
But for $A \in SU(N)$ with $N$ even, this means that $(-c)^{N/2} = +1$,
which is a contradiction in the two cases ($N = 4k$, $c = e^{2\pi i/ N}$)
and ($N = 4k+2$, $c=1$).
\medskip

{\em Class (d1):}\/  If $A=A^T \in SU(N)$, then [see the proof of case (c1),
Theorem \ref{thmB.4}]  $A=U U^T$ with $U\in SU(N)$.
Let us now consider $U,V \in SU(N)$ such that $U U^T = V V^T$.
Then $U^\dagger V = \overline{ U^\dagger V}$, i.e.\
$U^\dagger V$ is real and thus belongs to $SO(N)$.
Thus $V = UK$ with $K\in SO(N)$.
Conversely, if $V = UK$ with $K\in SO(N)$, then $U U^T = V V^T$.
Therefore, the manifold $F$ is in one-to-one correspondence with the cosets
$SU(N)/SO(N)$.
\medskip

{\em Class (d2):}\/  If $A= - A^T$, then [see the proof of case (c2),
Theorem \ref{thmB.4}]  $A= U J U^T$ or $A = e^{2\pi i /N} U J U^T$
with $U\in SU(N)$, and only one of these two cases holds for any given $A$.

Let us now consider $U,V \in SU(N)$ such that $U J U^T = V J V^T$.
Then $U^\dagger V J ( U^\dagger V)^T = J$, i.e.\
$U^\dagger V$ belongs to $\USp(N/2)$.
Thus $V = UK$ with $K\in \USp(N/2)$, and the converse also holds.
Therefore, the manifold $F_1$ is in one-to-one correspondence with the cosets
$SU(N)/\USp(N/2)$.
Clearly $F_2$ is isometric to $F_1$.

\qed

%
%

\clearpage
\end{document}